\documentclass[numberedappendix]{emulateapj}
\usepackage{amsmath,amssymb}
\usepackage{float}
\usepackage{booktabs}
\usepackage{color}
\usepackage{natbib}
\usepackage{graphicx}
\usepackage{textcomp}
\newcommand{\angstrom}{\mbox{\normalfont\AA}}

\newcommand{\ergs}{erg s\ensuremath{^{-1}}}

\def\gtrsim{\mathrel{\hbox{\rlap{\hbox{\lower4pt\hbox{$\sim$}}}\hbox{\raise2pt\hbox{$>$}}}}}

\newcommand{\hbeta}{H\ensuremath{\beta}}

\newcommand{\kms}{km~s\ensuremath{^{-1}}}

\newcommand{\lbol}{\ensuremath{L_{\mathrm{bol}}}}
\newcommand{\loiii}{\ensuremath{L_{\mathrm{[O{\tiny III}]}}}}

\newcommand{\lhb}{\ensuremath{L_{\mathrm{H{\beta}}}}}

\newcommand{\msun}{\ensuremath{M_{\odot}}}

\newcommand{\sii}{[\ion{S}{2}]}

\newcommand{\oiii}{[O~{\small III}]}
\newcommand{\OIII}{[O~{\small III}]}

\def\lax{{$\mathrel{\hbox{\rlap{\hbox{\lower4pt\hbox{$\sim$}}}\hbox{$<$}}}$}}
\def\gax{{$\mathrel{\hbox{\rlap{\hbox{\lower4pt\hbox{$\sim$}}}\hbox{$>$}}}$}}

\newcommand{\rv}{$R_{\mathrm{KDR}}$}
\newcommand{\riso}{$R_{\mathrm{NLR}}$}
\newcommand{\weighty}{$w_{80}$}
\newcommand{\weightyAVG}{$w_{80,\rm{AVG}}$}
\newcommand{\vmed}{$v_{\mathrm{med}}$}
\newcommand{\tdyn}{$t_{\mathrm{dyn}}$}
\newcommand{\Ekin}{$E_{\mathrm{kin}}$}
\newcommand{\dotEkin}{$\dot{E}_{\mathrm{kin}}$}
\newcommand{\dotEST}{$\dot{E}_{\mathrm{ST}}$}
\newcommand{\oiiil}{\OIII$\lambda$5007}
\newcommand{\oiiill}{\OIII$\lambda$4959}
\newcommand{\Leight}{$\nu L_{\nu,8}$}
\newcommand{\Lfifteen}{$\nu L_{\nu,15}$}
\newcommand{\Ltwentytwo}{$\nu L_{\nu,22}$}

\shorttitle{ENLR in Obscured AGN}
\shortauthors{A.-L. Sun et al.}

\begin{document}

\title{Sizes and Kinematics of Extended Narrow-Line Regions in Luminous Obscured AGN Selected By Broadband Images}

\author{Ai-Lei Sun\altaffilmark{1}\altaffilmark{*}, Jenny E. Greene\altaffilmark{1}, Nadia L. 
Zakamska\altaffilmark{2,3}}
\altaffiltext{1}{Department of Astrophysics, Princeton University, Princeton, NJ 08540, USA}
\altaffiltext{2}{Department of Physics and Astronomy, Bloomberg Center, Johns Hopkins University, Baltimore, MD 21218, USA}
\altaffiltext{3}{Institute for Advanced Study, Einstein Dr., Princeton, NJ 08540, USA}
\altaffiltext{*}{E-mail: aisun@astro.princeton.edu}

\begin{abstract}
To study the impact of active galactic nuclei (AGN) feedback on the galactic ISM, we present Magellan long-slit spectroscopy of 12 luminous nearby type 2 AGN (\lbol$\sim 10^{45.0-46.5}$ \ergs, $z\sim0.1$). These objects are selected from a parent sample of spectroscopically identified AGN to have high \oiiil{} and \emph{WISE} mid-IR luminosities and extended emission in the SDSS {\it r}-band images, suggesting the presence of extended \oiiil{} emission. We find spatially resolved \oiii{} emission (2-35 kpc from the nucleus) in 8 out of 12 of these objects. 
Combined with samples of higher luminosity type 2 AGN, we confirm that the size of the narrow-line region (\riso{}) scales with the mid-IR luminosity until the relation flattens at $\sim$ 10 kpc. 
Nine out of 12 objects in our sample have regions with broad \oiii{} linewidths (\weighty{} $>600$ \kms{}), indicating outflows. We define these regions as the kinematically-disturbed region (KDR). The size of the KDR (\rv{}) is typically smaller than \riso{} by few kpc but also correlates strongly with the AGN mid-IR luminosity. 
Given the unknown density in the gas, we derive a wide range in the energy efficiency $\eta=\dot{E}/\lbol=0.01\% - 30\%$. We find no evidence for an AGN luminosity threshold below which outflows are not launched. To explain the sizes, velocity profiles, and high occurrence rates of the outflows in the most luminous AGN, we propose a scenario in which energy-conserving outflows are driven by AGN episodes with $\sim 10^8$-year durations. Within each episode the AGN flickers on shorter timescales, with a cadence of $\sim 10^6$ year active phases separated by $\sim 10^7$ years.

\end{abstract}

\section{Introduction}

Feedback from active galactic nuclei (AGN) is a key ingredient in
modern models of galaxy evolution \citep{Silk1998,Springel2005}. It has been invoked to regulate star formation in massive galaxies \citep[e.g.,][]{Croton2006,Bower2006}, while the tight correlation between the
supermassive black hole (SMBH) masses and their host galaxy
properties \citep{Gebhardt2000,Ferrarese2000,Sun2013a,McConnell2013}
also suggests that feedback processes enforce the coevolution
between SMBHs and galaxies
\citep{DiMatteo2005,DeBuhr2010,Somerville2008}.

Supporting evidence for AGN feedback comes from observations of AGN 
outflows in both local and distant AGN.
These galactic outflows have a multi-phase structure, ranging from cold molecular
\citep{Feruglio2010,Sturm2011,Veilleux2013,Sun2014,Cicone2014} to warm
atomic and ionized gas
\citep{Alexander2010,Greene2011,Maiolino2012,Davis2012,
  Rupke2013,Cano-Diaz2012}, and could be related to nuclear X-ray
emitting outflows \citep{Gofford2013,Tombesi2015}. While we now have empirical evidence that AGN do host outflows, many questions remain
about how these outflows are driven, for example by jet, wind, or radiation pressure, and whether and how the outflow properties depend on the AGN luminosity. 

The warm ionized component of the outflow ($T\sim 10^4$ K) emits strong
forbidden emission lines, in particular \oiiil{}, which makes it
possible to detect and resolve AGN outflows via optical spectroscopy particularly at low-redshifts. 
At redshifts $\lesssim 0.5$, 
high velocity \oiii{} features indicative of outflows are commonly found in luminous AGN using spectroscopic surveys \citep[e.g.,][]{Greene2005,Mullaney2013,Zakamska2014,Woo2016,Harrison2016}.
Spatially resolved studies using long-slit and IFU spectroscopy have also identified a number of extended ionized outflows (few - 10 kpc) in luminous AGN (\lbol{}$\gtrsim 10^{46}$ \ergs) at these redshifts
\citep{Greene2012,Liu2013b,Liu2013,Liu2014a,Harrison2014,Hainline2014d},
particularly among obscured type 2 AGN \citep[][]{Zakamska2003,Reyes2008},
where the occultation of the active nucleus makes it easier to detect emission
lines from the extended ionized nebula. 

There are other studies that find outflows of much smaller sizes ($\lesssim 2$ kpc) and lower occurrence rates in samples with a wider range of luminosities and a diverse types of AGN \citep[e.g., ULIRG/Seyfert, type 1 and 2 AGN,][]{RodriguezZaurin2013,Husemann2013,Husemann2015,Karouzos2016}. 
The size of the outflow could indeed be strongly dependent on the luminosity of the AGN \citep{Karouzos2016}. Furthermore, these studies do not use uniform definitions of size. Some are based on intensities while others are based on kinematics. 
To understand the discrepancy between these results and to have a comprehensive picture of AGN outflow sizes, it is important to measure the AGN luminosity and to have a quantitative definition of the outflow size that reflects the extent of the kinematically disturbed region (KDR).

Compared to spatially resolved spectroscopy, broadband photometry could provide a much more efficient way to search for candidate extended outflows. 
Since the \oiiil{} line in obscured luminous AGN is bright and has high enough EW to be 
detectable in broadband images, optical photometric surveys, such as the 
Sloan Digital Sky Survey \citep[SDSS;][]{York2000},
have been used to find \oiiil{} emission in extended narrow-line regions \citep[e.g., ][]{Keel2012,Schirmer2013,Davies2015}. 
However, not all the extended narrow-line regions have disturbed kinematics. 
Some luminous AGN are capable of ionizing gas out to tens of kpc from the
host galaxy \citep{Fu2008, Villar-Martin2010}, including gas in small
companion galaxies and tidal debris left from a prior galaxy interaction, 
thus creating extended ionized regions that are kinematically quiescent. 
For this reason, we also need spectroscopy to confirm the kinematic state of 
the extended gas and identify outflows. 

To test if broadband images can help identify extended outflows, 
in this paper we select a sample of 12 SDSS-identified luminous obscured 
(type 2) AGN based on their extended emission in the broad band images. 
We observe them with Magellan IMACS long-slit spectroscopy to measure 
the extent and kinematic state of the ionized gas. 
We study the outflow occurrence rate, and constrain the outflow properties, 
including the sizes, velocities, and energetics, as well as their dependence 
on the AGN luminosity. In future work we will examine the correspondence
between the broadband images and the ionized gas nebula, and evaluate
the performance of the extended outflow selection.

In Section \ref{sec:data}, we describe the sample selection and
Magellan observations; in Section \ref{sec:sizes} we present the
Magellan spectra and measure the extents and the kinematics of the
ionized gas nebulae, and in Section \ref{sec:energetics} we infer the 
outflow properties, including the energetics, and analyze their dependence 
on the AGN luminosities.  We discuss the outflow occurrence rate and time scales in Section \ref{sec:timescale} and summarize in Section \ref{sec:summary}. We
use an $h=0.7, \Omega_m=0.3, \Omega_{\Lambda}=0.7$ cosmology
throughout this paper. We adopt vacuum wavelengths for the
analysis, the same as SDSS, but keep the line notations in air wavelengths, e.g. \oiiil. All error bars represent 1-sigma errors.

\section{Observations and Data Reduction}
\label{sec:data}

\subsection{Sample Selection}
\label{sec:data:selection}

\begin{figure}[H]
        \centering
        \hbox{
	\includegraphics[scale=0.45]{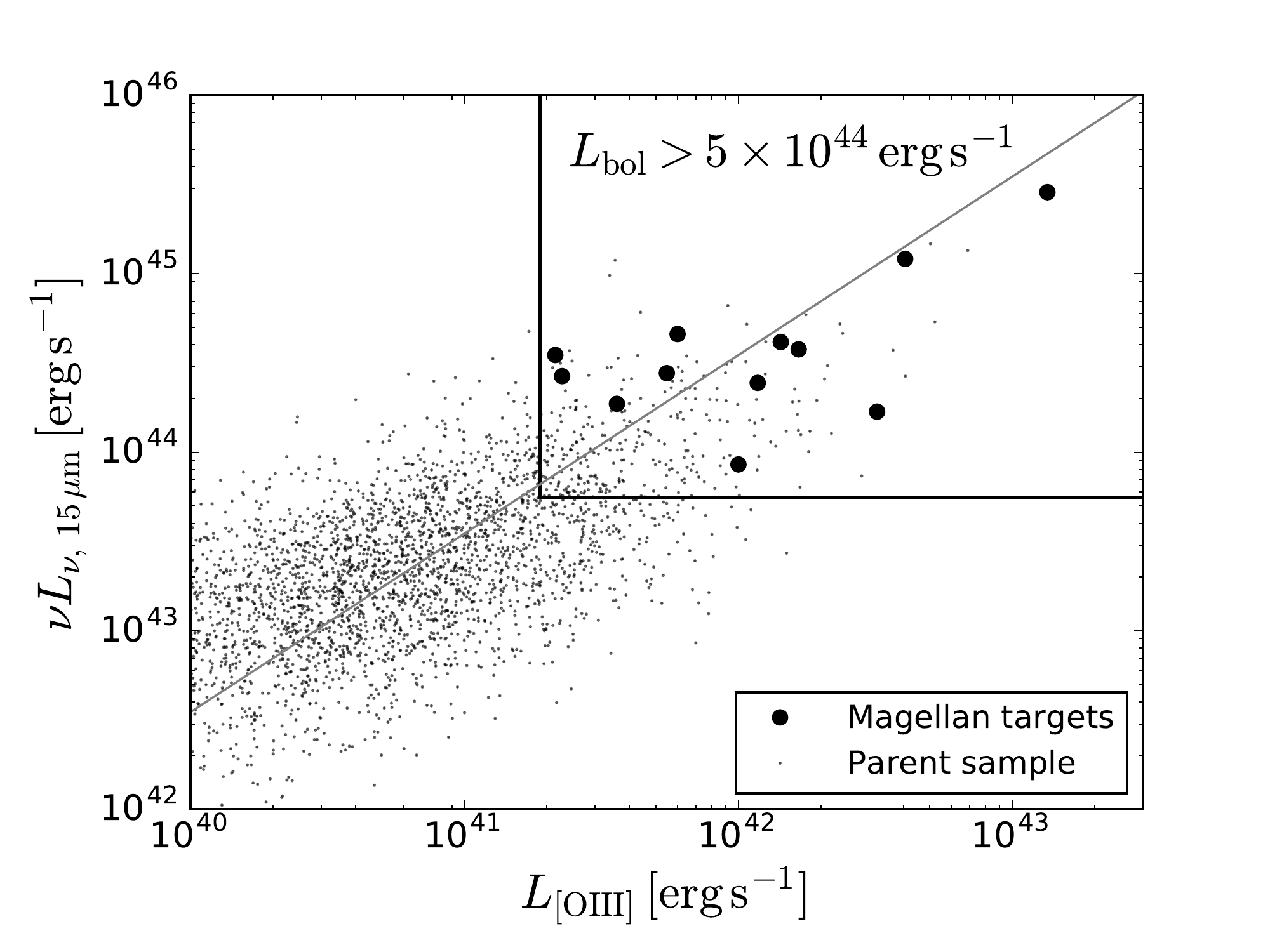}
	}
    \caption{
    Selection of the 12 luminous type 2 AGN studied in this paper (big dots) from the parent sample of SDSS spectroscopically identified type 2 AGN \citep[small dots,][]{Mullaney2013}. 
    Two luminosity indicators are used -- \oiiil{} luminosities from the SDSS spectra and mid-IR rest frame 15 \micron{} luminosities from {\emph WISE} (Sec. \ref{sec:data:WISE}). The gray line shows where the two indicators infer the same bolometric luminosity. The selected objects have bolometric luminosity higher than $5\times10^{44}$ \ergs{} as shown by the box on the upper-right. 
    The two most luminous objects are SDSS J1000+1242 and SDSS J1010+1413 shown in Fig. \ref{fig:spec2100} and \ref{fig:spec2101}.
    All the objects shown are at low-redshifts (z$<0.2$) and in the observable sky during the Magellan run (RA $<2$h or $>10$h, Dec $<15\deg$). 
    }
        \label{fig:selection}
\end{figure}

We select luminous AGN from the parent sample of SDSS spectroscopically identified AGN \citep{Mullaney2013} with $z<0.2$ and AGN luminosities above $L_{\rm{bol}} >5\times10^{44}~\rm{erg\/~s^{-1}}$ (Fig. \ref{fig:selection}). 
The AGN bolometric luminosity is inferred from two luminosity indicators --
the \oiii{} luminosity and the mid-infrared (mid-IR) luminosity (see Sec. \ref{sec:data:WISE}). 
We calculate the \oiiil{} luminosity as the sum of both kinematic components measured by \citet{Mullaney2013} from the SDSS spectra. 
To avoid introducing uncertainties\footnote{The correlation between the \oiiil{} luminosity and the mid-infrared 15 \micron{} luminosity (see Sec. \ref{sec:data:WISE}) disappears after the extinction correction, indicating that significant uncertainties could be introduced. }, we do not apply the extinction correction to the \oiiil{} luminosity, which has a median of 3 among our sample according to \citet{Mullaney2013}
The \oiiil{} luminosity is converted to the AGN bolometric luminosity with a correction factor of $10^{3.5}$ based on the empirical \loiii-\lbol\ relation of type 1 AGN \citep[Eq. 1,][]{Liu2009}. 
The mid-infrared luminosity is from the Wide-field Infrared Survey Explorer \citep[\emph{WISE};][]{Wright2010} and the conversion is described in Sec. \ref{sec:data:WISE}. 
The luminosities from the two indicators are correlated with a scatter of 0.5 dex. 

To maximize the chance of finding extended AGN outflows, we
looked at the SDSS images to identify the ones with extended
morphology. 
As the strong \oiii{} lines fall in the SDSS {\it r}-band, which has a green color in the composite images (Fig. \ref{fig:spec2100}, \ref{fig:spec2101}, and Appendix \ref{sec:append:objs}), we look for extended green-colored emissions in those images. 
In total, twelve type 2 AGN (narrow-lines only) and 
eight type 1 AGN (with nuclear blue continuum and broad Balmer lines) have successful observations with Magellan. 
While the type 1 AGN could be analyzed using methods that handle the nuclear 
emission \citep[e.g.,][]{Husemann2015}, it is beyond the scope of this paper. 
In this paper, we will focus on the sample of twelve type 2 AGN 
(Tab. \ref{tab:sample}), where the \oiiil{} line measurement is less affected
by the bright nuclei.

\subsection{Magellan Long-Slit Observations}

Our sample was observed with the Inamori-Magellan Areal Camera \&
Spectrograph (IMACS) spectrograph \citep{Dressler2011} at the
Magellan Baade telescope on Las Campanas on 23-24 June 2014. 
The seeing was between 0\farcs5 and 1\arcsec. 
We used
the Centerfield Slit-viewing mode with the 300 lines/mm grating on the f/4
camera.  We placed objects on the adjacent 1.0\arcsec{} and 1.3\arcsec{} 
slits\footnote{This widest 1\farcs3 slit, referred to in the IMACS
User Manual as the 1\farcs5 slit, was confirmed to have an actual
slit width of 1\farcs3, see Appendix \ref{sec:append:slit}.}, each about 17\arcsec{} long and separated by
1\arcsec, to simultaneously cover the central and extended regions
of our galaxies. 
The spectral resolutions are 5.1 and 6.7 $\angstrom$
(FWHM) for the two slits respectively, which corresponds to about 260
and 340 \kms{} for the \oiiil{} line measurements.  The
0\farcs75 slit is also used for background subtraction, but not for
measurements.  The wavelength coverage is 3800 to 9400 $\angstrom$
with three CCD chip gaps, each 75 $\angstrom$ wide. Each object is
observed for 15 to 60 minutes with one to three slit positions, as
listed in Table \ref{tab:sample}. The slit positions are chosen based
on the SDSS image to cover extended {\it r}-band emission. For each
object, there is at least one slit position along the major axis. 
The atmospheric dispersion corrector is used. 
Two flux calibrator
stars, Feige 110 and EG 274, and a set of velocity template stars
consisting of K to A giants/dwarfs are also observed with the 1\farcs3
slit. 

\subsection{Data Reduction}
\label{sec:data:reduction}

Basic data reduction, including bias subtraction, flat fielding,
wavelength calibration, rectification, and 2-D sky subtraction
\citep{Kelson2003} are performed using the Carnegie Observatories
reduction package
COSMOS\footnote{http://code.obs.carnegiescience.edu/cosmos}.  Cosmic
ray removal using LACosmic\footnote{http://www.astro.yale.edu/dokkum/lacosmic/} \citep{VanDokkum2001} is applied before
rectification.  We found an excess of red continuum background at
$\lambda >$ 8200 $\angstrom$ that was independent of slit width, which is most
likely due to scattered light in the spectrograph. This red background
excess can be well-subtracted by a 2-D sky subtraction if there are
emission-free regions on both sides of each slit.  In cases where
one slit is full of galaxy light, we subtract the
background by inferring the sky spectrum from the convolved 0\farcs75
slit and correcting for the red background excess. This excess background does not affect the \oiiil{} and \hbeta{} line measurements. 

The flux calibration and atmospheric extinction corrections are
performed using PyRAF\footnote{http://www.stsci.edu/institute/software$\_$hardware/pyraf} version 2.1.7.
We use the flux standard stars to determine the sensitivity functions
and the atmospheric extinction function.  
The calibrated fluxes are consistent with the SDSS spectra within a scatter of 20\%, taking into account that the SDSS and Magellan apertures are different\footnote{Because the SDSS fibers (3$\arcsec$) are wider than the Magellan slits (1\arcsec{} or 1\farcs3), the SDSS fluxes are higher than the Magellan fluxes by a factor of 1.7.}.
We adopt a fractional uncertainty on the flux calibration of
20\%. For the slit positions that have multiple exposures, we align and 
stack those spectra of the same position together
(the total exposure time is listed in Tab. \ref{tab:sample}).
The wavelength solution 
is applied after heliocentric-correction and air-to-vacuum conversion using PyAstronomy\footnote{https://github.com/sczesla/PyAstronomy}.

For the emission line measurements we subtract the stellar continuum
using a featureless 2-D model for the continuum spectrum. This model
is determined by smoothing and interpolating the line-free part of the
stacked 2-D spectra, excluding the contamination from the AGN emission lines
and sky lines. This method can operate at the outskirts of the galaxies
where the signal-to-noise ratio is low.
As the \hbeta{} emission line is affected by the stellar absorption, we
correct for this effect using the absorption line profiles obtained from the
pPXF stellar population synthesis fits described in Sec. \ref{sec:data:center}. 
The average \hbeta{} absorption correction is 12\%. 
Therefore the dominant uncertainty on \lhb{} is the flux calibration uncertainty of 20\%. 
Two systems have no \hbeta{} measurements (SDSS J0141$-$0945 and J2133$-$0712) due to chip gaps and strong sky lines.

\subsection{Position and Velocity References}
\label{sec:data:center}

The position and velocity measurements in this paper are defined
relative to the stellar component of the galaxies. The center position
is defined as the peak of the stellar continuum light profile
(nucleus), which has an uncertainty comparable to one pixel
(0.2\arcsec, or 0.3-0.6 kpc in our sample).

The systemic velocity of each galaxy is determined using the stellar
absorption features of the nuclear spectrum. 
To focus on the stellar absorption features, the emission lines, sky lines, galactic absorption, and chip gaps in the spectra are masked before the fitting. 
We fit the absorption lines with single stellar
population (SSP) templates from \citet{Bruzual2003}
using the stellar kinematics fitting code pPXF \citep{Cappellari2003}.
The templates include 10 solar-metallicity SSP spectra of ages
ranging from 5 Myr to 11 Gyr with a two degree additive and three
degree multiplicative polynomial. 
Two aperture sizes, 1\arcsec{} and 3\arcsec{}, are used to extract the nucleus spectra, which give consistent systemic velocities within $\sim 15$ \kms{}. 
The final systemic redshifts are taken as the average of the two apertures and are listed in Table \ref{tab:sample}. 

We adopt an uncertainty of 15 \kms{} on the systemic velocity\footnote{We run a Monte Carlo simulation and find that the root-mean-square uncertainty on the systemic velocity is $\sim$ 15 \kms{} for a Gaussian line with a dispersion of $\sigma=200$ \kms{} and a signal-to-noise ratio of 10. }. The average stellar velocity dispersion is 200 \kms{} and each object has a few absorption lines with signal-to-noise ratios $\gtrsim 10$.
While this systemic velocity is not used to measure the \oiiil{} linewidth \weighty{} in this paper (see Sec. \ref{sec:sizes:spec}), it is used as a reference to determine the velocity threshold for the high velocity emission (the blue and the red wings beyond $\pm 300$ \kms{}, Sec. \ref{sec:sizes:resolve}), which is used to measure the extent of the kinematically disturbed region \rv{} together with the \weighty{} linewidth profile (Sec. \ref{sec:sizes:rv}). Compared to the uncertainties due to the spatial PSF, the uncertainty in the systemic velocity is not the dominant source of error for \rv{}. 
Our redshifts agree with the SDSS redshifts within 285 \kms{} with an average discrepancy of $\langle{} |z-z_{SDSS}| \rangle{} = $95 \kms{}, while the latter is fitted to both the emission and the absorption lines. 

\subsection{AGN Luminosities from \emph{WISE}}
\label{sec:data:WISE}

In this paper, mid-infrared \emph{WISE} luminosity at rest-frame 15 \micron{} is used as the primary AGN luminosity indicator.
Mid-infrared luminosity traces hot dust heated by the AGN and has been found to correlate with the AGN hard X-ray luminosities \citep[e.g.,][]{Lutz2004,Matsuta2012}, which is presumably an isotropic AGN luminosity indicator (although see below). 

As mid-IR luminosity is independent of the properties of the narrow-line region and is presumably more robust against dust extinction compared to optical lines, it is commonly used in studies of the AGN narrow line regions and outflows \citep[e.g.][]{Hainline2014,Liu2013b,Liu2013}. 
The mid-IR \emph{WISE} luminosities correlate with the \oiiil{} luminosities among  type 2 AGN \citep{Rosario2013,Zakamska2014}, see also Sec. \ref{sec:data:selection} and Fig. \ref{fig:selection}.

We expect the mid-infrared luminosity of our sample to be AGN-dominated as opposed to star-formation dominated. AGN heated dust is much hotter ($T \gtrsim 100 $ K) and its emission peaks at shorter wavelengths ($\lambda_{\rm{peak}} \lesssim 30$ \micron{}) than dust heated by stars ($T \sim 25 $ K, $\lambda_{\rm{peak}} \sim 100$ \micron{}) \citep{Richards2006,Kirkpatrick2012,Zakamska2016,Sun2014}. In fact, most of our objects have AGN-like \emph{WISE} colors W1 $\langle 3.4 \rangle -$ W2 $\langle 4.6 \rangle >$ 0.8 in Vega magnitude \citep[][]{Stern2012}, 
indicating AGN-dominated luminosities. The two exceptions are SDSS J1419+0139 and SDSS J0141$-$0945 with only slightly bluer W1 $-$ W2 colors of 0.656 and 0.526, respectively.\footnote{The mid-infrared luminosities of these two objects are likely to be AGN-dominated as well. Their blue W1 $-$ W2 colors can come from the Rayleigh–Jeans tail of the old stellar population, while their W3 $-$ W4 colors are redder. Their high rest-frame 24 \micron{} luminosities require much higher star formation rates ($\sim$ 90 and 166 \msun{}/yr \citep{Rieke2009}) than typically seen in luminous type 2 AGN \citep[$< 18$ \msun{}/yr][]{Zakamska2016}. In the case of SDSS J1419+0139, it is also higher than the star formation rate of 55 \msun{}/yr inferred from the IRAS 60 and 100 \micron{} luminosities \citep{Kennicutt1998,Solomon1997}, which should be taken as an upper limit because AGN heated dust could also contribute to the 60 and 100 \micron{} fluxes. 
}
Therefore, the mid-infrared luminosities of our sample should be AGN-dominated and not significantly affected by star formation. 

In the mid-infrared, type 2 AGN are found to be redder and less luminous than their type 1 counterparts with the same \loiii{} luminosities \citep{Liu2013,Zakamska2016}, indicating that the mid-infrared may not be a perfectly isotropic indicator of the AGN bolometric luminosity. As discussed in Appendix \ref{sec:append:WISE}, we find that this discrepancy is more severe at shorter wavelengths, e.g., 8 \micron{}, than at longer wavelengths, e.g., 15 or 22 \micron{}.

To compare with other studies targeting higher redshift type 2 AGN \citep[z $\approx$ 0.5][]{Liu2013b, Liu2014a, Hainline2014d}, where the 
rest-frame 22 \micron{} flux is not available, we use the rest-frame 15
\micron{} luminosity as our AGN luminosity indicator with a bolometric correction of 9 \citep{Richards2006}, see Tab. \ref{tab:sample} and \ref{tab:sample_liu13}. 

These mid-infrared luminosities at rest-frame 8, 15, and 22 \micron{} are referred to as \Leight{}, \Lfifteen{}, and \Ltwentytwo{}. They are 
interpolated or extrapolated from the \emph{ALLWISE}
source catalog 3-band photometry at 4.6 (W2), 12 (W3), and 22 (W4)
\micron{} using a second-order spline in log-log space. 
 We ignore the
filter response function and adopt a magnitude to flux density
conversion of a flat spectrum, which may lead to a few percent error
depending on the source spectral shape \citep{Cutri2013}.

\section{Sizes of the Narrow-Line and the Kinematically Disturbed Regions}
\label{sec:sizes}

The goal of this section is to quantify the extent of the AGN influence on the interstellar medium
of the host galaxy. To evaluate the ionization state of the gas, it is important to measure the extent of the photo-ionized region, also called the narrow-line region (NLR; its radius \riso{}, see Sec. \ref{sec:sizes:riso})
. One possibility is to measure 
the extent of a bright emission line (e.g., \oiiil{}) down to a fixed surface brightness level. 
Another is to measure the extent of the AGN-ionized region based on ionization diagnostics, e.g., 
using the \oiiil{} to \hbeta{} ratio. 
These options have been used extensively in long-slit or IFU spectroscopy
\citep[e.g.,][]{Bennert2006,Fraquelli2003,Greene2011,Liu2013b,Hainline2013,Hainline2014d,Husemann2014},
narrow-band imaging \citep[e.g.,][]{Bennert2002,Schmitt2003}, or even
broad-band imaging studies \citep[e.g.,][]{Schirmer2013}.  

However, to understand whether the energy from
the AGN can be coupled kinematically to the interstellar medium or even drive
outflows, we need a kinematic measure of the extent of the AGN
influence. We define a kinematically disturbed region (KDR), where the ionized gas is kinematically disturbed, and a corresponding radius \rv{} ( see Sec. \ref{sec:sizes:rv}).
Together, \riso{} and \rv{} quantify the extent of the AGN influence on its host
galaxy through two different channels: photoionization and mechanical
feedback respectively.  It is important to investigate how these two radii
relate to each other and how they depend on the AGN luminosity.

In this section, we present \oiiil{} spectra of the twelve
type 2 AGN (Sec. \ref{sec:sizes:spec}), and measure both the 
narrow-line region radius \riso{} and the kinematically disturbed region radius
\rv{} (Sec. \ref{sec:sizes:riso} and \ref{sec:sizes:rv}). With these
radii, we can revisit the narrow-line region size-luminosity relation
with our lower luminosity objects, and explore the kinematic
size-luminosity relation
(Sec. \ref{sec:sizes:relation}). The kinematic size \rv\ will also be
used to estimate the outflow energetics
(Sec. \ref{sec:energetics:definition}) and to study the relationship between
AGN luminosity and outflow properties
(Sec. \ref{sec:energetics:regress}).

\subsection{Surface Brightness and Velocity Profiles of \oiii{}}
\label{sec:sizes:spec}

\begin{figure}[H]
    \centering
    \includegraphics[scale=0.5]{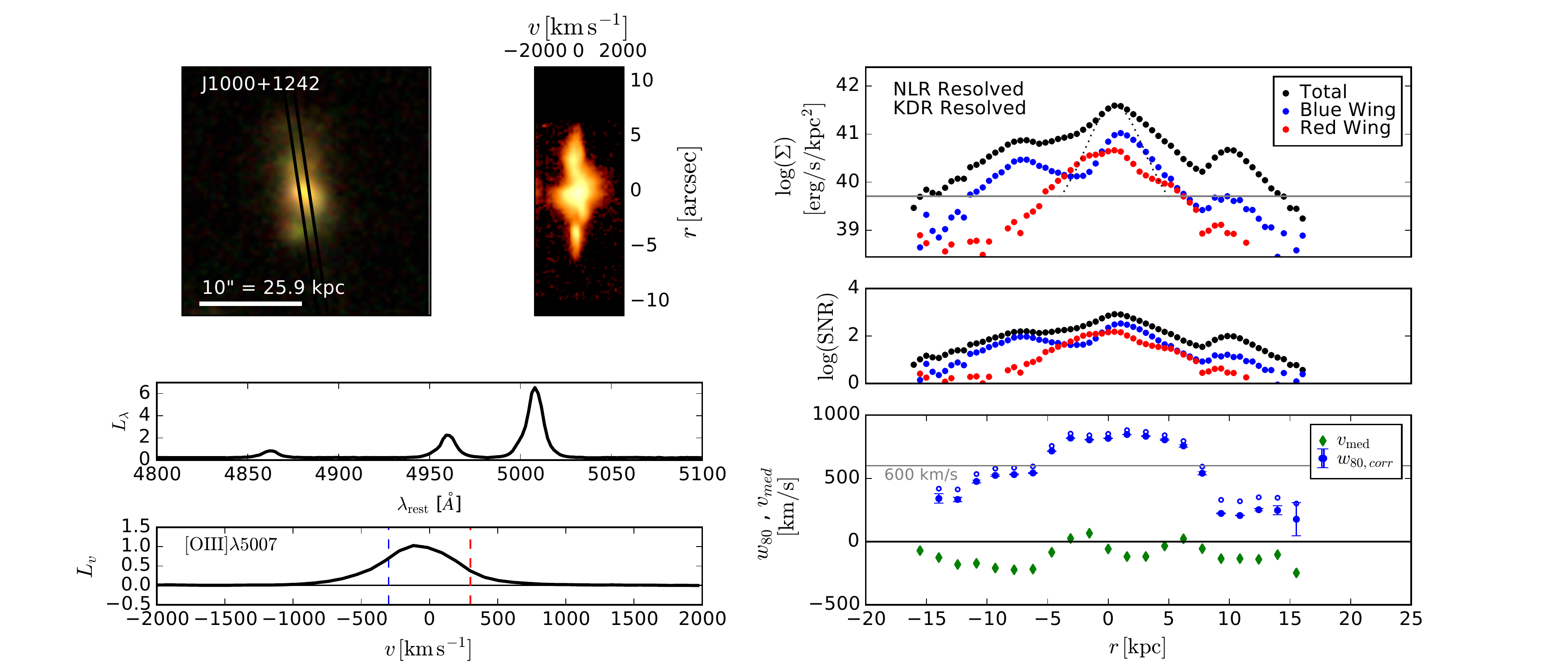}
    \caption{
    The \oiii$\lambda$5007 spectrum and measurements for SDSS J1000+1242.  {\it Top Left:} The SDSS image with the black lines showing the Magellan slit position (left) and the Magellan \oiiil{} 2-D spectrum (right). {\it Bottom Left:} Magellan nuclear spectrum extracted from the central 1\arcsec{} covering the \hbeta{}, \oiiill{}, and \oiiil{} lines in units of $10^{40}$ erg s$^{-1}$ $\angstrom^{-1}$, and the zoomed in \oiiil{} spectrum in units of $\mathrm{10^{40}~erg~s^{-1}~km^{-1}}$. The blue and red dashed lines mark the velocities of $\pm$ 300 \kms. 
    {\it Top Right:} The \oiiil{} surface brightness profile for the entire line (black), the blue wing of the line (blue, $v<-300$ \kms), and the red wing of the line (red, $v>300$ \kms) in units of $\mathrm{10^{41}~erg~s^{-1}~kpc^{-2}}$, overplotted with the scaled PSF (gray dotted line) and the \riso~isophotal cut (gray horizontal line). {\it Middle Right:} The signal-to-noise ratio of the \oiiil{} surface brightness profile. {\it Bottom Right:} The profile of the \oiiil{} line width \weighty{} (blue) and median velocity $v_{med}$ (green) in units of \kms{}, overplotted with the \rv~line width threshold of 600 \kms{} (gray horizontal line). The empty blue circles are the observed \weighty{} and the filled blue circles are the spectral PSF corrected \weighty{}.
    }
    \label{fig:spec2100}
\end{figure}

\begin{figure}[H]
    \centering
    \includegraphics[scale=0.5]{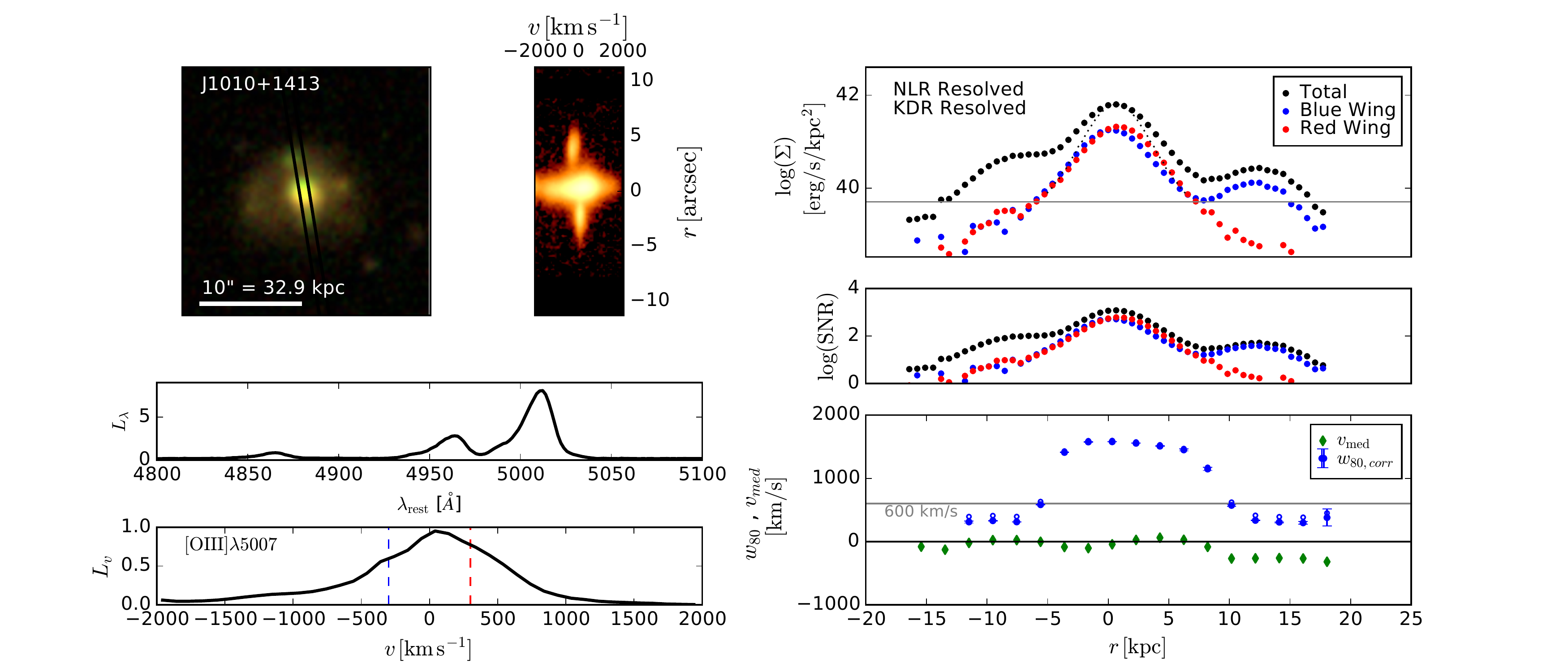}
    \caption{Same as Fig. \ref{fig:spec2100} but for SDSS J1010+1413. 
     }
    \label{fig:spec2101}
\end{figure}

In the upper left panels of Fig. \ref{fig:spec2100}, \ref{fig:spec2101}, and Appendix \ref{sec:append:objs}, we show the Magellan slit positions of the SDSS images and the continuum subtracted two-dimensional \oiiil{} spectra of our objects. 
The lower left panels show the extracted nuclear spectra of the central 1\arcsec{} covering the \hbeta, \oiii$\lambda$4959, and \oiiil{} lines.
Three objects were observed with multiple slit positions (J1055+1102, J1255$-$0339, and J2133$-$0712). For these objects, we choose the slit with the most extended \oiiil{} emission as the representative slit and derive all the measurements from this slit. The size variation between different slit orientations are $\lesssim$ 20\%. 

With the two-dimensional spectra, we can measure the \oiiil{} line
surface brightness profile (upper right panels), which is
integrated within a velocity range of $-2000$ to $2000$ \kms{} to cover
the entire line. The signal-to-noise ratios of these surface brightness measurements are all above 10 and can reach $\sim$10$^3$ at the nucleus (middle right panels). 
We find that the line-emitting gas is mostly AGN-ionized instead of 
star-formation ionized with an \oiiil{} to \hbeta{} line ratio between 3 and 10.
The only exception is part of the nuclear region of SDSS J1255$-$0339, where the ratio is close to two.

In addition to photoionization, we are also interested in the mechanical feedback that can disturb or accelerate the gas, which can be traced by the emission line profiles.  From the two-dimensional spectra, we can measure the line velocity and the line width as a function of position (upper and
middle right panels). 
To avoid the biases introduced by parametric fitting, we calculate the median velocity \vmed{} and the 80 percent linewidth \weighty{} in a non-parametric way. 
We take the cumulative integral of the original spectrum to find its 10th, 50th, and 90th percentile velocity. The integrated spectrum is spline interpolated to avoid discretization.  
The median velocity \vmed{} is the 50th percentile velocity and \weighty{} is the velocity difference between the 10th and 90th percentiles. 
\weighty{} roughly corresponds to the FWHM for Gaussian profiles, but is more sensitive to line wings and therefore suitable to capture high velocity motions \citep{Liu2013,Harrison2014}.  
Both \vmed{} and \weighty{} are measured in each 0.6\arcsec{} bin
to achieve a good signal-to-noise ratio. 

The \weighty{} measurement is used to derive other important quantities in this paper, such as \rv{}. However, it may be biased by the spectral PSF and affected by noise. To quantify these effects, we perform a series of simulations as described in Appendix \ref{sec:append:sim_w80}. Both the PSF biases and the uncertainties due to the noise are $\lesssim$ 10\% for lines wider than \weighty$>600$ \kms{} with a SNR above 30. The noise uncertainties become negligible ($\lesssim 10$ \kms{}) for typical lines with SNR $>100$. We correct for the PSF bias and assign the errors on \weighty{} according to the best fit results from the simulation (shown as the solid blue dots in the lower right panels of Fig. \ref{fig:spec2100}, \ref{fig:spec2101}, and Appendix \ref{sec:append:objs}). This correction has a minimal effect on the results of this paper, except in the case of SDSS J2154+1131, which is ruled out as a kinematically disturbed region after the correction (see Sec. \ref{sec:sizes:rv}). 
The \weighty{} measurement is not the major source of uncertainty for \rv{} and its derived quantities. 
For each object, we calculate \weightyAVG{} as the luminosity weighted quadratic mean of the \weighty{} profile, which is tabulated in Tab. \ref{tab:measurements}. We assign conservative errors of 20 \kms{} on \weightyAVG{} to encompass other unaccounted sources of errors.

\subsection{Spatial Resolution and Size Uncertainties}
\label{sec:sizes:resolve}

To measure the spatial extent of the ionized region and the kinematically disturbed region, it is important to consider the smearing effect of the spatial PSF, which can exaggerate these size measurements depending on the resolution. This effect is especially important when the nucleus outshines the extended emission. 

For type 1 AGN, one way to robustly recover the true size of the extended emission is to measure the point-spread function (PSF) from the broad emission lines and subtract it from the nucleus \citep[e.g,][]{Husemann2013,Husemann2015}. 
After the subtraction, \citet{Husemann2015} reveals that many objects in their type 1 AGN sample still retain their extended high velocity \oiii{} nebula. 
This technique cannot be easily applied to our type 2 sample. 
We instead quantify the effect of the spatial-PSF with a series of 2-D simulations described in Appendix \ref{sec:append:sim_pv}. The simulations consider a range of source kinematic structures, including a high velocity component and disk rotation.  
They also adopt empirical spectral and spatial PSFs measured from the data. 
Based on the simulations, we determine whether the narrow line region or the kinematically disturbed region is spatially resolved, and adopt representative errors on the size measurements.

The \oiiil{} line surface brightness profile is compared to the PSF to determine whether the narrow line region is spatially resolved, see the upper right panels of Fig. \ref{fig:spec2100}, \ref{fig:spec2101}, and Appendix \ref{sec:append:objs}. 
The fiducial PSF is conservatively taken from a flux calibration star with the 
worst seeing (FWHM = 1\farcs0), whereas the median seeing for the targets is only 0\farcs7.  
Four objects that have surface brightness profiles consistent with the PSF are determined to have unresolved NLR (SDSS J1419+0139, J2102$-$0647, J2133$-$0712, and J2142+0001, also see Tab. \ref{tab:measurements}). Their \riso{} are treated as upper limits. 

Even when the total \oiiil{} emission of an object is resolved, its high velocity gas may not be. 
To determine whether a kinematically disturbed region is resolved, we use the surface brightness profiles of the high velocity gas beyond $\pm$300 \kms{}. 
These velocity cuts are made with respect to the systemic velocities fitted to the stellar absorption features as described in Sec. \ref{sec:data:center}, and are higher than the typical galaxy rotation. 
Four objects are classified as having unresolved KDRs, as these surface brightness profiles are consistent with the PSF (J1419+0139, J2102$-$0647, J2133$-$0712, J2142+0001), and the other five KDRs are considered as resolved (J0141$-$0945, J1000+1242, J1010+1413, J1255$-$0339, J2333+0049). 
However, there can be cases where the light profiles deviate from the PSF  because of contamination from galaxy rotation, even if the KDR is compact (see Appendix \ref{sec:append:sim_pv}). SDSS J2154+1131 is one example where the light profiles could be affected by rotation, while it has intrinsically narrow linewidths \weighty{}$<600$ \kms{}. 
We visually inspect the 2-D spectra of these five KDRs and determine that they are resolved and that the broad high velocity surface brightness profiles are not due to rotation. 
In the end, among the twelve objects, three have no kinematically disturbed regions, four have unresolved ones, and five have resolved ones (see Tab. \ref{tab:measurements}). 

The PSF can also bias the \riso{} and \rv{} size measurements. 
According to the simulations in Appendix \ref{sec:append:sim_pv}, this bias is less than 1\arcsec{} with an average of $\lesssim 0\farcs5$, but the exact amount depends on the structure of the \oiiil{} line so cannot be easily corrected. To account for this uncertainty, we assign an error of $0\farcs5$, which is also about half of the PSF FWHM, to the \riso{} and \rv{} measurements. This is the dominant source of errors for these sizes.

We also incorporate studies of type 2 AGN at higher AGN luminosities 
\citep{Liu2013b, Hainline2014d}. 
\citet{Hainline2014d} determines that 5 out of 30 objects in their sample have unresolved narrow-line regions, while the other 25 are resolved. 
\citet{Liu2013b} determines that all of their narrow-line-regions are resolved based on either the surface brightness profile or structures in the velocity fields.
We will determine \rv{} for the \citet{Liu2013b} sample to extend our luminosity baseline, but we cannot use identical criteria to determine whether the kinematically disturbed regions are resolved. 
However, structures in the \weighty{} profile are seen in all of their objects, and the measured sizes of the kinematically disturbed regions (Sec. \ref{sec:sizes:rv}) are all larger than the PSF. 
Moreover, HST narrow-band images of these objects reveal resolved high velocity dispersion components on several kpc scales (Wylezalek et al in prep.). 
We thus treat the 13 kinematically disturbed regions in the \citet{Liu2013b} sample as spatially resolved. 
We do not apply seeing corrections to these object but adopt size errors on \riso{} and \rv{} equivalent to half of the PSF FWHM to encompass the potential size bias, which is $0\farcs35$ for \citet{Liu2013b} and $0\farcs5$ for \citet{Hainline2014d}. This error is larger than the one estimated by \citet[][]{Liu2013b}, 5-14\%, which does not take the uncertainties due to seeing into account.

\subsection{Narrow-Line Region Radius \riso{}}
\label{sec:sizes:riso}

The size of the influence of AGN photoionization can be quantified by
the narrow-line region radius \riso{}.  
We adopt a common definition of \riso{} as
the semi-major axis of the $10^{-15}~(1+z)^{-4}$ erg s$^{-1}$ cm$^{-2}$
arcsec$^{-2}$ isophote of the \oiiil{} line \citep{Liu2013b,Hainline2013,Hainline2014d}. 
This isophote corresponds to a fixed intrinsic surface brightness
($5.1\times10^{39}$ erg s$^{-1}$ kpc$^{-2}$), such that this measurement can be compared across studies and is independent of the redshift or the depth of the observation, provided the observations reach this depth. 
\riso{} is designed to measure the largest extent of the \oiii{} region along its most elongated axis. 
To match our measurements with the ones from IFU studies \citep[e.g.,][]{Liu2013b}, where the semi-major axis can be easily determined, we align our Magellan slits along the semi-major axis of the SDSS $r$-band images, which contains the \oiiil{} line. When multiple slit positions are available, we take the one with the largest \riso{} as the representative slit for all the measurements. 
\citet{Liu2013b} suggests that narrow-line regions are often round, in which case the size \riso{} may not depend strongly on the slit orientation, although some of our objects that have IFU observations show irregular \oiii{} morphology \citep[J1000+1242 and J1010+1413, ][]{Harrison2014}. 
The measured \riso{} are listed in Table
\ref{tab:measurements} and demonstrated on the upper right panels of
Fig. \ref{fig:spec2100} and \ref{fig:spec2101}, etc. The four
unresolved objects in our sample are treated as \riso{} upper limits.
One object SDSS J1255$-$0339 has a particularly large \riso{}$=33$ kpc 
because it has a pair of extended but kinematically cold tidal features 
(Appendix \ref{sec:append:objs}).  

We incorporate 14 \riso{} measurements of luminous type 2 AGN 
from \citet{Liu2013b}, as well as 20 \riso{} 
measurements and 5 unresolved upper limits from \citet{Hainline2014d} (Fig. \ref{fig:sizelum_whichR}). 
\citet{Liu2013b} uses the Gemini GMOS IFU while \citet{Hainline2014d} uses 
Gemini GMOS long-slit spectroscopy and both reach similar depths as ours. 
Five objects in \citet{Hainline2014d} are excluded due to duplication with 
\citet{Liu2013b} (4/10) or \emph{WISE} source confusion (1/10).

\subsection{Kinematically Disturbed Region Radius \rv}
\label{sec:sizes:rv}

The radius of the kinematically disturbed region \rv{} measures the spatial extent of the high velocity gas. 
We use two criteria to define the kinematically disturbed region. First, 
the \oiiil{} line width \weighty{} has to be larger than a threshold of 600 \kms{}. This is similar to the criterion used by \citet{Harrison2014} to identify high velocity non-virialized motions. 
While this value of 600 \kms{} is somewhat arbitrary, it is also conservative, 
since galaxy velocity dispersions rarely exceed 300 \kms. 
Typical ellipticals have velocity dispersion $\sigma \sim 200$ \kms{} 
\citep{Sheth2003}, thus their \weighty{} should be under 500 \kms{} 
assuming virialized motions with Gaussian profiles. 
The second criterion is to require the surface brightness of the high velocity gas (the red $>300$ \kms{} or the blue $<-300$ \kms{} side) to be higher than the isophotal threshold defined in Sec. \ref{sec:sizes:riso}. 
Without this surface brightness threshold, in some cases \rv{} could be severely biased by the spatial PSF when the PSF propagates high line widths to large radii, see Appendix \ref{sec:append:sim_pv}. 
\rv{} is taken as the largest measured radius where both criteria are met.

The resulting \rv{} are tabulated in Table \ref{tab:measurements} and
plotted in Fig. \ref{fig:sizelum_whichR}. Three objects, SDSS J1055+1102, J1351+0728, and J2154+1131, do not have kinematically disturbed regions as their \weighty{} are below 600 \kms{}. These are plotted as empty squares in
Fig. \ref{fig:sizelum_whichR}.
The four unresolved objects, SDSS J1419+0139, J2102$-$0647, J2133$-$0712, and J2142+0001, are treated as \rv{} upper limits, and plotted as down-facing triangles. The other five resolved KDRs, J0141$-$0945, J1000+1242, J1010+1413, J1255$-$0339, and J2333+0049, are treated as \rv{} measurements and are plotted as circles. We adopt an error of $0\farcs5$ on \rv{} as discussed in \ref{sec:sizes:resolve}. 

To increase the sample size, we include the \citet{Liu2013b} sample, where the median \weighty{} as a function of radius is also measured. 
Among the fourteen objects, only one (J0842+3625, the empty red square in Fig. \ref{fig:sizelum_whichR}), does not have \weighty{} above $600$ \kms{}. 
We calculate the \rv{} of the other thirteen objects based on their \weighty{} profiles (empty red squares in Fig. \ref{fig:sizelum_whichR}).  
We cannot apply the surface brightness requirement on the \rv{} measurements as the high velocity light profiles are not available for their sample. 
Without the surface brightness requirement, \rv{} can be largely overestimated when there is no spatially extended narrow component, such that the high \weighty{} of the broad component is propagated to large radius by the PSF, see Appendix \ref{sec:append:sim_pv}. 
The observed drop in \weighty{} at large radius in the \citet{Liu2013b} sample indicates that there is a narrow extended component, which in our simulations makes it very difficult for an unresolved broad component to impact \weighty{} at large scales and give strongly biased \rv{} measurement. 
We adopt an error of $0\farcs35$ on \rv{} (see Sec. \ref{sec:sizes:resolve}). 
These numbers are tabulated in Table \ref{tab:measurements_liu13}. 
The combination of these two samples covers a wide dynamic range in
AGN luminosity from $10^{45}$ to $10^{47}$ \ergs.

\subsection{The Size-Luminosity Relations}
\label{sec:sizes:relation}

In this section we investigate the relationship between AGN luminosity 
and our two size measurements - \riso, which depends on
photoionization, and \rv, which is based on kinematics. 
These two radii typically extend from a few to 
15 kpc (with the exception of SDSS J1255$-0339$ where \riso$=33$ kpc), 
corresponding to a light travel time of $\sim10^4$ years. 

\begin{figure*}[h]
	\vbox{
	\hbox{
	\includegraphics[scale=0.45]{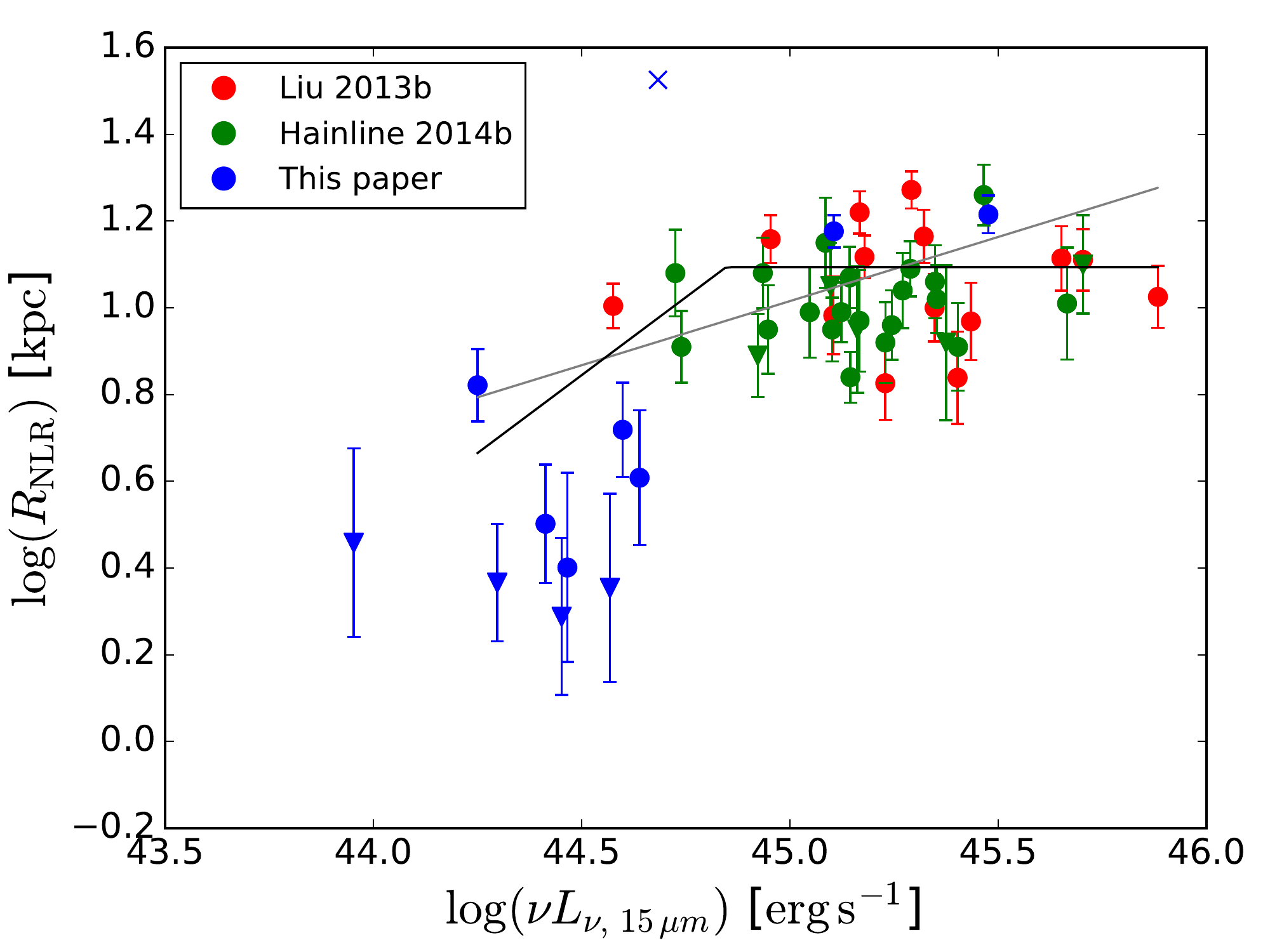}
	\includegraphics[scale=0.45]{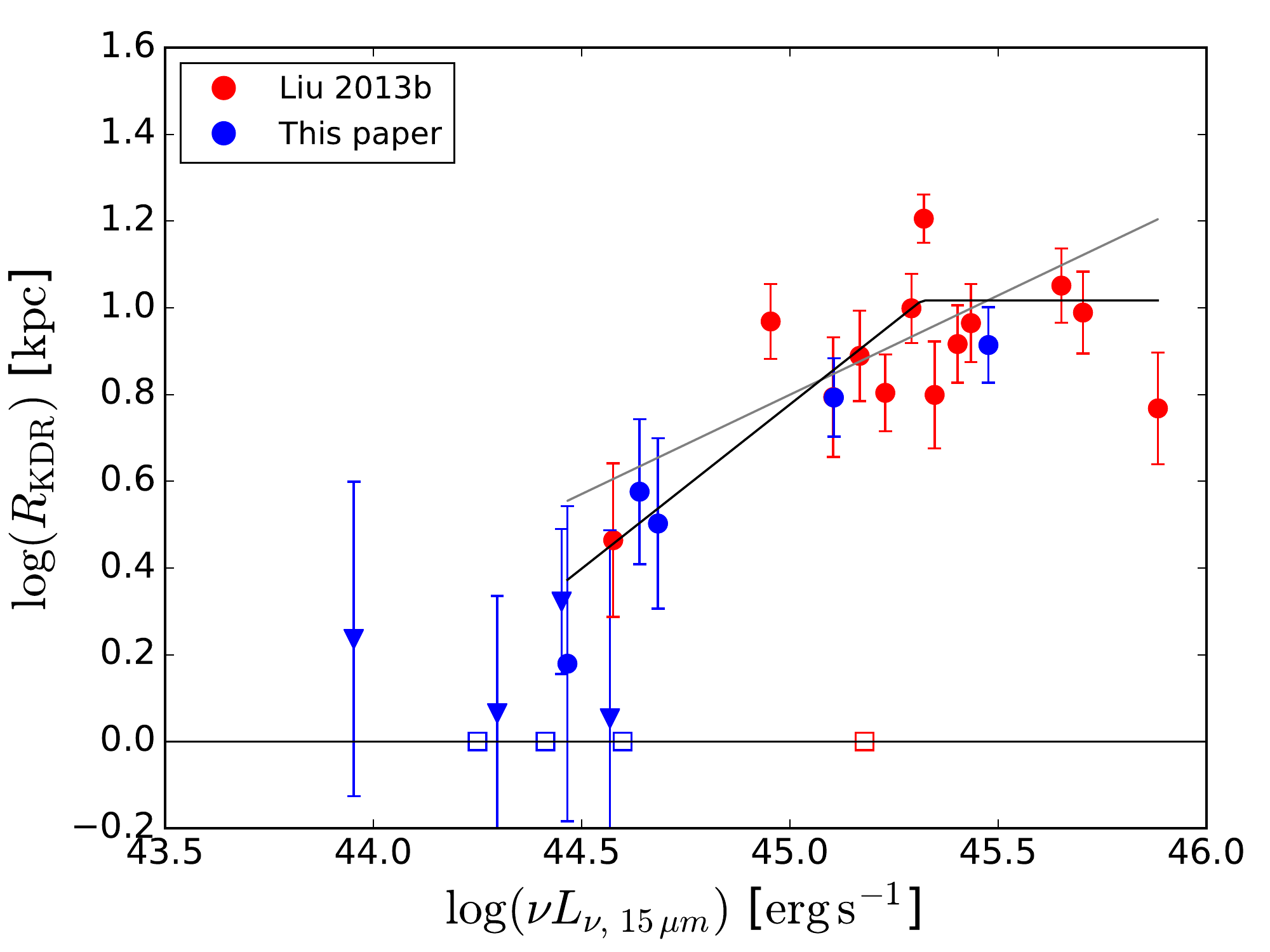}
	}
	}
    \caption{
    The comparison of two size luminosity relations, based on the radius of the narrow line region \riso{} defined by the surface brightness of the \oiiil{} line ({\it left}) and the radius of the kinematically disturbed region \rv{} defined by a high \oiiil{} linewidth of \weighty{}$>600$ \kms{} ({\it right}). 
    Three samples are used: this paper (blue), \citet{Liu2013} (red), and \citet{Hainline2014d} (green). 
    On the horizontal axis is the mid-IR rest frame 15 \micron{} luminosity \Lfifteen{} as a proxy of the AGN luminosity. 
    The solid circles are valid size measurements. The downward triangles are size upper-limits for unresolved objects. 
    The empty squares represent objects with no high velocity \oiiil{} features and thus \rv{} undefined.
    The blue cross is the source SDSS J1255$-$0339 with ionized tidal tails. 
    The error bars represent the size uncertainty of $\pm 0\farcs5$ for our and the \citet{Hainline2014d} sample and $\pm 0\farcs35$ for the \citet{Liu2013} sample. 
    The gray (black) lines show the best-fit single power-law (flattened power-law) models fitting only the solid circles. 
    Both relations prefer a flattened power-law (Sec. \ref{sec:sizes:relation}). 
    }
    \label{fig:sizelum_whichR}
\end{figure*}

The relation between \riso{} and the AGN luminosity has been studied 
extensively and
there are tentative signs that it flattens at high AGN luminosity
\citep{Hainline2013,Liu2013b,Hainline2014d,Liu2014a}. In Fig. \ref{fig:sizelum_whichR}, left, we revisit this relation, supplementing it with our new
high quality \riso{} measurements. Our objects populate a
lower luminosity range compared to previous studies, allowing us to extend 
the luminosity baseline. Furthermore, we use a different AGN
luminosity indicator -- 15 \micron{} luminosity \Lfifteen{}, which is arguably less sensitive to the anisotropy of the infrared emission (Appendix \ref{sec:append:WISE}). 

On the right-hand side of Fig. \ref{fig:sizelum_whichR} we show the dependence 
on AGN luminosity of our derived kinematically disturbed region radius \rv{}, which illustrates the effect of the AGN luminosity on the mechanical feedback operating on galaxy scales. 
Using only the valid size measurements (circles in Fig. \ref{fig:sizelum_whichR}, the \riso{} outlier, J1255$-$0339, is not included), we find that both radii are positively correlated with the 15 \micron{} luminosity, with Pearson's {\it r} correlation coefficient above 0.6 and the $p$-values below 0.01. 

An essential property of \riso{} is that it includes any
photoionized gas in the vicinity of the galaxy, independent of its origin,
including tidal features or illuminated companion galaxies. As an extreme example
of illuminated tidal features, SDSS J1255$-$0339 has a pair of extended
tidal tails emitting in \oiii. They can be seen in the SDSS {\it r}-band
image and they yield a very large \riso\ measurement. This object is a
distinct outlier in the \riso{} - \Lfifteen{} relation
(blue cross in the left panel of Fig. \ref{fig:sizelum_whichR}). But in the \rv{} - \Lfifteen{} space, 
this object follows the trend defined by other AGN because this extended feature has quiescent kinematics.

As our new sample can improve the constraints on the low end slope of
the \riso{} size-luminosity relation, and we are interested in
quantitatively comparing the \riso{} and \rv{} size-luminosity
relations, we fit these two relations with a single power-law (gray
lines in Fig. \ref{fig:sizelum_whichR}) and a flattened power-law
(black line). To determine whether the flattening of the size-luminosity 
relations is significant, we use the Bayesian Information
Criteria (BIC) to distinguish which model is preferred by the data
(Tab. \ref{tab:whichR}). Only objects with valid size measurements
(circles in Fig \ref{fig:sizelum_whichR}) rather than limits are included
for this analysis.  

We find that the BIC of both the \riso{} and \rv{} size-luminosity relations prefer a flattened power law ($\Delta$BIC = 13.8 and 12.6). Both relations saturate at a radius of about 10 kpc. But the saturation for \rv{} occurs at a higher luminosity (\Lfifteen{}$=10^{45.3}$ \ergs{}) compared to \riso{} (\Lfifteen{}$=10^{44.8}$ \ergs{}). \rv{} are in general lower than \riso{} by $\sim$ 0.5 dex below the saturation luminosity. So using a flux-based measurement can lead to overestimation of the outflow sizes, as suggested by \citet{Karouzos2016}.

Therefore, we confirm the findings of \citet{Liu2013b} and
\citet{Hainline2013,Hainline2014d} that, beyond a luminosity of about
\lbol$>10^{46}$ \ergs, the \riso{} -\Lfifteen{}
relation flattens such that \riso{} is a constant ($\sim10$ kpc) with respect to the AGN luminosity. 
One possible explanation of the observed limit to the narrow-line region size is a change in the ionization state of the gas at large radii. For example, as the density of the gas drops, the clouds can transition from an ionization-bounded to matter-bounded state, such that the O$^{2+}$ ions become ionized to O$^{3+}$ \citep{Liu2013b}.  
Our measured slope of the \riso{}-\Lfifteen{} relation at the low luminosity end (0.72) as
fitted by the flatted power law is steeper than the value of 0.47 found by
\citet{Hainline2013}, likely because our objects provide a better sampling of the relationship at lower luminosities. 

However, the flattening of \rv{} can also partly be due to the drop in the \oiiil{} intensity at $\sim$10 kpc, as the \rv{} measurement is also limited by the surface brightness of \oiiil{}. 
If we use only objects where \weighty{} drops below 600 \kms{} at large radii, such that \rv{} marks the edge of the high velocity gas, not just the \oiiil{} luminous gas, then the confidence for the flattened power law preference becomes lower (BIC = 5.0). 
More observations are needed to confirm whether the sizes of the kinematically disturbed regions indeed saturate at 10 kpc. 
Nonetheless, the size of the AGN-disturbed region \rv{} seems to scale with the AGN luminosity until a high luminosity of \Lfifteen{}$\sim10^{45}$ or \lbol{}$\sim10^{46}$ \ergs{}. 
The slope of the scaling at the low luminosity end is not well constrained and requires a larger sample. 

It is possible that the size of the kinematically disturbed region
continues to decrease in lower luminosity AGN. For example, NGC 1068, a local
type 2 AGN at a lower AGN luminosity of \lbol$\sim 10^{44-45}$ \ergs{}
(\citealt{Goulding2010,Alonso-Herrero2011,Garcia-Burillo2014} and
references therein), hosts ionized outflows with deprojected velocities 
as fast as $\sim$ 1300 \kms{}, but with a much smaller
outflow size on the scale of $\sim$ 200 pc \citep{Cecil1990,Crenshaw2000}.

In summary, we find that both \riso{} and \rv{} correlate with the AGN luminosity and both saturate at about 10 kpc at high AGN 
luminosities. \rv{} is in general lower than \riso{} by $\sim$ 0.5 dex before
the saturation and saturates at a higher luminosity (\Lfifteen{}$=10^{45.3}$ \ergs{} versus $10^{44.8}$ \ergs{}). 
\rv{} is also less affected by the presence of tidal tails or companions.

\section{Outflow Properties and Energetics}
\label{sec:energetics}

The large \oiiil{} linewidths (\weighty{} $= 600-1500$ \kms) commonly seen 
in our sample suggest that many of these systems have high velocity 
non-virialized gas motions, likely outflows. 
While AGN outflows on galactic scales are thought to be an important agent to regulate star-formation, their size distributions, energy efficiencies, 
and dependence on the AGN luminosity are not well-understood. 
Therefore, it is important to measure the properties of these outflows, 
including size, velocity, and energetics, and study their dependence on
the AGN luminosity.

In Sec. \ref{sec:energetics:vprof}, we discuss kinematic models to explain the observed \oiiil{} velocity profiles. 
In Sec. \ref{sec:energetics:definition}, we define and calculate the outflow properties including the sizes, velocities, and time scales. We also use two methods to estimate the outflow kinetic power. In Sec. \ref{sec:energetics:regress}, we study the correlation between the outflow properties and the AGN bolometric luminosities and discuss the outflow energy efficiency.  

We find that the outflow size, velocity, and energy are correlated with the
AGN luminosity. Although the actual outflow efficiency cannot be
constrained with high accuracy (we estimate 
$\eta=\dot{E}/\lbol=0.01\% - 30\%$),
our results are consistent with a hypothesis that the energy efficiencies of 
AGN outflows are roughly constant for AGN in the luminosity range of 
$10^{45-47}$ \ergs.

\subsection{Velocity Profiles and a Kinetic Model of the Outflow}
\label{sec:energetics:vprof}

With only three exceptions, the objects in our sample have gas velocities (\weighty$=600-1500$ \kms)
faster than virialized motions in a typical galactic potential 
(\weighty$<500$ \kms, see Sec. \ref{sec:sizes:rv}). 
In this section, we discuss possible outflow models that can explain the 
observed linewidth and its profile. 

\citet{King2005} and \citet{King2011} found that for an energy-conserving (no radiative loss) 
spherical outflow propagating in a galaxy with an isothermal potential and 
gas distribution, the outflow's shock front expands at a constant velocity, which for black holes on the $M-\sigma$ relation accreting at their Eddington rate is of order 1000 \kms. At the same time, gas at large radii that has not yet been shocked and accelerated by the outflow should remain at its original velocity. Therefore, there should be a sharp drop in the gas velocity profile 
corresponding to the shock front.

The resolved KDRs in our sample show \weighty{} profiles that are consistently high (\weighty$\sim 600-1500$ \kms{}) within the central few kpc  (Fig. \ref{fig:spec2100}, \ref{fig:spec2101}, and Appendix \ref{sec:append:objs}). 
Those that still have high signal-to-noise ratio measurements at large radii 
(SDSS J1000+1242, J1010+1413, J1255$-$0339) do show a sudden velocity drop at radii of 5-10 kpc. 
Such a high linewidth plateau followed by a sudden drop in \weighty{} are also 
commonly seen in other studies of type 2 AGN \citep[e.g.][]{Greene2011,Liu2013,Karouzos2016}. 
\citet{Liu2013} suggests that flat \weighty{} profiles correspond to a constant outflow velocity $v\sim$ \weighty$/1.3=460 -1100$ \kms{}, if the outflow is spherical/quasi-spherical with a power-law intensity profile. 
Such spherical morphology is seen in their IFU spectroscopic data, but whether the outflows in our objects are spherical cannot be verified with our long-slit data alone\footnote{Some objects show irregular morphologies of the emission lines from the SDSS images (e.g, J1000+1242 and J1010+1413), but those irregular morphologies could be due to the extended narrow emission and do not necessarily reflect the morphology of the outflow. 
}. This plateau-shaped velocity profile is broadly consistent with the prediction of the \citet{King2011} model, and the kinematically disturbed region radius \rv{}, defined based on the line width threshold of 600 \kms, is able to capture the location of the velocity drop. 
We adopt this constant-velocity spherical outflow model as a simplified framework to interpret our observations and use the linewidth \weightyAVG{} as a measure of the outflow velocity and \rv{} as the outflow size.

\subsection{Outflow Properties Definition}
\label{sec:energetics:definition}

In this section, we define the outflow properties, including the radius, 
velocity, dynamical time scale, and energetics. 

As discussed in Sec. \ref{sec:energetics:vprof}, we use \rv{} -- the radius of the kinematically disturbed region where \weighty $> 600$ \kms{} -- as the radius of the outflow. 
The errors in \rv{} are taken to be half of the seeing FWHM, which is $0\farcs5$ for our sample and $0\farcs35$ for the \citet{Liu2013b} sample
(see Sec. \ref{sec:sizes:resolve}).
Following \citet{Liu2013}, the outflow velocity $v$ is taken as $v=$\weightyAVG/1.3, where the factor of 1.3 is the projection correction for quasi-spherical outflows. 
As described in Sec. \ref{sec:sizes:spec}, for our objects,
\weightyAVG{} is represented by the luminosity-weighted quadratic mean of \weighty{} (spectral-PSF corrected, see Appendix \ref{sec:append:sim_w80}), and a conservative error of 20 \kms{} is assumed. 
The \weightyAVG{} for the \citet{Liu2013b} sample is the \weighty{} measured from their SDSS fiber spectrum \citep[Column 7, Table 1][]{Liu2013b}. These are not corrected for the spectral resolution so we adopt conservative errors of 75 \kms{} corresponding to half of the SDSS spectral FWHM. 
We then derive the outflow dynamical time scale as \tdyn =\rv$/v$.  
All of these quantities and their errors are tabulated in Table \ref{tab:measurements}.

As discussed in \citet{Greene2012}, measuring the mass of the outflow can be 
challenging, which is the biggest uncertainties in estimating the energetics. 
As the emissivity scales with density squared, strong emission lines, such as 
\oiiil{} and \hbeta, trace only the densest ionized gas clouds. 
These clouds occupy only a small fraction of the total volume 
($\sim10^{-2}$), and there can be a large amount of diffuse ionized gas 
unaccounted for. Parts of the outflows could even be in different phases, 
such as molecular or hot plasma, that are not traced with these lines. 

We adopt two methods to bracket the range of possible kinetic power of the outflow. Assumptions in gas densities are made for both methods using reasonable values for type 2 AGN. 
While the exact values of energy depend on these assumptions, these measurements are only order-of-magnitude estimations and we focus on trends in outflow properties with AGN luminosity, which do not depend as strongly on these assumptions.

For the first method, we estimate the mass of the dense ionized gas
from the \hbeta{} luminosity assuming case B recombination\footnote{As SDSS J0141$-$0945 and J2133$-$0712 don't have
\hbeta{} measurements from the Magellan spectra, their $M_\mathrm{H}$
and \dotEkin{} estimates are not available.}. We follow \citet{Osterbrock2006} and use the case B recombination equation at $10^4$ K 
\begin{equation}
   M_\mathrm{H} = 6.8 \times 10^8~L_{\mathrm{H\beta,43}}~n^{-1}_{e,100}~\msun, 
\end{equation}
where $L_{\mathrm{H\beta,43}}$ is the \hbeta{} luminosity in units of
$10^{43}$ erg s$^{-1}$, and $n_{e,100}$ is the electron density in
units of 100 cm$^{-3}$. 

The electron densities inferred from the {\sii} $\lambda6716/\lambda6731$ ratios are typically a few hundred cm$^{-3}$ for AGN outflows \citep[e.g.,][]{Nesvadba2006,Villar-Martin2008,Greene2011}, while much higher densities up to $10^5$ cm$^{-3}$ have been measured using other diagnostics \citep[e.g.,][]{Holt2010}. 
Such measurements are likely biased to the densest gas clumps of small volume-filling factors in the outflow \citep{Greene2011}. Studies of extended AGN scattered light, which is not biased to the dense gas, infers much lower densities $<$ 1 cm$^{-3}$ \citep{Zakamska2006}. 
For the purpose of an order-of-magnitude estimation, we assume an electron density of 100 cm$^{-3}$, which represents the dense clumps in the outflow. 
\lhb{} is measured from the Magellan slits assuming the \hbeta{} surface brightness profile is azimuthally symmetric.
We then calculate the kinetic power as
\begin{equation}
  \dot{E}_{\mathrm{kin}} = \frac{\frac{1}{2} M_{H} v^2}{t_{dyn}} = 
\frac{1}{2} M_{H} R_{\mathrm{KDR}}^{-1} v^3, 
\end{equation}
where \tdyn{} is the dynamical time scale, \rv{} is the size of the kinematically disturbed region, and $v=$\weightyAVG/1.3 is the deprojected velocity of the outflow.
Here $M_{H}$ includes the mass of all the velocity components, not just the high velocity parts, so $M_{H} v^2$ represent the total kinetic energy of the dense ionized gas. 
We assume the kinetic energy of the outflow dominates the total kinetic energy such that $\dot{E}_{\mathrm{kin, outflow}} \sim \dot{E}_{\mathrm{kin, total}}$. This value can still be an underestimate of the true kinetic energy of the outflow as the \hbeta{} emission line traces only the densest ionized gas.

The second method is similar to the Sedov-Taylor solution for
a supernova remnant where a spherical bubble is expanding into a medium
of constant density.  This method is motivated by 
observations of such organized outflows in similar 
type 2 AGN, e.g. SDSS J1356+1026 \citep{Greene2012}.  We
adopt a simple definition of \dotEST{} as
\begin{equation}
    \dot{E}_{\mathrm{ST}} = \frac{1}{2} \dot{M} v^2 = 2 \pi \rho_0 R_{\mathrm{KDR}}^2 v^3, 
\end{equation}
where
\begin{equation}
    \dot{M}= 4 \pi \rho_0 R_{\mathrm{KDR}}^2 v
\end{equation}
is the rate at which ambient gas enters the outflow, $\rho_0$ is the ambient gas density, \rv{} is the size of the kinematically disturbed region, and $v=$\weightyAVG/1.3 is the deprojected velocity of the outflow.
The ambient gas
density $\rho_0$ is assumed to be a constant
$\rho_0=m_p\times(0.5~\mathrm{cm}^{-3})$. Such a density is supported
by scattering measurements of type 2 AGN by \citet{Zakamska2006}. This
definition is within 20\% of the Sedov-Taylor solution described in
eq. 39.9 of \citet{Draine2011}, and e.q. 7.56 of \citet{Dyson1980},
and about 30\% lower than the one adopted by \citet{Nesvadba2006} and
\citet{Greene2012}. This method likely overestimates the kinetic power, 
as it assumes that all of the ambient gas is entrained in the outflow.
Indeed, the resulting \dotEST{} is higher than \dotEkin{} by 1 to 3 orders of magnitude (Sec. \ref{sec:energetics:regress}).

All of these quantities -- \rv{}, \weightyAVG{}, \tdyn{}, \dotEkin{}, and \dotEST{} -- as well as their errors are tabulated in Tables \ref{tab:measurements} and \ref{tab:measurements_liu13}. 
The errors on \rv{} and \weightyAVG{} are 0\farcs5 [0\farcs35] and 20 \kms{} [75 \kms{}] for our sample [the \citet{Liu2013b} sample]. 
The errors on \tdyn{}, \dotEkin{}, and \dotEST{} are propagated from the input quantities, where the errors on \lhb{} and the gas densities, $n_{e,100}$ and $\rho_0$, are assumed to be 20\% and 50\%, respectively. 
The size upper limits for unresolved objects are also propagated to the derived quantities. 
The absolute values of \dotEkin{} and \dotEST{} should be taken as order-of-magnitude estimations and are only used to bracket the true value of the outflow kinetic power.

\subsection{Relation between the Outflow Properties and the AGN Luminosities}
\label{sec:energetics:regress}

In this section, we investigate how outflow size, velocity, dynamical time-scale, and energy correlate with AGN luminosity.  We adopt the 15 $\micron$
luminosity \Lfifteen{} as the AGN luminosity indicator, as discussed in
Sec. \ref{sec:data:WISE}. The outflow size, velocity, dynamical time
scale, and energetics are defined in section
\ref{sec:energetics:definition}.

\begin{figure*}[h]
	\vbox{
	\hbox{
	\includegraphics[scale=0.45]{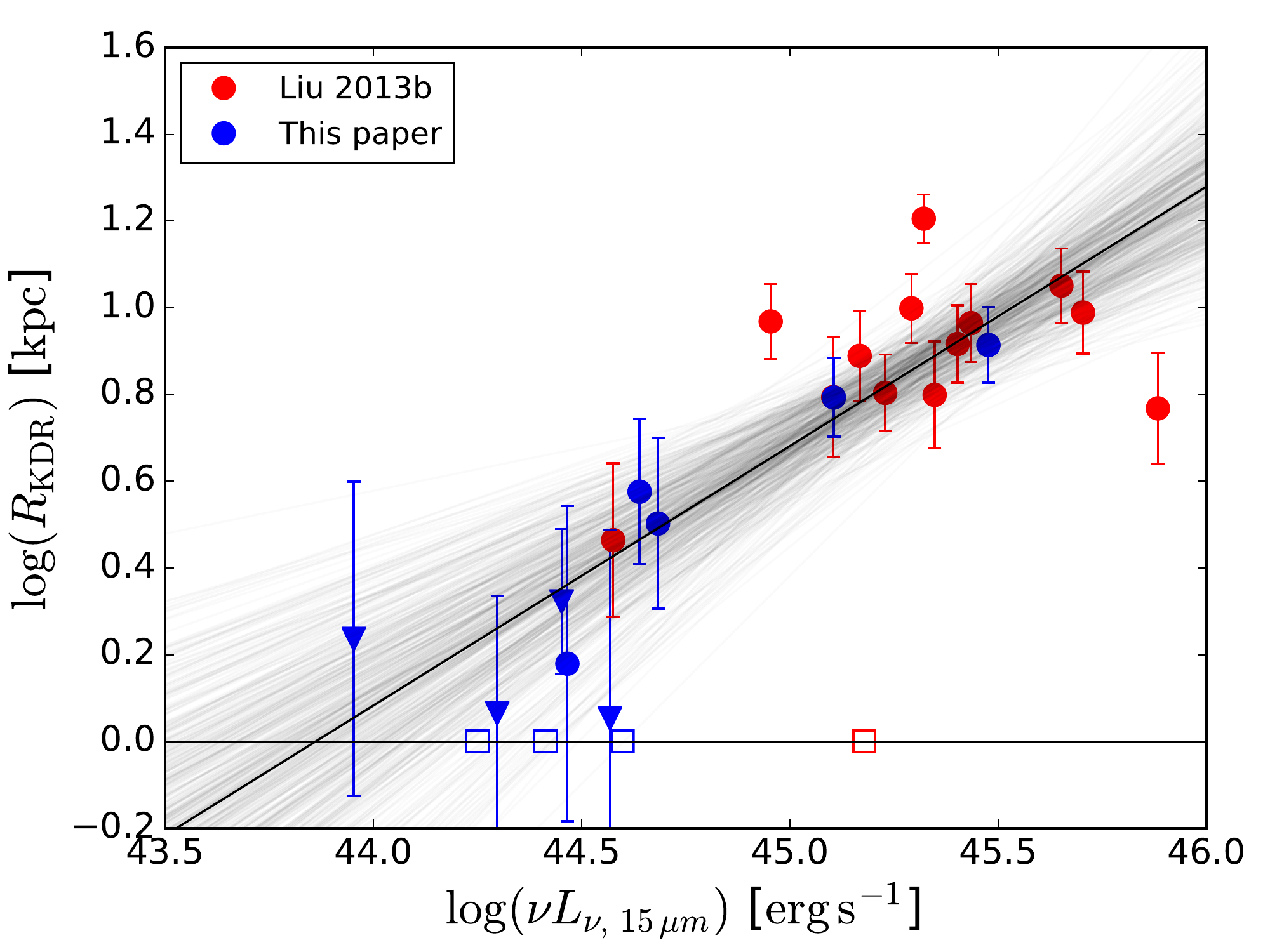}
	\includegraphics[scale=0.45]{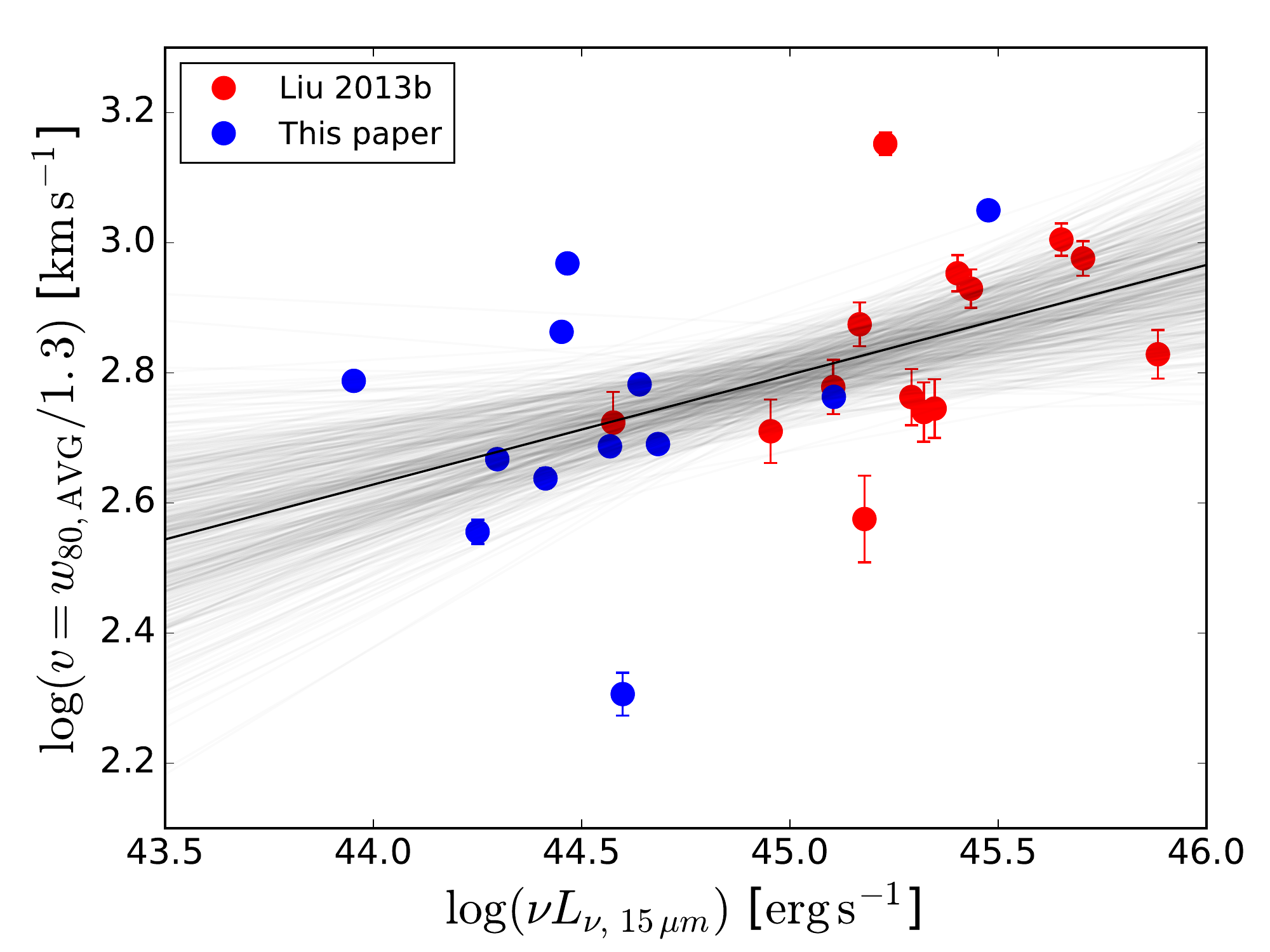}
	}
	\hbox{
	\includegraphics[scale=0.45]{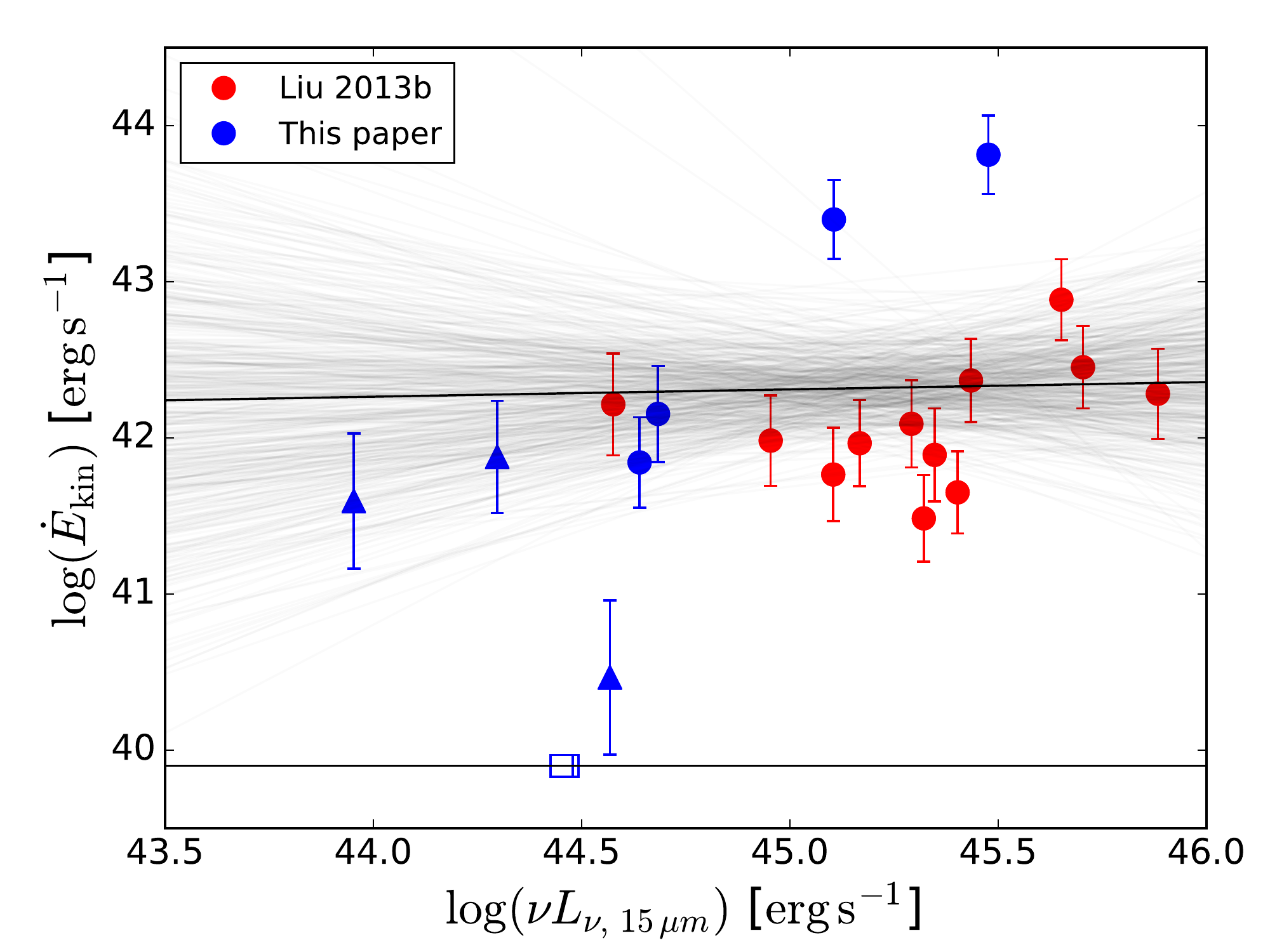}
	\includegraphics[scale=0.45]{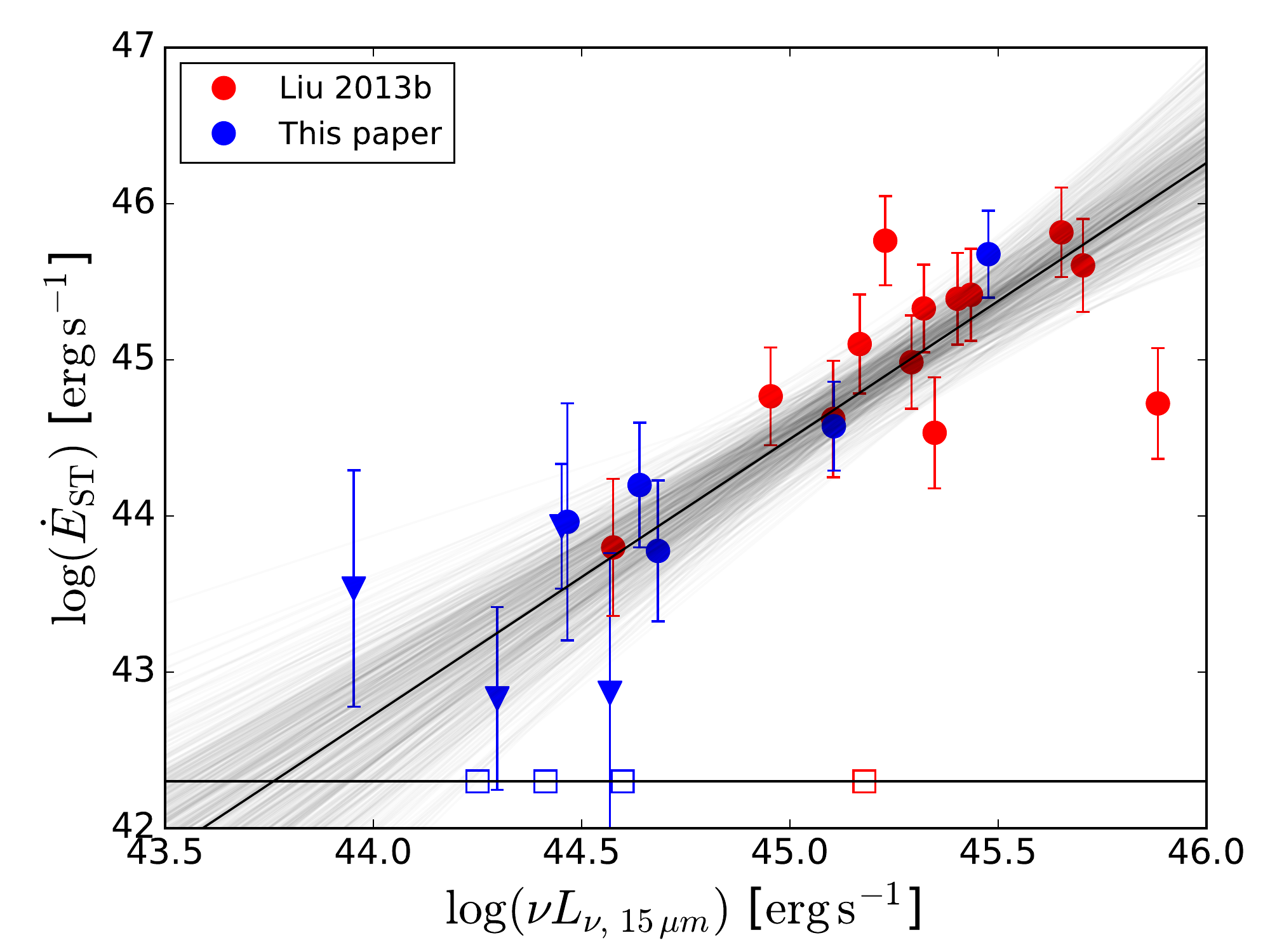}
	}
	\hbox{
	\includegraphics[scale=0.45]{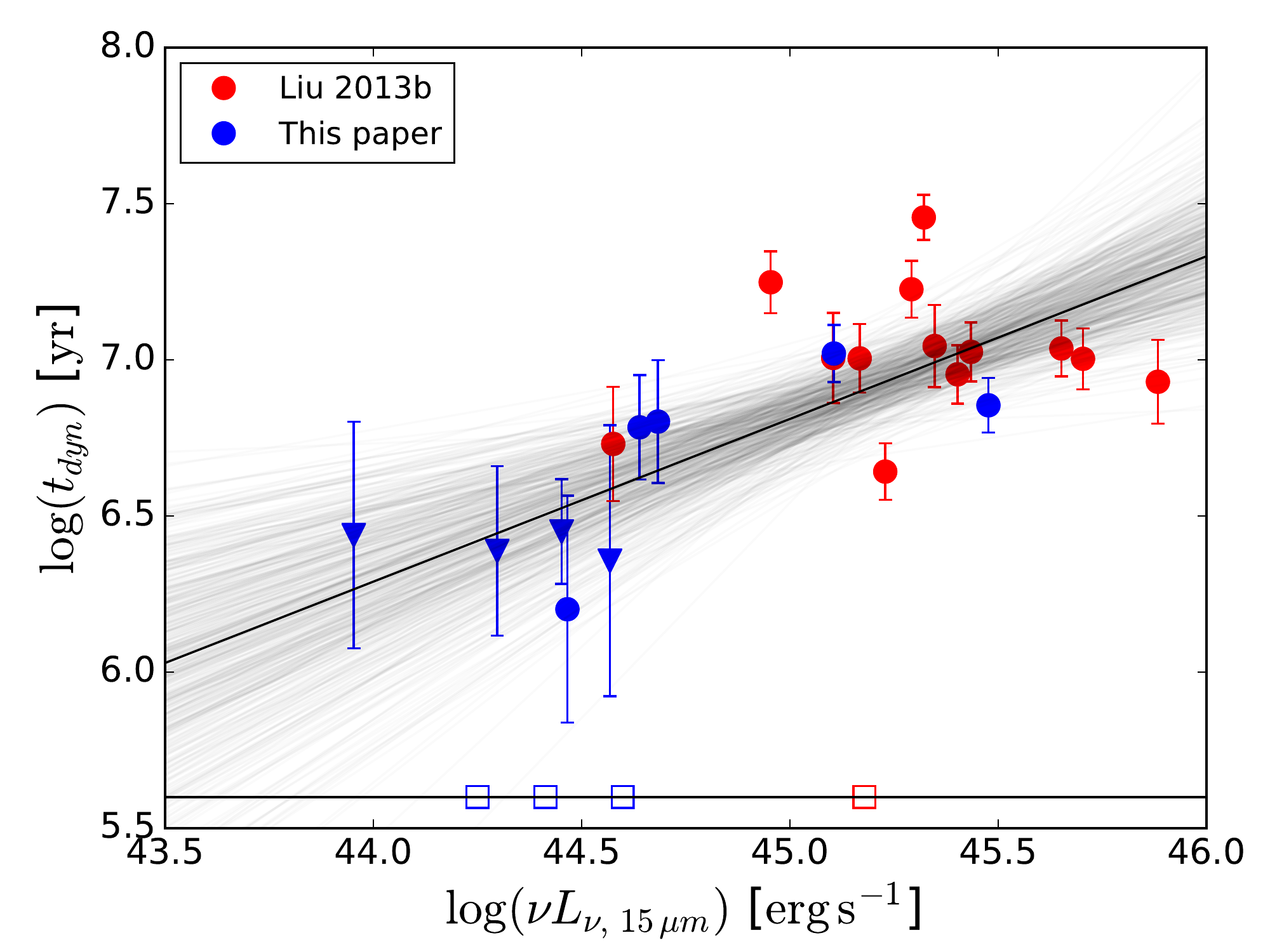}
	}
	}
    \caption{
    The outflow properties versus the AGN luminosity. The outflow properties are: radius \rv{} ({\it upper left}), velocity $v=$\weighty$/1.3$ ({\it upper right}), ionized gas kinetic power \dotEkin{} ({\it middle left}), Sedov-Taylor kinetic power \dotEST{} ({\it middle right}), and dynamical time \tdyn{} ({\it bottom}). 
    On the horizontal axis is the mid-IR rest frame 15 \micron{} luminosity \Lfifteen{} used as a proxy of the AGN luminosity. 
    Except for \dotEkin{}, all the properties are positively correlated with the AGN luminosity. 
    The symbols and their colors are as described in Fig. \ref{fig:sizelum_whichR}. Galaxies with no \hbeta{} measurements are not shown in the \dotEkin{} plot. The gray lines sample the posterior distribution of the power-law model, and the black line shows the model corresponding to the posterior mean. The model parameters and correlation coefficients are listed in Tab. \ref{tab:regress} and discussed in \ref{sec:energetics:regress}. 
 	}
    \label{fig:kimlum_regress}
\end{figure*}

The relations between these outflow quantities ($y$) and the AGN 
luminosity indicator \Lfifteen{}, are 
quantified by a single power law, 
\begin{align}
\log(y) = \alpha + \beta \times \log(\nu L_{\nu, 15}). 
\end{align}
We adopt a Bayesian linear regression approach developed by \citet{Kelly2007} 
using a Markov chain Monte Carlo sampling method, which accounts for the 
measurement errors, intrinsic scatter, and upper or lower limits. 
The measurement errors are as summarized in Sec. \ref{sec:energetics:definition}. We assume that there is no error in the AGN luminosity indicator \Lfifteen{}. 
The intrinsic scatter is fitted as a hidden variable. 
The upper- and lower limits are included in the fits as censored data. 
Three objects in our sample and one object from \citet{Liu2013b} have no kinematically disturbed region. They are only used for the \weighty{} - luminosity relation. 
Two objects in our sample and one object in the \citet{Liu2013b} sample have no \hbeta{} measurement and are unavailable for the \dotEkin{} relation, see Tables \ref{tab:measurements} and \ref{tab:measurements_liu13} for details. 
To access the statistical significance of the correlations, 
we calculate the Pearson's $r$ correlation coefficient and its $p$-value using only the valid measurements (solid circles in Fig. \ref{fig:kimlum_regress}). 
The results are shown in Fig. \ref{fig:kimlum_regress} and tabulated in Table \ref{tab:regress}. 

We find that the outflow radius \rv{} correlates strongly with the AGN luminosity with a Pearson's $r$ $p$-value of $3\times10^{-4}$ and a power-law index of $0.60 ^{+0.13}_{-0.13}$. The correlations of the \weightyAVG{} - luminosity and the \tdyn{} - luminosity relations are not as strong, with $p$-values of only about $10^{-2}$ and power-law indices of $0.17^{+0.06}_{-0.07}$ and $0.52 ^{+0.16}_{-0.16}$, respectively. 
The Sedov-Taylor power estimate \dotEST{} also correlates with the AGN luminosity with a power-law index of $1.76 ^{+0.31}_{-0.31}$ but the kinetic power estimate \dotEkin{} shows no strong correlation with the luminosity. 
The errors represent 1-$\sigma$ errors. 

\begin{figure}[h]
	\includegraphics[scale=0.5]{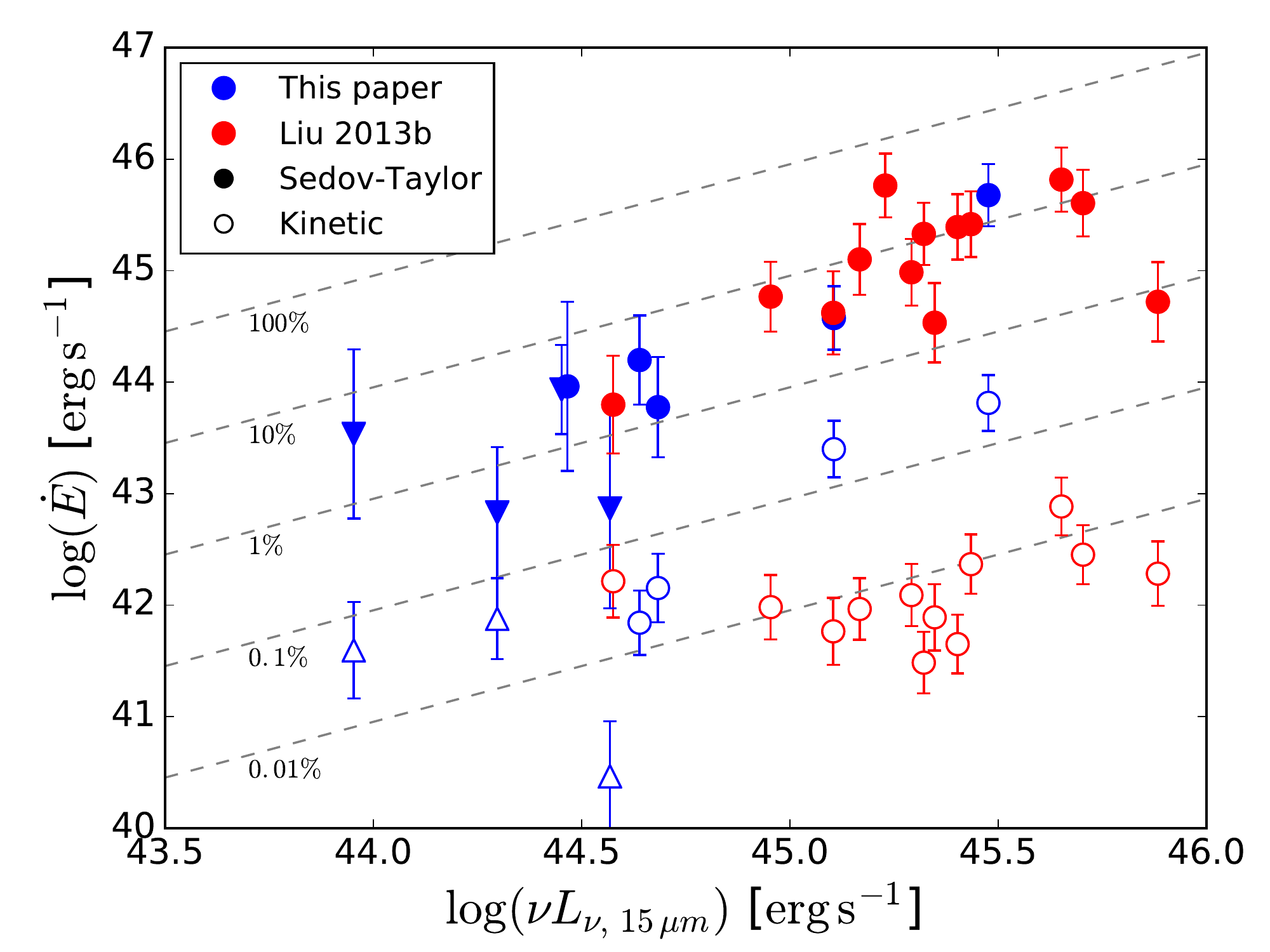}
    \caption{The outflow kinetic power versus the AGN luminosity. 
    The outflow kinetic power is bounded by two estimates: the ionized gas kinetic power \dotEkin{} (empty symbols), and the supernova-like Sedov-Taylor kinetic power \dotEST{} (solid symbols).
    On the horizontal axis is the mid-IR rest frame 15 \micron{} luminosity \Lfifteen{} used as a proxy of the AGN luminosity. 
    The symbols and their colors are as described in Fig. \ref{fig:sizelum_whichR}. 
    \dotEkin{} of galaxies with no \hbeta{} measurements are not shown.
    The diagonal dashed lines show different energy efficiencies $\eta=\dot{E}/\lbol$, where $\lbol=9\times$\Lfifteen{}. 
    The outflow energy efficiency is bracketed within the range of $\eta=\dot{E}/\lbol=0.01\% - 30 \%$.
 	}
    \label{fig:kimlum_energetics}
\end{figure}

We compare the two energy estimates in Fig. \ref{fig:kimlum_energetics}.
The \dotEST{} are typically 1 to 3 orders of magnitudes higher than \dotEkin,
meaning that we cannot constrain the outflow energetics
precisely. These two methods bracket a very large range of
feedback energy efficiency $\eta=\dot{E}/\lbol=0.01\% - 30\%$,
reflecting big uncertainties in the outflowing mass. 
The dependence of this energy efficiency on the AGN luminosity also cannot be constrained precisely. Our data do not rule out the scenario where $\eta$ is a constant within the luminosity range of \lbol $\sim 10^{45-47} $ \ergs{}. It is possible that most AGN in this luminosity range are capable of driving outflows with energy proportional to their AGN luminosity.

We find that the outflow properties, including the radius and velocity, correlate and increase with the AGN bolometric luminosity. An AGN outflow should be a common phenomenon within the
luminosity range of $L_{\nu,~\rm{15 \micron}} \sim 10^{44-46}$
\ergs{} or $\lbol \sim 10^{45-47} $ \ergs{}.  
If there is a 
critical luminosity threshold for AGN feedback, below which 
outflows cannot be driven, it must occur at yet lower AGN luminosities.

\section{Outflow Occurrence Rates and Timescales}
\label{sec:timescale}

In this section we discuss the occurrence rates and the sizes of the extended ionized outflows in luminous type 2 AGN and implications for
characteristic timescales and variability of accretion.

Kpc-scale ionized outflows are found to be common among luminous type 2
AGN. 
If we focus on objects with \lbol{} $\sim10^{46}$ \ergs, 13 of the 14 
objects in the combined \citet{Liu2013b} plus our sample host 10-kpc scale 
extended outflows based on our kinematic requirement. This gives a high occurrence rate of extended outflows $\gtrsim 90\%$. At a lower luminosity of \lbol$\sim10^{45-46}$ \ergs, a high fraction of these objects also host outflows (9/12), but the typical sizes of these outflows are smaller $\sim 1-3$ kpc. 
Using Gemini GMOS IFU to study luminous type 2 AGN (\lbol$=10^{45-46}$ \ergs{}) at $z=0.1-0.2$, \citet{Harrison2014} also finds all 16 of their AGN have outflows $> 6$ kpc. 
The \citet{Liu2013b} sample is selected purely based on \oiii{} luminosity, but the \citet{Harrison2014} sample could be biased by their high \oiiil{} line width selection. 
Likewise, our broadband image selection could potentially bias our sample. 
But the luminous type 2 AGN (\lbol$>10^{46}$ \ergs{}) in the parent \citet{Mullaney2013} sample also have a high fraction 
(59\%)\footnote{
To measure this fraction, we take the double Gaussian fits from \citet{Mullaney2013} and measure \weighty{} from the profiles. The fraction of \oiiil{} \weighty{} $> 600$ \kms{} objects is a strong function of the luminosity cut, which is 18\% (38\%) for \lbol$>10^{44}$ ($10^{45}$) \ergs{}. 
} 
of objects with high linewidths (\weighty{} $> 600$ \kms{}), likely indicating outflows. 

While it is a concern that beam smearing could lead to an overestimation of the outflow sizes, the occurrence rate of extended outflows is still high after such effects are taken into account. 
After subtracting the unresolved nuclear component,
\citet{Husemann2015} still recover high line widths (\weighty $>600$ \kms)
in the extended nebula in 
seven out of twelve ($60\%$) type 1 AGN from \citet{Liu2014a}. 
In the $z\sim 0.5$ sample of \citet{Liu2013b}, where the effect can be most 
severe, if we conservatively take out all four objects that could be 
considered as being marginally resolved\footnote{SDSS J0841+2042 and J1039+4512 have \oiiil{} surface brightness profiles
close to the PSF; SDSS J0149$-$0048, J0841+2014, and J0210$-$1001 have flat \weighty{} 
profiles that could be dominated by the nuclear component.}
 we still arrive at a occurrence rate of 60\%. 
Therefore, while most type 2 AGN studies suggest a high extended outflow 
occurrence rate of $\sim 90$\% among luminous AGN ($\lbol\gtrsim 10^{46}$ \ergs{}), we can place a conservative lower limits of $60$\% accounting for beam-smearing effects. 

To maintain such a high occurrence rate, each AGN outflow episode must be
much longer than the outflow dynamical timescale, to reduce the
probability of catching undersized outflows as they grow.  As it takes 
\tdyn$\sim 10^7$ years (Sec. \ref{sec:energetics:regress} and
Fig. \ref{fig:kimlum_regress}) to inflate a 10 
kpc-scale bubble with an observed
velocity of $\sim 1000$ \kms{}, these extended outflows have to be
launched at least $\sim 10^7$ yr in the past.  If $80\%$ of the luminous
AGN were active $\sim 10^7$ yr ago, the entire outflow episode has to
last for $\gtrsim 5\times10^7$ years.

It seems unlikely that the AGN stay luminous (\lbol $> 10^{46}$ \ergs{}) throughout the entire $\sim 10^8$ yr episode, as this
timescale is very similar to the total growth time of a massive black
hole $\sim 10^{7-8}$ yr \citep[e.g.,][inferred from quasar clustering and black hole mass density]{Soltan1982,Martini2001,Yu2002}.  Also, with this constant
energy supply, the outflow would continue to expand at a velocity of
$\sim 1000$ \kms{} and eventually reach a size of $\sim$ 100 kpc in
$10^8$ years, if the outflow is described by the energy-conserving model of
\citet{King2011}.  However, most systems in our sample and the
\citet{Liu2013b} sample with good signal-to-noise ratio at large radii do not 
show signs of extended outflows
beyond $\sim 10$ kpc, but rather have clear velocity drops on these
scales. 

Instead, we suggest it is far more natural that the AGN flickers on and off 
throughout this $\sim 10^8$ yr episode.  
In an analytical model by \citet{King2011} of an energy conserving outflow 
expanding in an isothermal potential of $\sigma = 200$ \kms{}, 
when the AGN is accreting close to its Eddington rate, the 
outflow will expand at a constant velocity $\sim 2000$ \kms{}. 
If the AGN is shut off after $10^6$ years, the outflow will still continue to 
expand due to its internal thermal energy, but it will slowly decelerate, 
until $\sim 10^7$ years later the velocity will drop below, say, 300 \kms{} 
and then stall. 
At this point the outflow has reached a size of $\sim$ 10 kpc, as calculated by 
\citet{King2011}. 
Therefore, to maintain the high observed duty cycle of outflows with sizes of a few to 10 kpc and
velocities about $1000$ \kms, there should be several AGN bursts, each $\sim 10^6$ 
year long with $\sim 10^7$ year intervals between them, 
so that if we observe a high luminosity AGN, often times it lights up the 
extended bubble driven by the previous AGN burst. 
Each AGN burst may even be shorter \citep[e.g., $10^5$ years,][]{Schawinski2015} 
and more frequent, as long as it supplies enough energy to sustain extended 
outflows throughout the episode. 

There are other reasons to favor such an AGN flickering model.
Theoretically, it is expected due to the episodic nature of 
gas cooling and feedback \citep{Novak2011}.  We have posited an AGN cadence
of $\sim 10^6$-year bursts with $10^7$-year intervals to explain one
particular system with multi-scaled ionized and molecular outflows
\citep{Sun2014}.  AGN variability on timescales $\lesssim 10^{7}$ years has
also been proposed to statistically tie star formation and AGN
activity, in a model that can successfully reproduce observed AGN
luminosity functions \citep{Hickox2014}.  Therefore, short-term AGN
variability ($\lesssim 10^7$ yr) over a long-term episode ($\sim 10^8$ yr)
appears to be a feasible scenario to explain the sizes and the
occurrence rate of extended outflows.

If the type 2 AGN studies \citep[this paper, ][]{Liu2013,Harrison2014} underestimate the impact of seeing 
and the occurrence rate is actually 60\% or lower, 
long outflow episodes with $\sim 10^8$-year duration would no longer be 
required. 
Furthermore, flickering may be in conflict with the energy requirements 
inferred from SZ observations of luminous AGN \citep{Crichton2016}. Finally, we note 
that these objects are all selected by virtue of their high \oiii{} 
luminosities, so we may be biased to objects in an outflow-dominated phase.

In summary, we estimate that extended (few - 10 kpc) ionized outflows are present in $> 60\%$ and possibly $90\%$ of all luminous type 2 AGN (\lbol $> 10^{46}$ \ergs{}).  
Given that the outflow formation times are $\sim 10^7$
years, such a high occurrence rate implies a long duration for each outflow
episode of $\sim 10^8$ years.  It is unlikely that
the AGN maintains a high luminosity (\lbol$>10^{46}$ \ergs) throughout this $10^8$-year episode.  
Instead, our observations suggest that AGN flicker on a
shorter time scale ($\lesssim 10^7$ years) and spend only $\sim$ 10\%
of their time in such a high luminosity state, and still maintain a high
occurrence rate of extended outflows.

\section{Summary}
\label{sec:summary}

We observe twelve luminous (\lbol$\sim10^{45.0-46.5}$ \ergs{}) nearby ($z\sim0.1$) type 2 (obscured) AGN with the Magellan IMACS long-slit spectrograph to study their ionized outflow properties using primarily the \oiiil{} line. 
These objects are selected from a parent sample of $\sim 24~000$ $z<0.4$ spectroscopically identified AGN from SDSS \citep{Mullaney2013} to have high \oiii{} and \emph{WISE} mid-IR luminosities as well as extended emission in SDSS images signaling extended ionized nebula. 

To increase the sample size for statistical and correlation analysis, we include two external samples from \citet{Liu2013b} and \citet{Hainline2014d} of luminous type 2 AGN to cover AGN luminosities from \lbol$=10^{45}$ to $10^{47}$ \ergs. The AGN luminosities in this paper are inferred from \emph{WISE} mid-IR luminosity at rest-frame 15 \micron. 

\medskip
The main results are as follows: 

\medskip  
(i) The radii of the narrow-line regions \riso{}, as defined by the
\oiiil{} isophotal radius, are 2 - 16 kpc in our sample. The exceptions are four unresolved objects and one that has a particularly large \riso{} of 33 kpc, which is most likely an ionized tidal feature. We find that \riso{} increases with the AGN luminosity at low AGN luminosities but flattens beyond a radius of $\sim$ 10 kpc, possibly due to change in the ionization state (Sec.
\ref{sec:sizes:relation}; Fig. \ref{fig:sizelum_whichR}).  \riso{} is
sensitive to the presence of gas at large radii such as extended tidal
features.

\medskip
(ii) A large fraction (9/12) of our objects have high \oiiil{} line-widths
(\weighty$>600$ \kms) indicating disturbed motions that are most likely
outflows, five of which are spatially resolved. To quantify the size of these outflows, we define \rv{} as the radius of the kinematically disturbed region 
where the \oiiil{} line-width \weighty{} is higher than $600$ \kms{} and the high velocity component ($|v| > 300$ \kms{}) is brighter than an isophotal threshold (see Sec. \ref{sec:sizes:rv}).
The resolved \rv{} are between 2 and 8 kpc and are typically smaller than \riso{} by a few kpc. \rv{} correlates strongly with the AGN luminosity. 
It is possible that the \rv{}-\Lfifteen{} relation also follows a flattened power-law that saturates at about 10 kpc at a higher luminosity, but more observations are needed to confirm. 
The best-fit power-law index of the \rv{}-\Lfifteen{} relation is $0.60 ^{+0.13}_{-0.13}$ assuming a single power law.

\medskip
(iii) Both the velocities and the dynamical time scales of the outflows show  correlations with AGN luminosity (Sec. \ref{sec:energetics:regress}, Fig. \ref{fig:kimlum_regress}). The outflow velocities range from a few hundred to 1500 \kms{} and scale with luminosity to a small power of $0.17 ^{+0.06}_{-0.07}$ and a large scatter. The dynamical time-scales are about \tdyn$\sim 10^{6.5-7}$ years and have a steeper scaling with luminosity of $0.52 ^{+0.16}_{-0.16}$. 

\medskip
(iv) The outflow masses and energetics are uncertain due to the unknown clumping factor of the \oiii{} emitting gas. 
We use two methods, which provide upper and lower limits, to constrain the energetics and the energy efficiency. The constraint on the efficiency is loose ($\eta=\dot{E}/\lbol=0.01\% - 30\%$) and there is no evidence that the outflow energy efficiency depends on the AGN luminosity (Sec. \ref{sec:energetics:regress}, Fig. \ref{fig:kimlum_energetics}).

\medskip
(v) There are three objects in our sample that have a high \oiiil{} linewidth
plateau of \weighty$\sim 600-1500$ \kms{} followed by a sudden linewidth drop at a few kpc (SDSS J1000+1242, J1010+1413, and J1255-0339). Such a \weighty{} profile is consistent with a constant-velocity outflow. The location of the velocity drop, which is captured by \rv{}, could correspond to the edge of the outflow where the shock fronts encounter the undisturbed galactic medium (Sec. \ref{sec:energetics:vprof}).

\medskip
(vi) The occurrence rate of extended outflows is high among luminous type 2 AGN ($> 60\%$, \lbol{}$\gtrsim 10^{46}$ \ergs{}).
Given the outflow dynamical time scales of $\sim 10^7$ years, to have such a
high occurrence rate, each outflow episode should last for $\sim 10^8$ years. 
While the AGN is unlikely to remain at a high luminosity the
entire time, the AGN could flicker on shorter time scales. For example, it could have several $\sim 10^6$-year-long bursts with $\sim 10^7$-year intervals between them. 
If the outflows are energy-conserving, each burst may drive a kpc-scale outflow that lasts for $\sim 10^7$ years \citep{King2011} to reproduce the high occurrence rate (Sec. \ref{sec:timescale}).

\medskip In this paper, we find that extended ionized outflows are common among
luminous type 2 AGN, with their sizes positively
correlated with the AGN luminosities.  It is important to extend these
measurements to lower luminosity AGN to test if this relation continues.  On
the other hand, the extended outflows identified in this paper (e.g., SDSS
J1000+1242, SDSS J1010+1413) provide good candidates for multi-wavelength
follow-up, e.g. in the sub-millimeter and X-ray, that can probe the other
relevant phases of the outflow (e.g., cold molecular and hot plasma) and
provide a more complete picture of the feedback processes. 
This work also confirms that optical broadband images can help identify extended ionized nebula. It is important to explore the potential of broadband imaging selection to find extended outflows in large imaging surveys, e.g., SDSS, HSC, or in the future LSST. Such a technique could help us explore the demographics of the most energetic AGN feedback systems.

\acknowledgments
{\bf Acknowledgments:}

A.-L. Sun is thankful for A. Dressler, D. Kelson, and E. Villanueva for assistance with the Magellan data reduction. A.-L. Sun thanks G. Liu and D. Wylezalek for communicating their research results. 
J.E. Greene acknowledges funding from the National Science Foundation under Grant No. AAG: \#1310405. 

This paper includes data gathered with the 6.5 meter Magellan Telescopes located at Las Campanas Observatory, Chile.
This publication makes use of data products from the Wide-field Infrared Survey Explorer, which is a joint project of the University of California, Los Angeles, and the Jet Propulsion Laboratory/California Institute of Technology, funded by the National Aeronautics and Space Administration.
This research made use of Astropy, a community-developed core Python package for Astronomy\footnote{http://www.astropy.org}.
This work made use of packages PyAstronomy.
PyRAF is a product of the Space Telescope Science Institute, which is operated by AURA for NASA

\appendix
\label{sec:append}

\section{Individual Objects}
\label{sec:append:objs}

The Magellan spectroscopic data for all of our sources are displayed in this appendix (Fig. \ref{fig:spec2200} to \ref{fig:spec3005}), except for the two objects SDSS J1000+1242 and J1010+1413 that are shown in Fig. \ref{fig:spec2100} and \ref{fig:spec2101}, as they are described in the main text. 
The majority of the objects have not been studied in detail in the literature, except for SDSS J1000+1242, SDSS J1010+1413, and SDSS J1419+0139, described below. Another object, SDSS J1255$-$0339 is also discussed here for its abnormally extended narrow line region.

\medskip
{\it SDSS J1000+1242. } This system was observed by \citet{Harrison2014} with Gemini GMOS IFU. 
Their IFU observation reveals regions of broad line widths (up to \weighty{} of 850 \kms{}) with kinematic size of 14 kpc, roughly consistent with our observation. 
As this broad component shows a clear velocity gradient, they suggest that it is a pair of bi-polar super-bubbles. 

The more extended narrow-line component is also partly seen in this IFU data, but is limited to the central 3-4\arcsec{} due to its small field-of-view. Our observation confirms that this component extends to about 10\arcsec{}, roughly the same as the SDSS optical image.

\medskip
{\it SDSS J1010+1413.} As J1000+1242, this system was observed by \citet{Harrison2014} with the Gemini GMOS IFU, which reveals a very broad \oiii{} component of \weighty{}=1450 \kms{}, an unambiguous sign of high velocity outflows. The size of outflow was not constrained by \citet{Harrison2014} due to the limited field-of-view, but is measured in our Magellan data to have a radius of \rv{} = 8 kpc. 

Our Magellan slit is placed along the minor axis of the galaxy to capture the two bright green blobs in the SDSS image, which signal \oiii{} emission. \citet{Harrison2014} observed the inner parts of these two features and found narrow \oiii{} emissions separated by $\sim 350$ \kms{} in velocity. Our Magellan spectra confirms that these narrow emission clouds extent to $\sim$ 16 kpc each from the nucleus. They could be galactic medium being passively illuminated by ionization cones or parts of bipolar outflows. 

While \citet{Harrison2014} selects targets based on broad \oiii{} line widths and ours are based on the \oiii{} extent, it is interesting that both samples pick up the two powerful outflows J1000+1242 and J1010+1413. Possibly both the high velocity and extended \oiii{} are results of powerful AGN feedback.

\medskip
{\it SDSS J1255$-$0339. } 
This object has received little attention in the literature, but it has a spectacular pair of extended green spiral features of size about 60 kpc in the SDSS image, most likely tidal tails. 
Our Magellan spectra reveal narrow \oiii{} of width \weighty $\lesssim$ 300 \kms{} all along these features, making it the most extended narrow line region in the sample. 
These tidal features are likely ionized by the central AGN, as the \oiii{} to \hbeta{} ratios are about 10.
The system's high infrared luminosity (classified as a ULIRG by \citet{KilerciEser2014}) and complex nuclear morphology also suggest that it may be in the late stages of a merger.

\medskip
{\it SDSS J1419+0139.} This target was observed by \citet{McElroy2014} with AAT's SPIRAL IFU, which finds a spatially resolved \oiii{} emitting region with a moderate line width \weighty{}$_\mathrm{, max}$ of 529 \kms{}, consistent with our observations. Its SDSS image reveals a extended tidal tail indicating merging activities.

\section{Slit Widths of the Magellan IMACS Centerfield Slit-viewing Spectrograph}
\label{sec:append:slit}

We inspect the slit widths of the Magellan IMACS Centerfield Slit-viewing Spectrograph, and find that the widest of its five slits, referred to as the 1.5\arcsec{} slit in the IMACS User Manual, has an actual slit width of 1.3\arcsec{}. 

This result is confirmed by comparing the line widths of the calibrating arc lamp observed through these five slits. 
As shown in  Fig. \ref{fig:slitwidths}, the line widths of the first four slits follow the relation 
\begin{equation}
   w_l^2 = w_0^2 + r W_s^2, 
\end{equation}
where $w_l$ is the observed arc line width, and $W_s$ is the slit width --
 0.25\arcsec{}, 0.50\arcsec{}, 0.75\arcsec{}, and 1.0\arcsec{}. 
The intercept $w_0^2$ and the slope $r$, are fixed by linear regression of these four slits. 
However, the fifth slit has an arc line width narrower than expected if the slit width were 1.5\arcsec{}. It is instead consistent with a slit width of 1.3\arcsec{}.

\section{\emph{WISE} Luminosities of Type 1 and Type 2 AGN}
\label{sec:append:WISE}

\emph{WISE} mid-IR luminosities have been used to determine the AGN 
bolometric luminosities. 
However, type 2 AGN in general have redder \emph{WISE} 
colors compared to their type 1 counterparts \citep{Yan2013,Liu2013,Zakamska2016}, such 
that the inferred bolometric luminosities for type 2 AGN can be 
underestimated compared to type 1 at shorter mid-IR wavelengths. Therefore, 
one should be cautious when using mid-IR to compare the luminosities between 
type 1 and type 2 AGN. 

We investigate the difference in \emph{WISE} mid-IR luminosities between 
type 1 and type 2 AGN at three different wavelengths -- rest-frame 8 \micron{}
, 15 \micron{}, and 22 \micron{} -- using the sample of SDSS spectroscopically selected luminous AGN from \citet{Mullaney2013}. 
We use the luminous AGN at redshifts $0<z<0.2$ that have \oiiil{} luminosities above $L_{\rm{[OIII]}}>5\times10^{41}$ \ergs{}, similar to our Magellan sample. 365 of these objects are type 1 and 546 are type 2. As shown on the lower right of Fig. \ref{fig:KStest}, the type 1 and type 2 AGN have similar $L_{\rm{[OIII]}}$ distributions that are indistinguishable by a KS test with a high $p$-value of 0.41.

As shown in Fig. \ref{fig:KStest}, at fixed \oiii{}
luminosities, we find that the 8 \micron{} luminosities of the type 1 AGN are higher than the type 2 AGN by 0.2 dex. This difference is statistically significant with a KS-test $p$-value of $4\times10^{-10}$. This discrepancy is much smaller at 15 \micron{} (0.07 dex, $p$-value of 0.02), and negligible at 22 \micron{} (0.002 dex, $p$-value of 0.67). 
At a fixed X-ray luminosity, such a discrepancy has also been found between type 1 and type 2 AGN \citep{Burtscher2015}. 
These tests suggest that at a given intrinsic luminosity, the mid-IR luminosity of an AGN depends on its spectral type. Such an effect is especially severe at lower wavelengths, e.g. 8 \micron, and grows less significant for longer wavelengths, e.g. 15 - 22 \micron. 

With a sample of both type 1 and type 2 AGN, \citet{Liu2014a} find a
flattening at the high luminosity end of the \oiii{} nebula size - 8 \micron{} luminosity relation. However, they suspect that the flattening is an artifact caused by the higher mid-IR luminosity of type 1 AGN. 
We revisit this relation with a larger sample of objects from this paper, \citet{Liu2013b}, \citet{Liu2014a}, and \citet{Hainline2014d}. We also include the eight type 1 AGN observed in the same Magellan run as in this paper. 
As shown in Fig. \ref{fig:sizelum_whichWISE}, we find that the type 1 and type 2 AGN follow different nebula size - 8 \micron{} luminosity relations, such that adding luminous type 1 AGN to a sample of type 2 AGN can indeed result in or exaggerate the apparent flattening of the relation. 
However, if we use longer mid-IR wavelengths, say, 15 \micron{} (right panel), where the effect is less significant, the separation between the size - luminosity relations of the type 1 and type 2 AGN becomes smaller, and the flattening becomes less obvious. 

Therefore, combining type 1 and type 2 AGN samples to study their nebula size - mid-IR luminosity relations can be misleading, especially at shorter wavelengths such as 8 \micron{}. To use mid-IR luminosities as an AGN luminosity indicator, longer wavelengths, such as 15 \micron{} can be more robust against variations in AGN spectral types.


\section{Simulations of Bias and Uncertainty in \weighty{}}
\label{sec:append:sim_w80}
The \weighty{} measurement on a 1-D line spectrum could be affected by the instrumental spectral PSF and the noise. To quantify the biases and the uncertainties in \weighty{} due to these effects, we perform a series of 1-D simulations.

We simulate the 1-D spectrum of the \oiiil{} line with double Gaussian profiles with a range of line widths ($\sigma_{narrow}=$ 100 - 500 \kms{}), flux ratios ($F_{broad}/F_{narrow}= 0.1-10$), and width ratios ($\sigma_{broad}/\sigma_{narrow}= 1.5-3$). 
These simulated lines are convolved with the empirical spectral PSF measured from the arc frames. Gaussian noise is then inserted into the convolved double Gaussian line profiles.

We measure the \weighty{} of the original spectrum ($w_{80, \mathrm{model}}$), the one convolved with the PSF ($w_{80, \mathrm{convl}}$), and the one with noise ($w_{80, \mathrm{noise}}$), using the same method as described in Sec. \ref{sec:sizes:spec}. 
As shown on the left panel of Fig. \ref{fig:sim_w80}, the bias of \weighty{} due to the PSF is not a strong function of the detailed line shape but just depends on the line width. This relation is well fitted by the quadratic mean function
\begin{equation}
    w_{80, \mathrm{convl}}^2 = w_{80, \mathrm{model}}^2 + w_{80, \mathrm{inst}}^2,
\end{equation}
where $w_{80, \mathrm{inst}}$ is the constant instrumental resolution, which is 243 \kms{} for the 1\farcs0 slit and 282 \kms{} for the 1\farcs3 slit. 
The random noise introduces random uncertainties and a bias to the \weighty{} measurements. The bias is negligible for SNR $> 10$ but can be significant for low signal-to-noise data. 
We define 
\begin{equation}
 w_{80,\/ \mathrm{err}} = \langle w_{80,\/ \mathrm{noise}} - w_{80,\/ \mathrm{convl}}\rangle_{\mathrm{RMS}}
\end{equation}
to encompass both effects. 
We find that $w_{80,\/ \mathrm{err}}$ depends on both the signal-to-noise ratio and the width of the line, which can be fitted by a 2-dimensional 3rd-order polynomial function (right panel of Fig. \ref{fig:sim_w80}), and used to assign uncertainties to our \weighty{} measurements.

According to these results, for a typical line with a measured \weighty{} of 600 \kms{} and a peak signal-to-noise ratio of 30, both the bias and the random uncertainty on \weighty{} are about 10 \% (60 \kms). For wider lines or higher signal-to-noise ratios the correction and the noise level are even lower. 
We apply this spectral PSF correction and assign the errors for the \weighty{} measurements using the best-fit functions described above. The corrected \weighty{} profiles and their errors are shown in Fig. \ref{fig:spec2100}, \ref{fig:spec2101}, and Appendix \ref{sec:append:objs}. Those corrections do not affect our conclusions. 

\section{Simulations for Size Biases due to the spectral and spatial PSF}
\label{sec:append:sim_pv}
The finite spatial and spectral resolution could lead to overestimation of the \riso{} and \rv{} measurements, or lead us to tag an object as resolved that is not.
To quantify this effect, we perform a series of 2-D spectrum ($pv$-diagram) simulations, see Fig. \ref{fig:sim_pv}.
The components of the galaxies are modeled as 2-D Gaussians. But the spectral and spatial PSF are empirically measured from the data, not Gaussian functions. A flux calibration star with a seeing of FWHM=1\farcs0 is used to measure the spatial PSF. 
We use the results of these simulations to determine the criteria for whether the narrow line region or the kinematically disturbed region is spatially resolved and to estimate any bias in the size measurements. 

To cover the wide variety of kinematic structures 
measured in our sources, the simulated $pv$-diagram consists of four components: a narrow nuclear component, a blue-shifted broad nuclear component ($-200$ \kms{}), and a pair of narrow rotating components on the blue and red sides ($\pm200$ \kms{}; to represent typical edge-on galaxies). Each component is represented by a 2-D Gaussian. The velocity widths of the narrow and broad components are fixed to be $\sigma=$100 \kms{} and 600 \kms{}, respectively. The rotating components have symmetric spatial offsets from the nucleus. 
The rotating and the broad nuclear components, when used, have fluxes 20\% and 50\% of the narrow nuclear component, respectively. The sizes of all the components and the spatial offsets in the rotating components can take a range of values from 0\farcs1 to 5\arcsec. 
We then convolve this simulated $pv$-diagram with the empirical spatial and spectral PSF, and compare the changes in the total light profiles, the red and blue wing light profiles, the \weighty{} profiles, and the measured \riso{} and \rv{}. 

We find that the narrow nuclear and the rotating components alone cannot produce \weighty{} $>$ 600 \kms{}. So the \weighty{} = 600 \kms{} cut is a good discriminant for the presence of the broad component, independent of its size. 
Compact objects that are of sizes $\sigma \lesssim$ 0\farcs3 have their light profiles consistent with the PSF, independent of its velocity structure.
So the total light profile is a good indicator for whether the narrow line region is resolved, see panel (a.) of Fig. \ref{fig:sim_pv}. 
For the kinematically disturbed region, we find that when the broad component is compact ($\sigma \lesssim$ 0\farcs3), the core of its blue ($v<-300$ \kms{}) and red ($v>300$ \kms{}) wing light profiles are consistent with the PSF, even if the narrow components are extended or have rotation. But the extended rotation features can affect the wing light profiles at a fainter level ($<10^{-1}$ of the core) to make them deviate from the PSF, see panel (b.) of Fig. \ref{fig:sim_pv} .
This could be the reason why SDSS J2154+1131 appears to have a resolved broad component from its red wing light profile while its \weighty{} is low. 
So for a kinematically disturbed region to be determined as unambiguously resolved, it has to have the main core of its red or blue light profiles deviated from the PSF or mismatched with each other. 

For the sizes, we mimic the methods described above to measure \riso{} and \rv{} for the simulated data. 
For \riso{}, we adopt an isophotal threshold a factor of 10 lower than the peak intensity, which is comparable to the real measurements. 
The PSF convolved \riso{} is always about 1\arcsec{} for compact unresolved objects ($\sigma \lesssim$ 0\farcs3), so it is important to treat the \riso{} measurements of those objects as upper limits. For resolved objects, the bias in \riso{} due to the PSF is between 0\arcsec and 0\farcs5, and becomes negligible for large objects of \riso{}$>4$\arcsec, but the level of bias depends on the detailed shape of the light profile and cannot be easily corrected. 

For the \rv{} measurements here, we also correct for the \weighty{} bias according to Appendix \ref{sec:append:sim_w80} before measuring the \rv{}. 
The \rv{} of an unresolved ($\sigma \lesssim$ 0\farcs3), kinematically disturbed region is over-estimated, so it is also important to treat those numbers as upper limits. 
In general, the \rv{} of resolved objects can also be over-estimated by up to 1\arcsec{} with an average of $\lesssim$ 0\farcs5 (which corresponds to $1/2\times$ seeing FWHM) for both the 1\farcs0 and the 1\farcs3 slit, see panel (c.) of Fig. \ref{fig:sim_pv}. This amount also depends on the object and cannot be easily corrected. 
The only exception is when the size of the broad component is comparable to or larger than the narrow component, in which case the same broad line shape is propagated to large radii by the PSF, making the \weighty$>$600 \kms{} region unrealistically large, see panel (d.) of Fig. \ref{fig:sim_pv}. 
But this issue can be resolved by adding a surface brightness constraint to the kinematically disturbed region. 

The bias in \riso{} and \rv{} sizes due to the PSF, which is estimated to be $\sim$ 0\farcs5, should dominate over the noise to be the main uncertainties on \riso{} and \rv{}.
But the exact amount of the bias depends on the structure of the 2-D spectrum and thus cannot be easily quantified. 
We do not apply PSF correction but assign $\pm$ 0\farcs5 error to our size measurements to encompass this uncertainty.

\bibliography{local}

%

%

%

%
%

%
%
%

\clearpage

\begin{deluxetable*}{cccccccccc}
\tablecaption{The Sample}
\tablehead{\colhead{Name} & \colhead{RA} & \colhead{Dec} & \colhead{z} & \colhead{$L_{\mathrm{[OIII],}~42}$} & \colhead{$L_{bol,~\mathrm{[OIII],}~45}$} & \colhead{$\nu L_{\mathrm{\nu, 15,}~44}$} & \colhead{$L_{bol,~\mathrm{IR},~45}$} & \colhead{P.A.} & \colhead{$t_{\mathrm{exp}}$}\\ 
\colhead{ } & \colhead{ } & \colhead{ } & \colhead{ } & \colhead{[\ergs{}]} & \colhead{[\ergs{}]} & \colhead{[\ergs{}]} & \colhead{[\ergs{}]} & \colhead{[degree]} & \colhead{[minutes]}\\
\colhead{(1)} & \colhead{(2)} & \colhead{(3)} & \colhead{(4)} & \colhead{(5)} & \colhead{(6)} & \colhead{(7)} & \colhead{(8)} & \colhead{(9)} & \colhead{(10)}
}
\startdata
J0141$-$0945 & 01:41:25 & $-$09:45:16 & 0.143(5) & 0.5 & 1.4 & 2.9 & 2.6    & 360$^a$ & 30  \\
J1000+1242 & 10:00:13 & +12:42:26     & 0.148(2) & 4.1 & 10.4 & 12.7 & 11.5 & 180 & 15  \\
J1010+1413 & 10:10:23 & +14:13:01     & 0.198(7) & 13.4 & 34.1 & 30.0 & 27.0& 360 & 15  \\
J1055+1102 & 10:55:55 & +11:02:52     & 0.145(6) & 3.2 & 8.3 & 1.8 & 1.6    & 180$^b$ & 15  \\ 
J1255$-$0339 & 12:55:48 & $-$03:39:10 & 0.169(6) & 0.6 & 1.6 & 4.8 & 4.3    & 310$^b$ & 15  \\
J1351+0728 & 13:51:11 & +07:28:46     & 0.150(1) & 1.7 & 4.3 & 4.0 & 3.6    & 323 & 30  \\
J1419+0139 & 14:19:26 & +01:39:36     & 0.076(7) & 0.4 & 0.9 & 2.0 & 1.8    & 188 & 30  \\
J2102$-$0647 & 21:02:05 & $-$06:47:56 & 0.126(4) & 0.2 & 0.6 & 3.7 & 3.3    & 150 & 30  \\
J2133$-$0712 & 21:33:33 & $-$07:12:49 & 0.086(4) & 0.2 & 0.6 & 2.8 & 2.5    & 130$^b$ & 30  \\
J2142+0001 & 21:42:48 & +00:01:57     & 0.168(2) & 1.0 & 2.6 & 0.9 & 0.8    & 237$^a$ & 60  \\
J2154+1131 & 21:54:26 & +11:31:29     & 0.108(8) & 1.2 & 3.1 & 2.6 & 2.3    & 335 & 30  \\
J2333+0049 & 23:33:13 & +00:49:12     & 0.170(0) & 1.4 & 3.7 & 4.4 & 3.9    & 143 & 15  \\
\enddata
\tablecomments{Details of the Magellan sample of luminous type 2 AGN. 
Column 1: Object SDSS Name. 
Column 2, 3: Object coordinate from SDSS DR7 (J2000). 
Column 4: Systemic redshift from the stellar absorption features in the Magellan spectra, see Sec. \ref{sec:data:selection}. The last digit is uncertain by $\pm 1$ given the uncertainty on the systemic velocity of 15 \kms{}. 
Column 5, 6: The total \oiiil{} luminosity in units of $10^{42}$ \ergs{} from the SDSS spectrum \citep{Mullaney2013} and its corresponding bolometric luminosity in units of $10^{45}$ \ergs{} converted using the bolometric correction from \citet[][]{Liu2009}. 
Column 7, 8: The mid-IR rest frame 15 \micron{} luminosity in units of $10^{44}$ \ergs{} from WISE and the corresponding bolometric luminosity in units of $10^{45}$ \ergs{} with a bolometric correction factor of 9, see Sec. \ref{sec:data:WISE}. 
Column 9, 10: The position angle and exposure time of the Magellan long-slit exposure. 
\\
$^a$ The object is observed with the 1.3\arcsec slit instead of the 1.0\arcsec slit. \\
$^b$ Multiple slit positions are taken and only the representative slit, which is used for the measurements, is listed, see Sec. \ref{sec:sizes:spec}. \\
}
\label{tab:sample}
\end{deluxetable*}

\begin{deluxetable}{ccccccccc}
\tablecaption{The Measurements}
\tablehead{
\colhead{Name} & \colhead{NLR} & \colhead{KDR} & \colhead{\riso} & \colhead{\rv} & \colhead{\weightyAVG} & \colhead{$\log($\tdyn$)$} & \colhead{$\log($\dotEkin$)$} & \colhead{$\log($\dotEST$)$}
\\ 
\colhead{ } & \colhead{ } & \colhead{ } & \colhead{[$\mathrm{kpc}$]} & \colhead{[$\mathrm{kpc}$]} & \colhead{[\kms{}]} & \colhead{[$\mathrm{Myr}$]} & \colhead{[$\mathrm{erg\,s^{-1}}$]} & \colhead{[$\mathrm{erg\,s^{-1}}$]}
\\
\colhead{(1)} & \colhead{(2)} & \colhead{(3)} & \colhead{(4)} & \colhead{(5)} & \colhead{(6)} & \colhead{(7)} & \colhead{(8)} & \colhead{(9)}
}
\startdata
J0141$-$0945 & R & R &  $2.5\pm1.3$      & $1.5\pm1.3$      &  $1208\pm20$  & $6.2\pm0.4$      & \nodata$^d$       & $44.0\pm0.8$ \\
J1000+1242   & R & R &  $15.0\pm1.3$     & $6.2\pm1.3$      &   $753\pm20$  & $7.0\pm0.1$      & $43.4\pm0.3$      & $44.6\pm0.3$ \\
J1010+1413   & R & R &  $16.4\pm1.6$     & $8.2\pm1.6$      &  $1458\pm20$  & $6.9\pm0.1$      & $43.8\pm0.3$      & $45.7\pm0.3$ \\
J1055+1102   & R & N &  $6.6\pm1.3$      & \nodata$^c$      &   $467\pm20$  & \nodata$^c$      & \nodata$^c$       & \nodata$^c$ \\
J1255$-$0339 & R & R &  $33.5\pm1.4$$^a$ & $3.2\pm1.4$      &   $637\pm20$  & $6.8\pm0.2$      & $42.2\pm0.3$      & $43.8\pm0.5$ \\
J1351+0728   & R & N &  $5.2\pm1.3$      & \nodata$^c$      &   $263\pm20$  & \nodata$^c$      & \nodata$^c$       & \nodata$^c$ \\
J1419+0139   & U & U & $<2.3\pm0.7$$^b$  & $<1.2\pm0.7$$^b$ &   $604\pm20$  & $<6.4\pm0.3$$^b$ & $>41.9\pm0.4$$^b$ & $<42.8\pm0.6$$^b$ \\
J2102$-$0647 & U & U & $<2.3\pm1.1$$^b$  & $<1.1\pm1.1$$^b$ &   $632\pm20$  & $<6.4\pm0.4$$^b$ & $>40.5\pm0.5$$^b$ & $<42.9\pm0.9$$^b$ \\
J2133$-$0712 & U & U & $<1.9\pm0.8$$^b$  & $<2.1\pm0.8$$^b$ &   $948\pm20$  & $<6.5\pm0.2$$^b$ & \nodata$^b$$^d$  & $<43.9\pm0.4$$^b$ \\
J2142+0001   & U & U & $<2.9\pm1.4$$^b$  & $<1.7\pm1.4$$^b$ &   $797\pm20$  & $<6.4\pm0.4$$^b$ & $>41.6\pm0.4$$^b$ & $<43.5\pm0.8$$^b$ \\
J2154+1131   & R & N &  $3.2\pm1.0$      & \nodata$^c$      &   $564\pm20$  & \nodata$^c$      & \nodata$^c$       & \nodata$^c$ \\
J2333+0049   & R & R &  $4.1\pm1.4$      & $3.8\pm1.4$      &   $787\pm20$  & $6.8\pm0.2$      & $41.8\pm0.3$      & $44.2\pm0.4$

\enddata
\tablecomments{
Measurements of the Magellan sample of luminous type 2 AGN. 
Column 1: SDSS Name. 
Column 2: Whether the narrow line region is resolved (R) or unresolved (U). 
Column 3: Whether the kinematically disturbed region (outflow) is resolved (R), unresolved (U), or non-existent (N). 
Column 4: Radius of the narrow line region. 
Column 5: Radius of the kinematically disturbed region, that has high \oiiil{} line widths \weighty$>600$ \kms. 
Column 6: Averaged \oiiil{} line width \weighty{} (the width that encloses 80\% of the flux). It is used to infer the outflow velocity $v=$\weighty$/1.3$. The error is conservatively assumed to be 20 \%. 
Column 7: Outflow dynamical time scale \tdyn$=$\rv$/v=1.3\times$\rv$/$\weighty. 
Column 8: Outflow kinetic power \dotEkin{} based on the ionized gas mass inferred from the \hbeta{} intensity. These numbers are underestimates as the mass in the more diffuse gas is not accounted for. 
Column 9: Sedov-Taylor kinetic power \dotEST{} for a supernova-like bubble of size \rv{} and velocity $v=$\weighty$/1.3$. This is an overestimation of outflow kinetic power as it assumes all the enclosed gas mass participates in the outflow, see Sec. \ref{sec:energetics:definition}.
\\ \\
$^a$ The \riso{} measurement is affected by extended tidal features. It is marked as a cross in Fig. \ref{fig:sizelum_whichR}. \\
$^b$ The source is unresolved, so the \riso{}, \rv{}, \tdyn{}, and \dotEST{} are treated as upper-limits, and \dotEkin{} as lower-limits. These are plotted as triangles corresponding to the limits. \\
$^c$ There is no kinematically disturbed region with \weighty $> 600$ \kms{}. These are plotted as empty squares. \\
$^d$ \Ekin{} is not available as the \hbeta{} measurement is affected by observational defects. These points are not plotted.  \\
}
\label{tab:measurements}
\end{deluxetable}

\begin{deluxetable}{cccc}
\tablecaption{The Liu 2013 Sample}
\tablehead{
\colhead{Name} & \colhead{z} & \colhead{$\nu L_{\mathrm{\nu, 15,}~44}$} & \colhead{$L_{bol,~\mathrm{IR,}~45}$} \\ 
\colhead{ } & \colhead{ } & \colhead{[\ergs]} & \colhead{[\ergs]} \\
\colhead{(1)} & \colhead{(2)} & \colhead{(3)} & \colhead{(4)} 
}
\startdata
J0149$-$0048     & 0.5670 & 25.3 & 22.7 \\
J0210$-$1001     & 0.5400 & 9.0  & 8.1  \\
J0319$-$0019     & 0.6350 & 16.9 & 15.2 \\
J0319$-$0058     & 0.6260 & 12.7 & 11.4 \\
J0321$+$0016     & 0.6430 & 14.7 & 13.2 \\
J0759$+$1339     & 0.6490 & 50.5 & 45.5 \\
J0841$+$2042     & 0.6410 & 22.3 & 20.0 \\
J0842$+$3625     & 0.5610 & 15.1 & 13.6 \\
J0858$+$4417     & 0.4540 & 76.4 & 68.8 \\
J1039$+$4512     & 0.5790 & 27.2 & 24.5 \\
J1040$+$4745     & 0.4860 & 44.8 & 40.3 \\
J0224$+$2750$^a$ & 0.3110 & 3.8  & 3.4  \\
J0807$+$4946$^a$ & 0.5750 & 21.0 & 18.9 \\
J1101$+$4004$^a$ & 0.4570 & 19.6 & 17.6 \\
\enddata
\tablecomments{The redshifts and luminosities of the \citet{Liu2013} sample. 
The attributes are as described in Tab. \ref{tab:sample}. 
\\
$^a$ Radio loud or intermediate. 
}
\label{tab:sample_liu13}
\end{deluxetable}

\begin{deluxetable}{ccccccccc}
\tablecaption{The Measurements for the Liu 2013 Sample}
\tablehead{
\colhead{Name} & \colhead{NLR} & \colhead{KDR} & \colhead{\riso} & \colhead{\rv} & \colhead{\weightyAVG}  & \colhead{$\log($\tdyn$)$} & \colhead{$\log($\dotEkin$)$} & \colhead{$\log($\dotEST$)$} \\
\colhead{ } & \colhead{ } & \colhead{ } & \colhead{[kpc]} & \colhead{[kpc]} & \colhead{[\kms]} & \colhead{[Myr]} & \colhead{[\ergs]} & \colhead{[\ergs]}\\
\colhead{(1)} & \colhead{(2)} & \colhead{(3)} & \colhead{(4)} & \colhead{(5)} & \colhead{(6)} & \colhead{(7)} & \colhead{(8)} & \colhead{(9)} 
}
\startdata
J0149$-$0048     & R & R & $6.9  \pm 1.7$ & $ 8.2 \pm 1.7$     & $1167 \pm 75$ & $7.0\pm0.1$ & $41.7\pm0.3$ & $45.4\pm0.3$ \\
J0210$-$1001     & R & R & $14.4 \pm 1.8$ & $ 9.3 \pm 1.8$     & $ 667 \pm 75$ & $7.2\pm0.1$ & $42.0\pm0.3$ & $44.8\pm0.3$ \\
J0319$-$0019     & R & R & $6.7  \pm 1.3$ & $ 6.4 \pm 1.3$     & $ 1845\pm 75$ & $6.6\pm0.1$ & \nodata$^b$ & $45.8\pm0.3$ \\
J0319$-$0058     & R & R & $9.6  \pm 2.0$ & $ 6.2 \pm 2.0$     & $ 780 \pm 75$ & $7.0\pm0.1$ & $41.8\pm0.3$ & $44.6\pm0.4$ \\
J0321$+$0016     & R & R & $16.6 \pm 1.9$ & $ 7.7 \pm 1.9$     & $ 974 \pm 75$ & $7.0\pm0.1$ & $42.0\pm0.3$ & $45.1\pm0.3$ \\
J0759$+$1339     & R & R & $12.9 \pm 2.1$ & $ 9.8 \pm 2.1$     & $1230 \pm 75$ & $7.0\pm0.1$ & $42.5\pm0.3$ & $45.6\pm0.3$ \\
J0841$+$2042     & R & R & $10.0 \pm 1.8$ & $ 6.3 \pm 1.8$     & $ 723 \pm 75$ & $7.0\pm0.1$ & $41.9\pm0.3$ & $44.5\pm0.4$ \\
J0842$+$3625     & R & N & $13.1 \pm 1.5$ &       \nodata$^a$  & $ 489 \pm 75$ & \nodata$^a$ & \nodata$^a$ & \nodata$^a$ \\
J0858$+$4417     & R & R & $10.6 \pm 1.7$ & $ 5.9 \pm 1.7$     & $ 876 \pm 75$ & $6.9\pm0.1$ & $42.3\pm0.3$ & $44.7\pm0.4$ \\
J1039$+$4512     & R & R & $9.3  \pm 1.9$ & $ 9.2 \pm 1.9$     & $1105 \pm 75$ & $7.0\pm0.1$ & $42.4\pm0.3$ & $45.4\pm0.3$ \\
J1040$+$4745     & R & R & $13.0 \pm 2.2$ & $ 11.2\pm 2.2$     & $1315 \pm 75$ & $7.0\pm0.1$ & $42.9\pm0.3$ & $45.8\pm0.3$ \\
J0224$+$2750$^c$ & R & R & $10.1 \pm 1.2$ & $ 2.9 \pm 1.2$     & $ 688 \pm 75$ & $6.7\pm0.2$ & $42.2\pm0.3$ & $43.8\pm0.4$ \\
J0807$+$4946$^c$ & R & R & $14.6 \pm 2.1$ & $ 16.1\pm 2.1$     & $ 714 \pm 75$ & $7.5\pm0.1$ & $41.5\pm0.3$ & $45.3\pm0.3$ \\
J1101$+$4004$^c$ & R & R & $18.7 \pm 1.8$ & $ 10.0\pm 1.8$     & $ 753 \pm 75$ & $7.2\pm0.1$ & $42.1\pm0.3$ & $45.0\pm0.3$
\enddata 
\tablecomments{Measurements of the \citet{Liu2013} sample. 
The attributes are as described in Tab. \ref{tab:measurements}. 
Column 4, 5, and 6 are taken from \citet{Liu2013b}: 
column 4 is directly from $R_\mathrm{int}$ of Table 2; 
column 5 \rv{} is read from Fig. 5; 
column 6 is $\langle w_{80} \rangle$ from Table 1. 
The later columns are derived from these two using the same formulas as for the main sample. 
The \hbeta{} luminosities for \dotEkin{} are from the \oiiil{} luminosity and \oiiil{}-\hbeta{} line ratios from \citet{Liu2013b}. 
The errors are as described in Sec. \ref{sec:energetics:definition}. 
\\
$^a$ There is no kinematically disturbed region with \weighty $> 600$ \kms{}. These are plotted as empty squares. \\
$^b$ \Ekin{} is not calculated as the \oiiil{} to \hbeta{} ratio is not available. \\
$^c$ Radio loud or intermediate. 
}
\label{tab:measurements_liu13}
\end{deluxetable}

\begin{deluxetable*}{cccccccccccc}
\tablewidth{0pc}
\tablecaption{Comparison of the \riso{} and \rv{} size-luminosity relations}
\tablehead{
& &\multicolumn{3}{c}{Single Power-Law} &\multicolumn{5}{c}{Flattened Power-Law} &\multicolumn{2}{c}{Comparison}\\
\cmidrule(r){3-5} \cmidrule(r){6-10}  \cmidrule(r){11-12}
\colhead{$x$} & \colhead{$y$} & \colhead{$\alpha$} & \colhead{$\beta$} & \colhead{$\rm{BIC}$} & \colhead{$\alpha$} & \colhead{$\beta$} & \colhead{$\log(L)$} & \colhead{$\log(R)$} & \colhead{$\rm{BIC}$} & \colhead{$\Delta \rm{BIC}$} & \colhead{Preferred Model} \\
\colhead{} & \colhead{} & \colhead{(1)} & \colhead{(2)} & \colhead{(3)} & \colhead{(4)} & \colhead{(5)} & \colhead{(6)} & \colhead{(7)} & \colhead{(8)} & \colhead{(9)} & \colhead{(10)}
}
\startdata
\Lfifteen & \riso{} & $-$12.29 & 0.30 & 305.99 & $-$31.12 & 0.72 & 44.84 & 1.09 & 292.24 & 13.75 & Flattened Power-Law \\
\Lfifteen & \rv{}   & $-$19.81 & 0.46 & 118.73 & $-$33.31 & 0.76 & 45.32 & 1.02 & 106.15 & 12.58 & Flattened Power-Law
\enddata
\tablecomments{The best-fit model parameters of the \riso{} and \rv{} size-luminosity relations, see Fig. \ref{fig:sizelum_whichR}. 
Column 1-3: The parameters and the Bayesian information criterion of the power-law model $\log(y)=\alpha+\beta \log(x)$. Column 4-8: Same as Column 1-3 but for a power-law that flattens beyond the point [$\log(L), \log(R)$]. Here $L$ stands for \Lfifteen{}. Column 9: The difference between the Bayesian information criteria of the two models, where positive means that the data prefers the flattened power-law and negative means otherwise, as shown in Column 10. Only valid size measurements (circles in Fig. \ref{fig:sizelum_whichR}, not limits, and not the \riso{} outlier J1255$-$0339) are used in these fits. The units are \ergs{} for the luminosity and kpc for the radius. 
}
\label{tab:whichR}
\end{deluxetable*}

\begin{deluxetable*}{cccccc}
\tablewidth{0pc}
\tablecaption{Relation between the Outflow Properties and the AGN Luminosities}
\tablehead{\colhead{x} & \colhead{y} & \colhead{Pearson's $r$} & \colhead{$p$-value} & \colhead{$\alpha$} & \colhead{$\beta$} \\
\colhead{} & \colhead{} & \colhead{(1)} & \colhead{(2)} & \colhead{(3)} & \colhead{(4)}
}
\startdata
\Lfifteen & \rv       & 0.75 & $3.75\times10^{-4}$ & $-26.25^{+ 6.02}_{- 6.03}$ & $0.60 ^{+0.13}_{-0.13}$ \\
\Lfifteen & \weighty  & 0.48 & $1.24\times10^{-2}$ & $- 4.64^{+ 2.95}_{- 2.91}$ & $0.17 ^{+0.06}_{-0.07}$ \\
\Lfifteen & \tdyn     & 0.50 & $3.67\times10^{-2}$ & $-16.58^{+ 7.40}_{- 7.30}$ & $0.52 ^{+0.16}_{-0.16}$ \\
\Lfifteen & \dotEkin  & 0.25 & $3.50\times10^{-1}$ & $ 39.61^{+18.25}_{-18.26}$ & $0.06 ^{+0.40}_{-0.40}$ \\
\Lfifteen & \dotEST   & 0.78 & $1.53\times10^{-4}$ & $-34.53^{+13.97}_{-13.91}$ & $1.76 ^{+0.31}_{-0.31}$ \\
\enddata
\tablecomments{
Statistics of the relation between the outflow properties and the mid-IR luminosity \Lfifteen. The outflow properties and their units are as described in Tab \ref{tab:measurements} and Sec. \ref{sec:energetics:definition}. 
Column 1 and 2: The Pearson's $r$ correlation coefficient between $\log(x)$ and $\log(y)$ and its $p$-value, calculated using only the valid measurements (solid circles in Fig. \ref{fig:kimlum_regress}) not the upper-/lower-limits. Except for \dotEkin{}, the outflow properties are positively correlated with the luminosity in a statistically significant sense ($p$-value $<$ 0.05). 
Column 3 and 4: The posterior mean and the 68\% confidence interval of the parameters of the power-law model $\log(y)=\alpha+\beta \log(x)$. All the measurements and upper/lower-limits (solid circles and triangles in Fig. \ref{fig:kimlum_regress}) are used with error bars taken into account. 
}
\label{tab:regress}
\end{deluxetable*}

\clearpage

{\bf Appendices: }

\setcounter{figure}{0}
\renewcommand{\thefigure}{A\arabic{figure}}
%
%
%
\begin{figure}[H]
    \centering
    \includegraphics[scale=0.48]{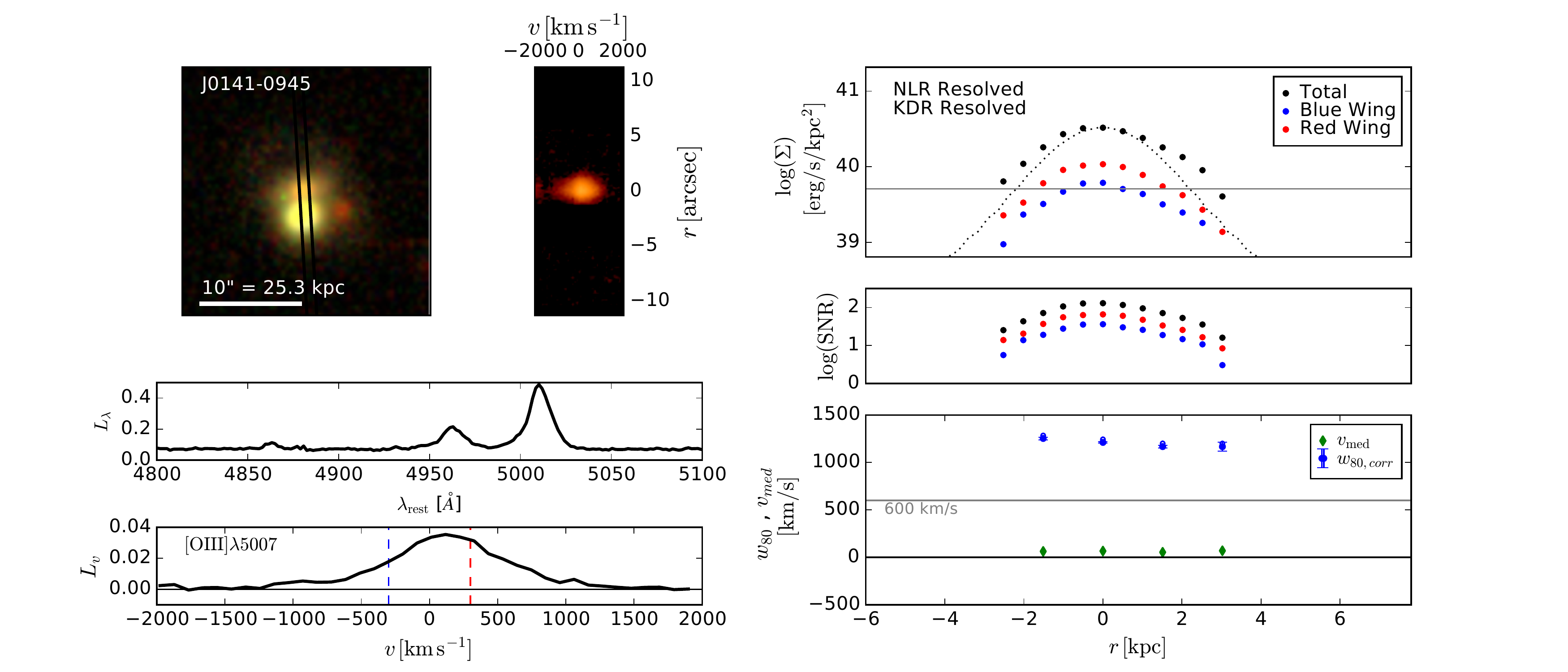}
    \caption{Same as Fig. \ref{fig:spec2100} but for SDSS J0141$-$0945. 
     }
    \label{fig:spec2200}
\end{figure}

\vspace{-20pt}

\begin{figure}[H]
    \centering
    \includegraphics[scale=0.48]{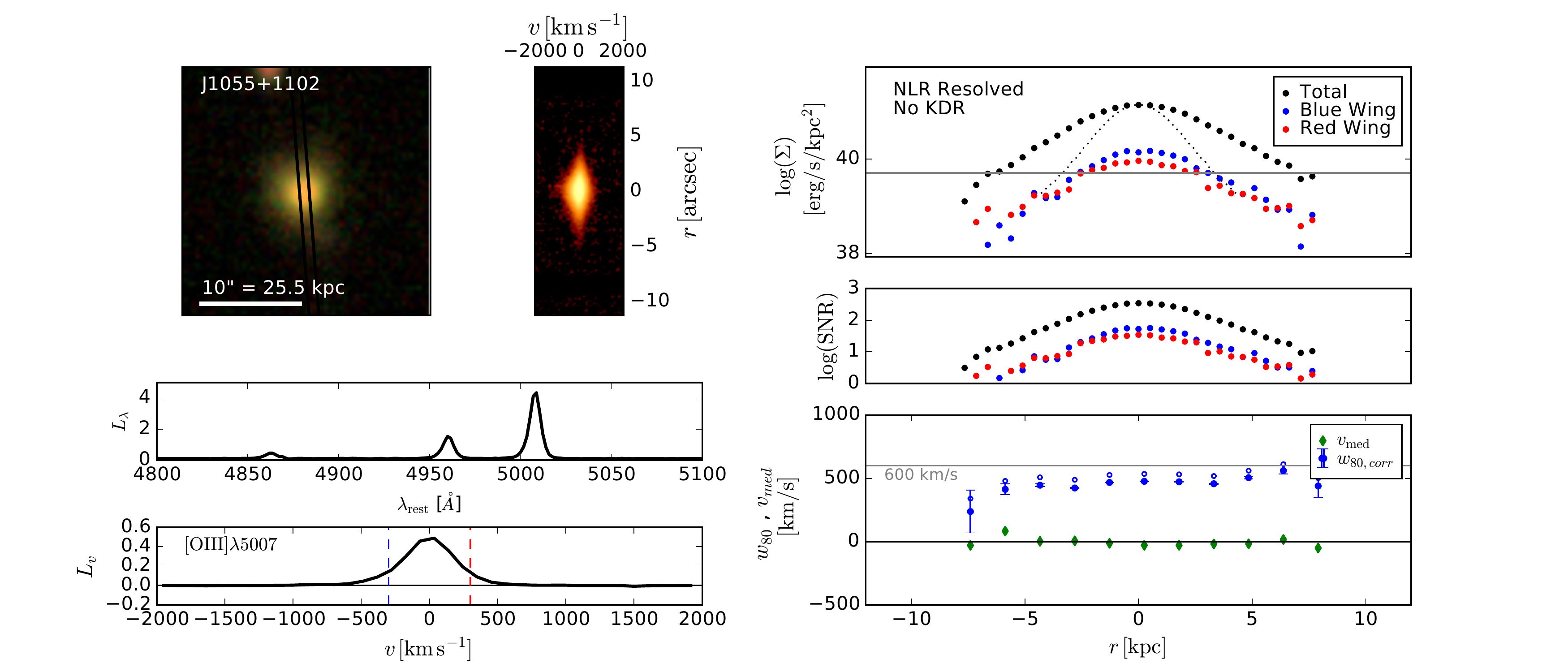}
    \caption{Same as Fig. \ref{fig:spec2100} but for SDSS J1055+1102.
	 }
    \label{fig:spec2102}
\end{figure}

\vspace{-20pt}

\begin{figure}[H]
    \centering
    \includegraphics[scale=0.48]{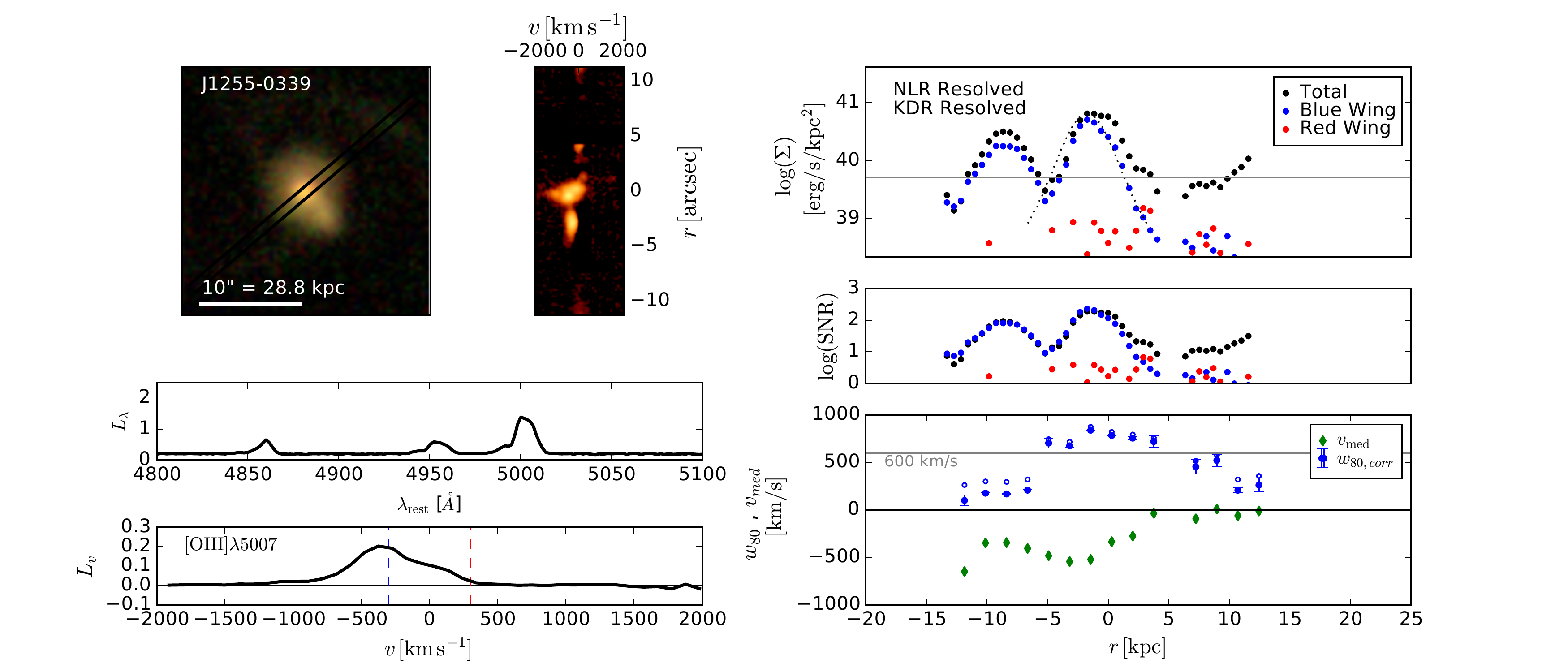}
    \caption{Same as Fig. \ref{fig:spec2100} but for SDSS J1255$-$0339.
     }
    \label{fig:spec2105}
\end{figure}

\vspace{-20pt}

\begin{figure}[H]
    \centering
    \includegraphics[scale=0.48]{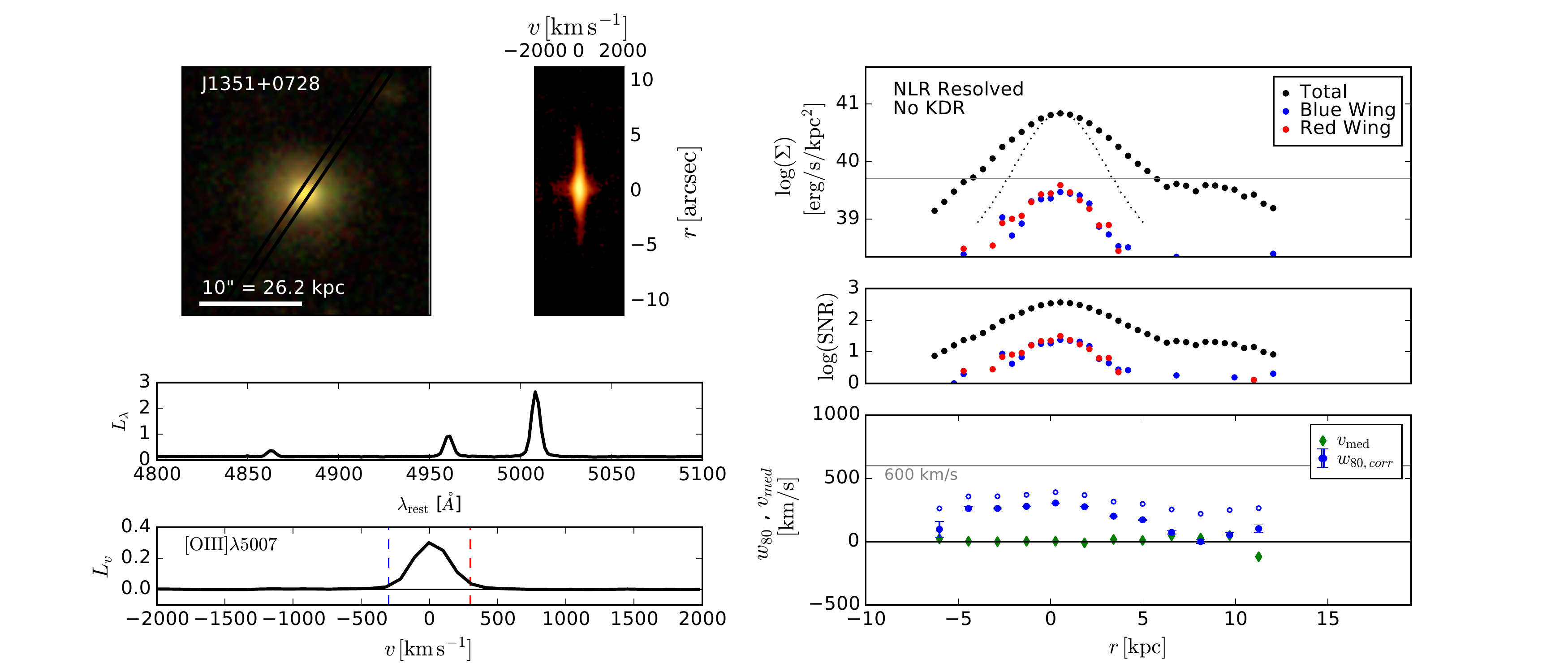}
    \caption{Same as Fig. \ref{fig:spec2100} but for SDSS J1351+0728. 
     }
    \label{fig:spec2106}
\end{figure}

\vspace{-20pt}

\begin{figure}[H]
    \centering
    \includegraphics[scale=0.48]{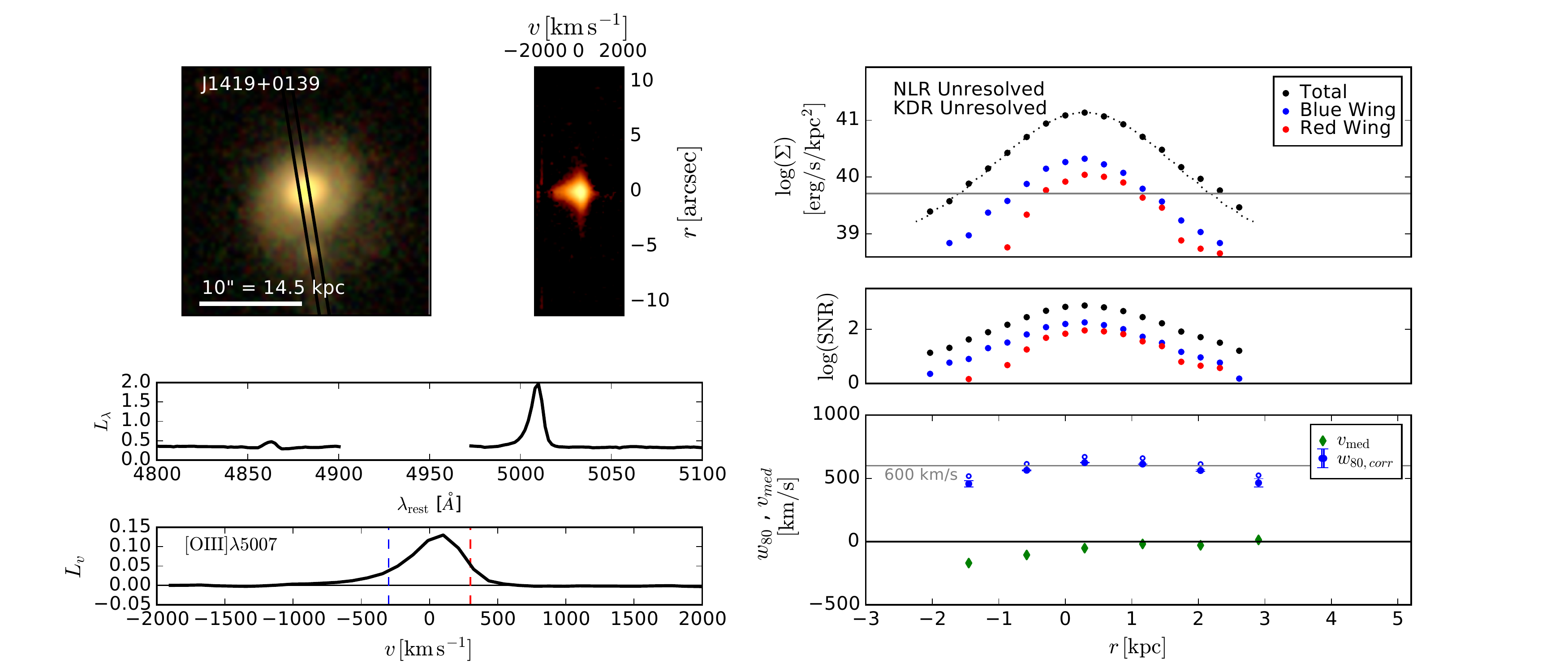}
    \caption{Same as Fig. \ref{fig:spec2100} but for SDSS J1419+0139. 
     }
    \label{fig:spec2107}
\end{figure}

\vspace{-20pt}

\begin{figure}[H]
    \centering
    \includegraphics[scale=0.48]{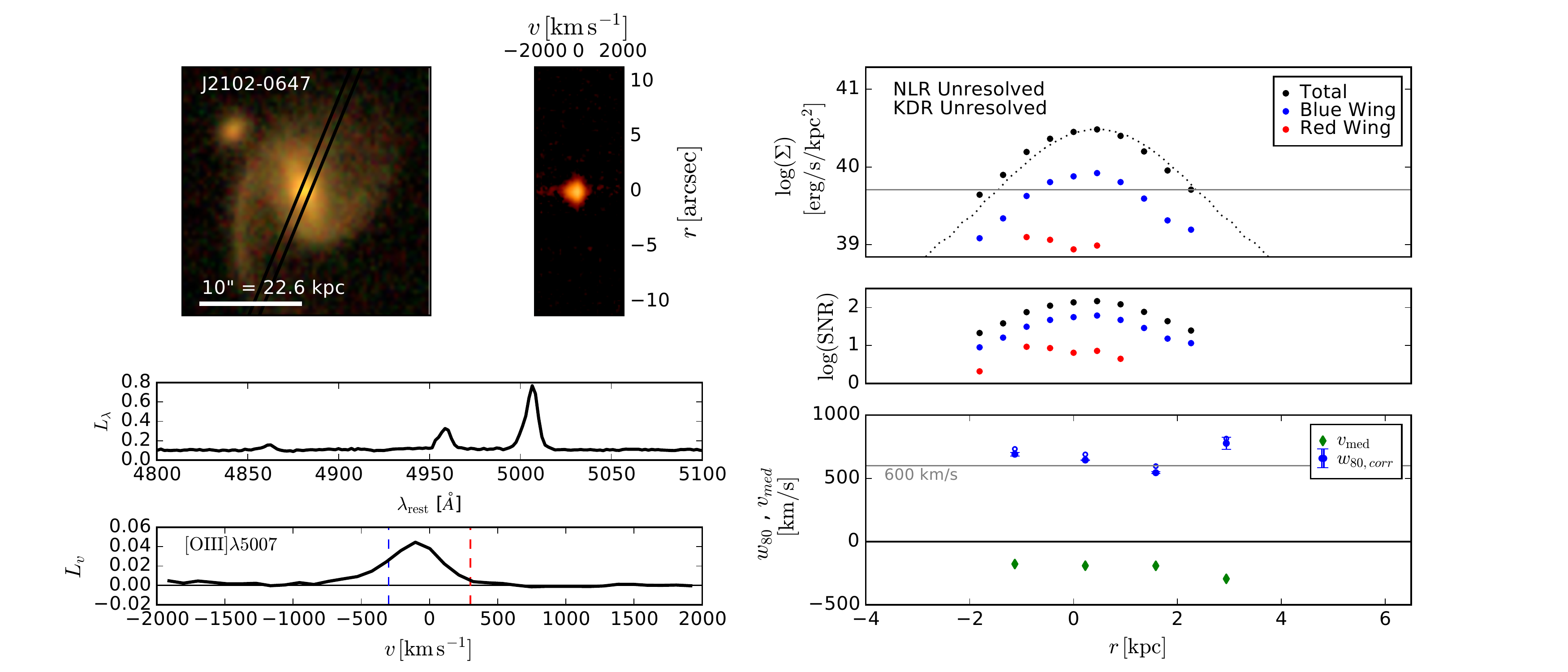}
    \caption{Same as Fig. \ref{fig:spec2100} but for SDSS J2102$-$0647. 
     }
    \label{fig:spec2337}
\end{figure}

\vspace{-20pt}

\begin{figure}[H]
    \centering
    \includegraphics[scale=0.48]{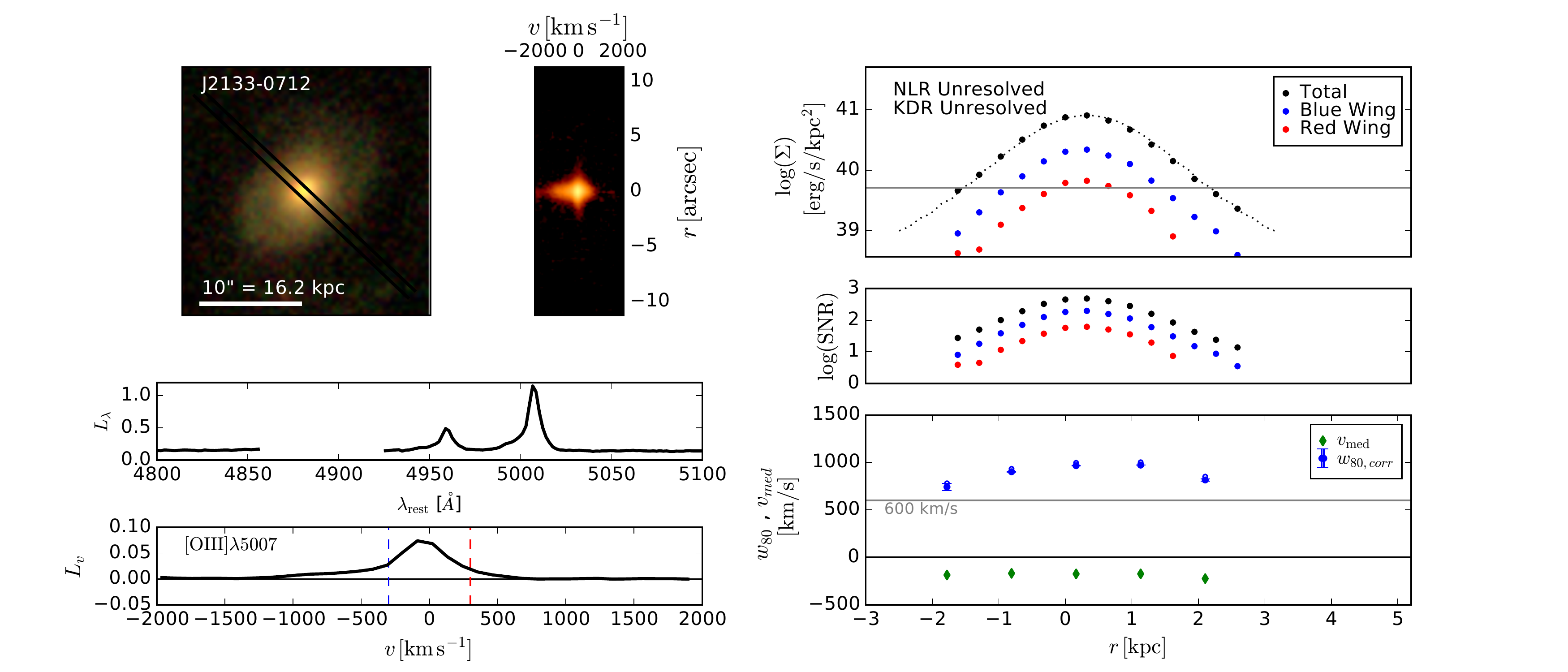}
    \caption{Same as Fig. \ref{fig:spec2100} but for SDSS J2133$-$0712.The \hbeta{} line is not observed as it falls in the chip gap. 
     }
    \label{fig:spec2215}
\end{figure}

\vspace{-20pt} 

\begin{figure}[H]
    \centering
    \includegraphics[scale=0.48]{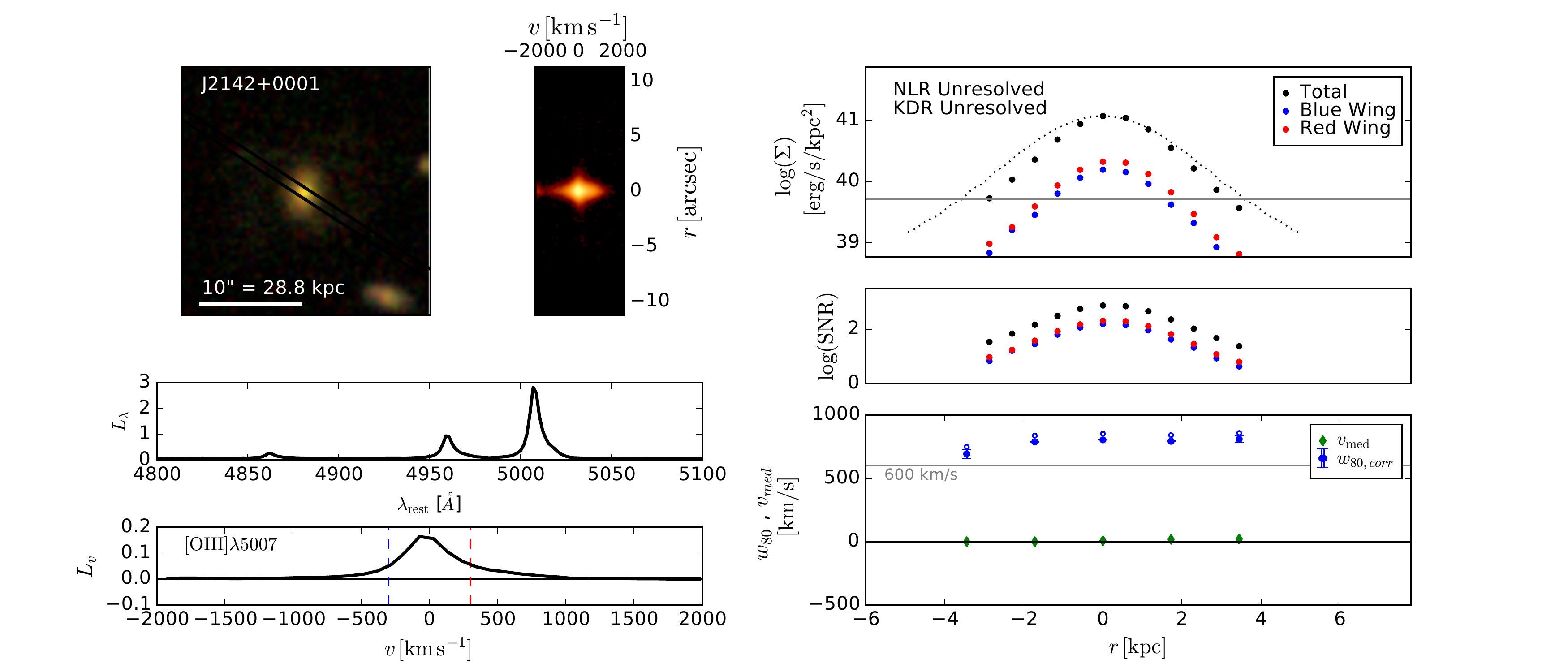}
    \caption{Same as Fig. \ref{fig:spec2100} but for SDSS J2142+0001. 
     }
    \label{fig:spec2108}
\end{figure}

\vspace{-20pt} 

\begin{figure}[H]
    \centering
    \includegraphics[scale=0.48]{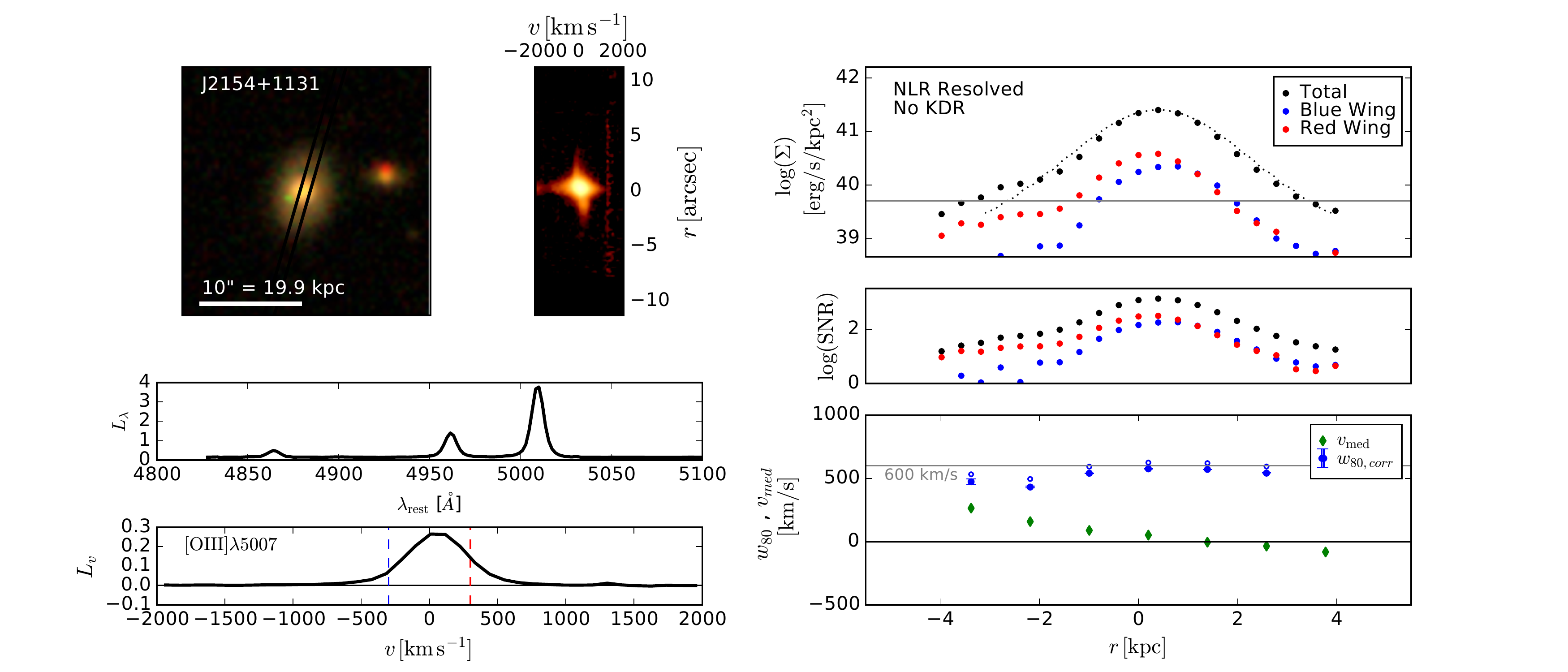}
    \caption{Same as Fig. \ref{fig:spec2100} but for SDSS J2154+1131. 
     }
    \label{fig:spec3001}
\end{figure}

\vspace{-10pt}

\begin{figure}[H]
    \centering
    \includegraphics[scale=0.48]{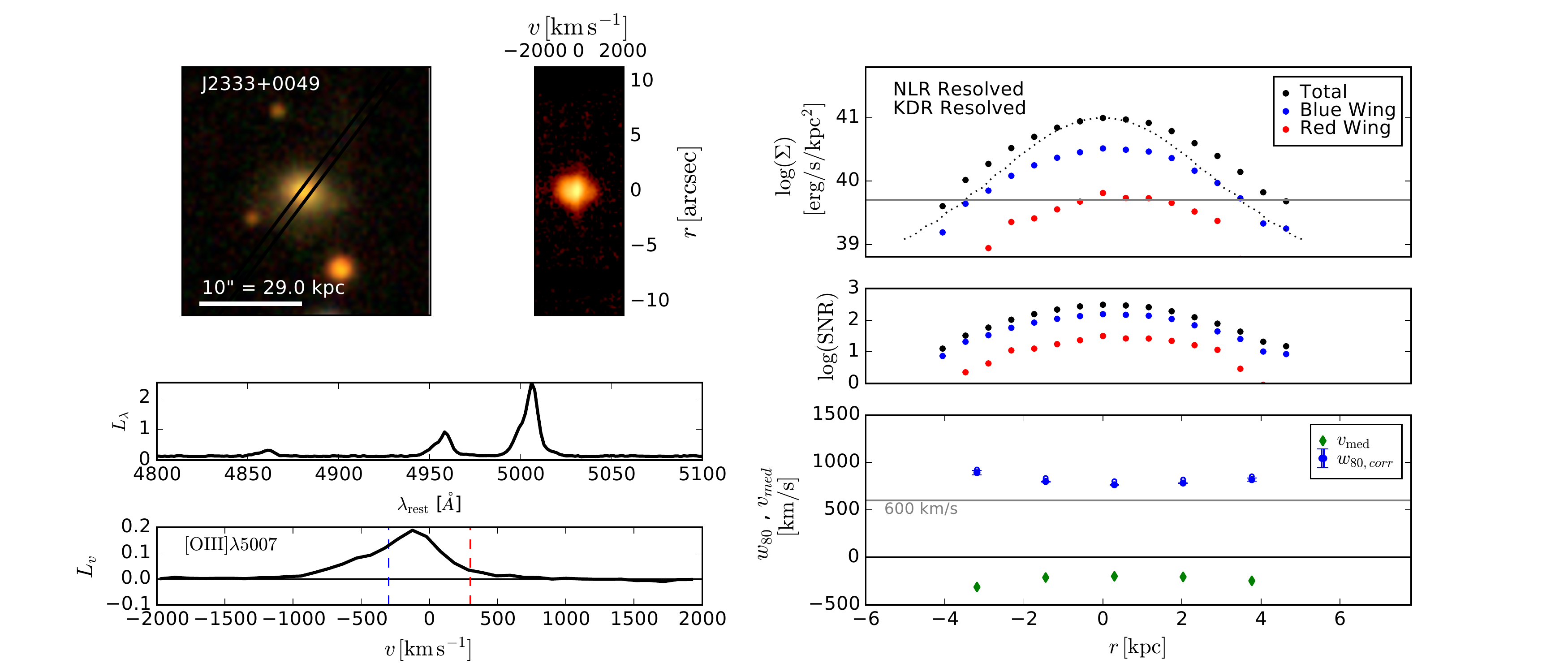}
    \caption{Same as Fig. \ref{fig:spec2100} but for SDSS J2333+0049. 
     }
    \label{fig:spec3005}
\end{figure}

\clearpage

\setcounter{figure}{0}
\renewcommand{\thefigure}{B\arabic{figure}}
%
%
%
\begin{figure}[H]
    \centering
    \includegraphics[scale=0.45]{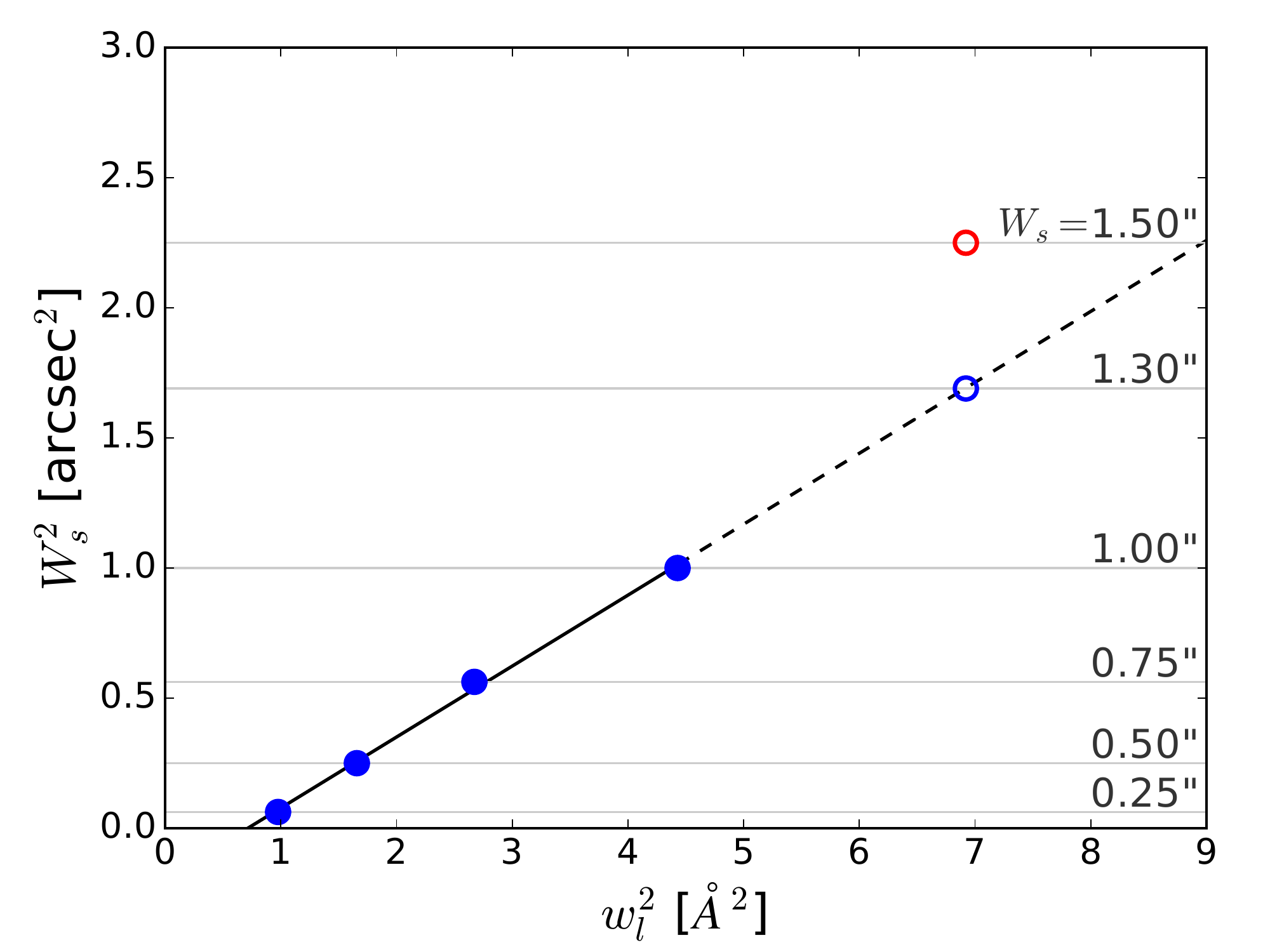}
    \caption{To examine the slit widths of the Magellan IMACS Centerfield Slit-viewing Spectroscopy, we plot the slit widths $W_s$ against the observed arc line widths $w_l$ through those slits. The four slits of widths 0.25\arcsec{}, 0.50\arcsec{}, 0.75\arcsec{}, and 1.0\arcsec{} (solid blue dots) follow the relation $w_l^2 = w_0^2 + r W_s^2$ (black solid line) as expected, whereas the fifth slit would lie on the same relation only if its width were 1.3\arcsec{} (empty blue dot) instead of 1.5\arcsec{} (empty red dot). 
        }
    \label{fig:slitwidths}
\end{figure}

\clearpage

\setcounter{figure}{0}
\renewcommand{\thefigure}{C\arabic{figure}}
%
%
%
\begin{figure*}[h]
	\hbox{
	\includegraphics[scale=0.45]{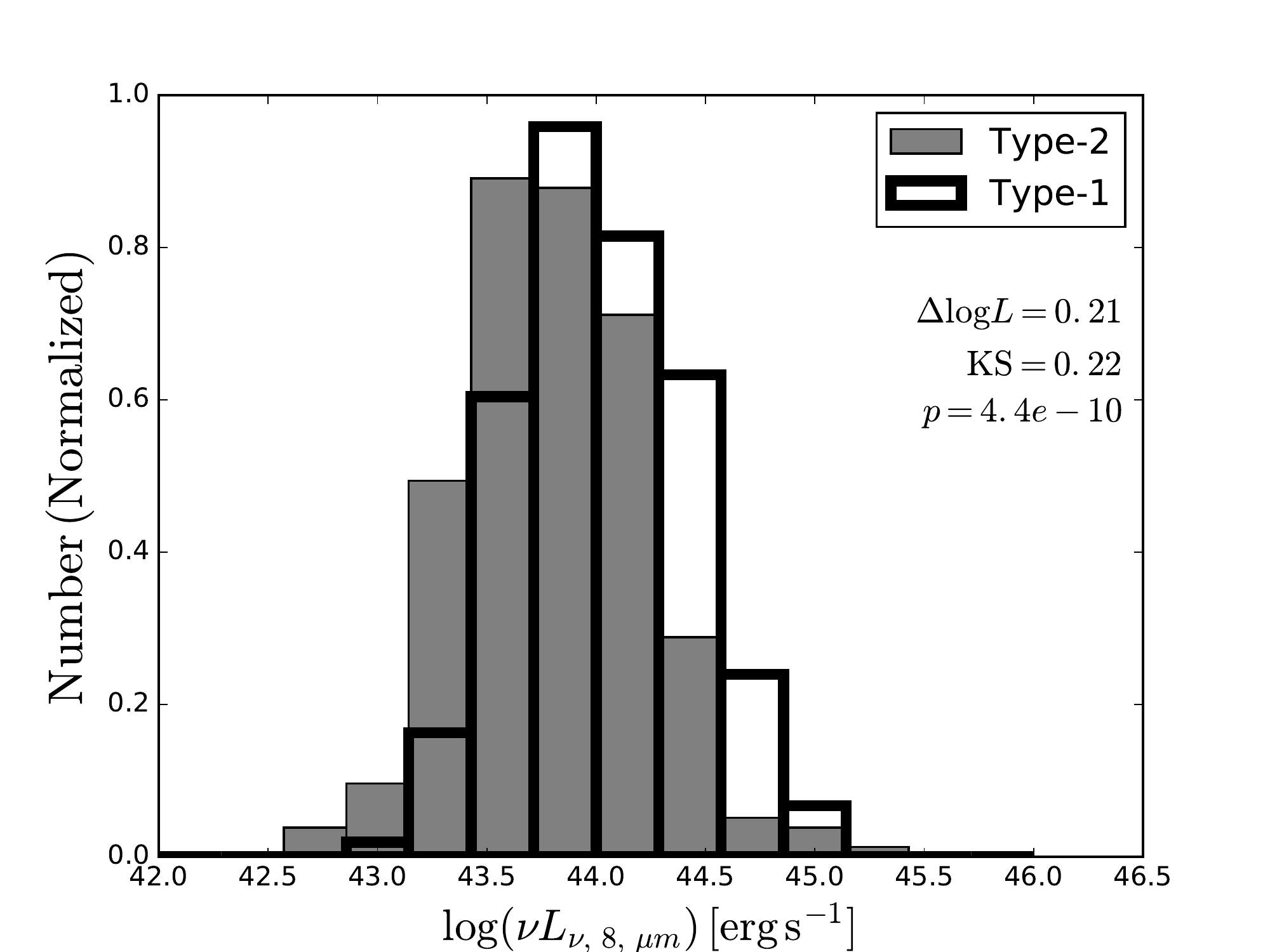}
	\includegraphics[scale=0.45]{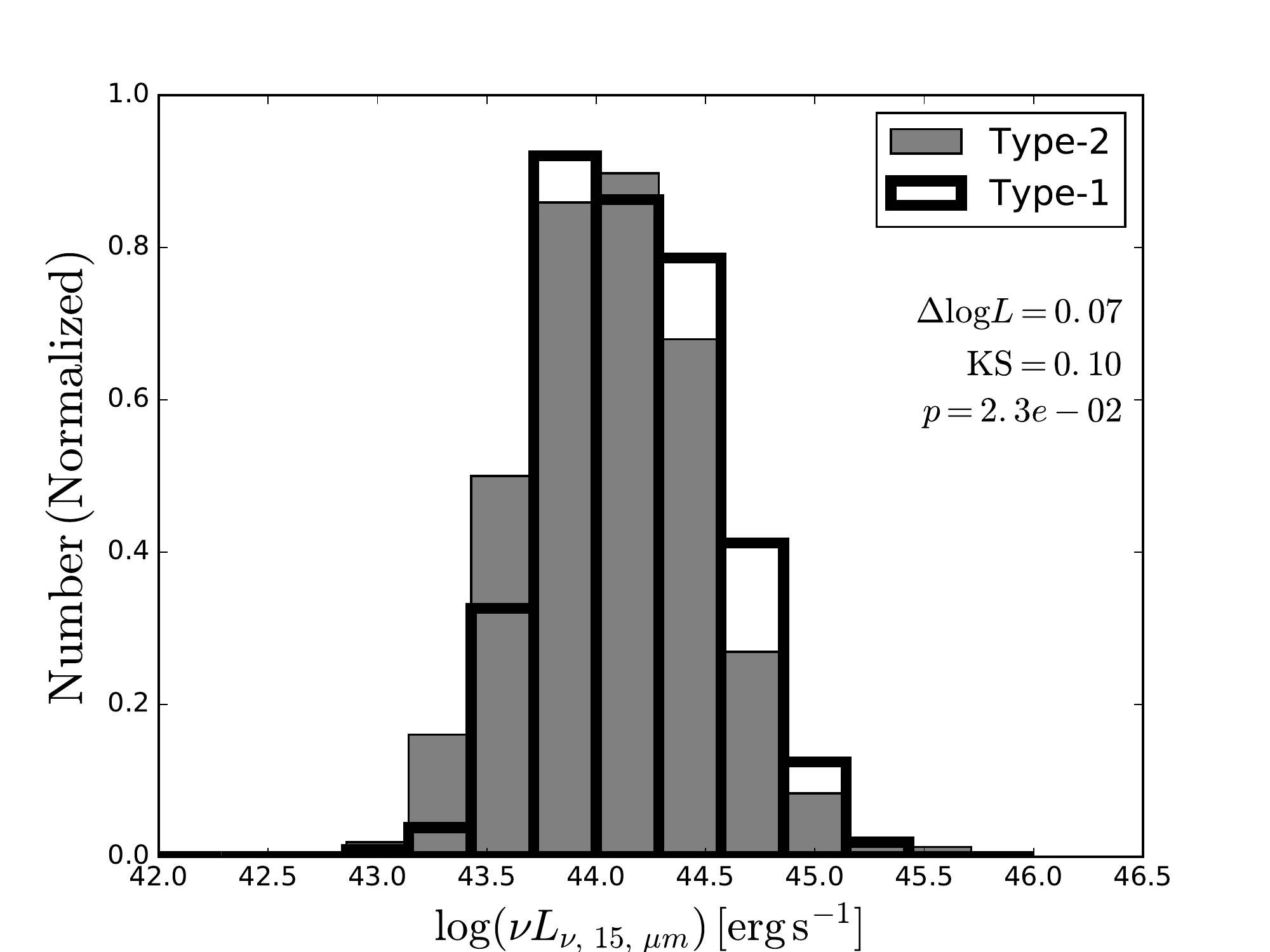}
	}
	\hbox{
    \includegraphics[scale=0.45]{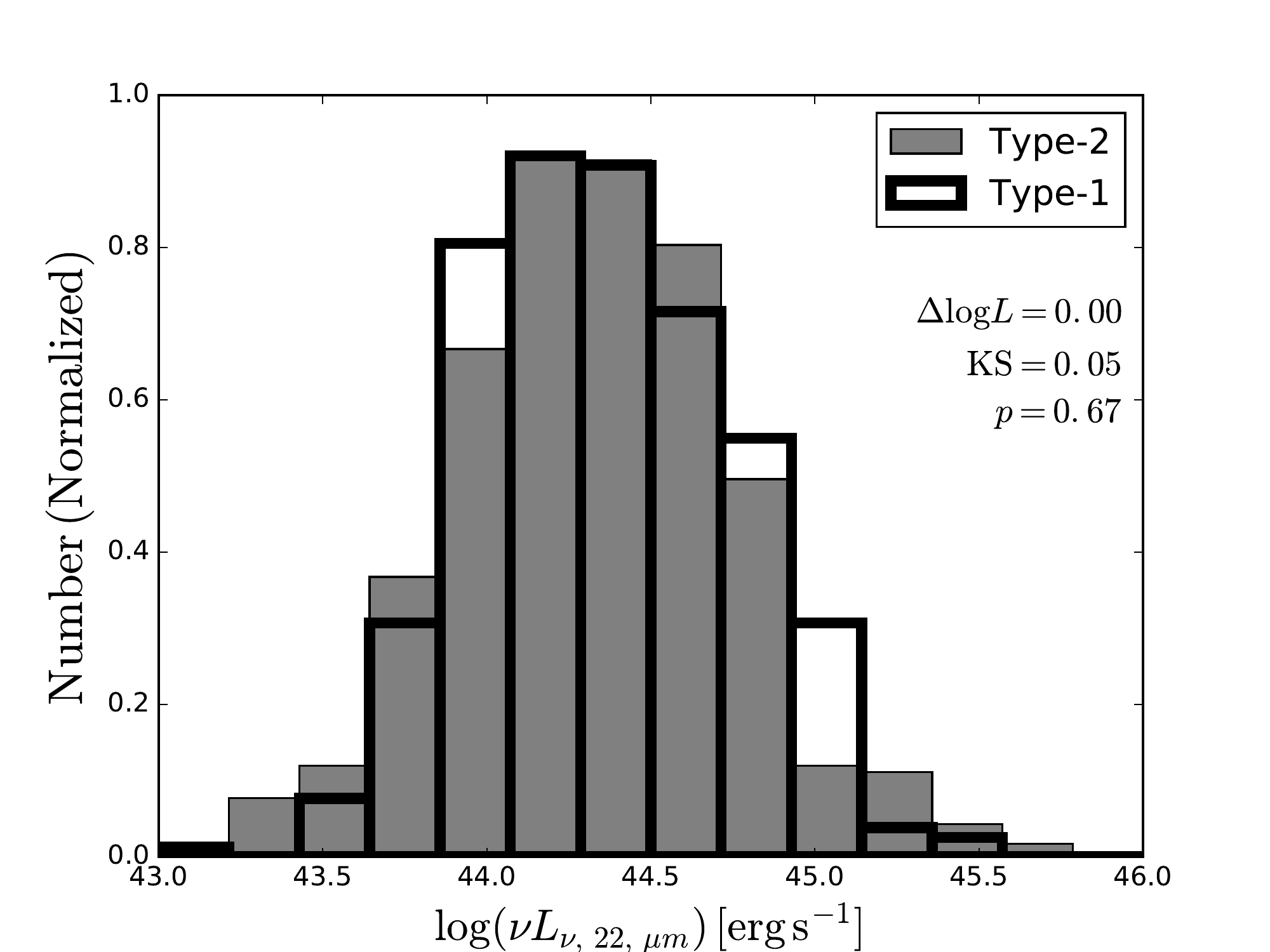}
	\includegraphics[scale=0.45]{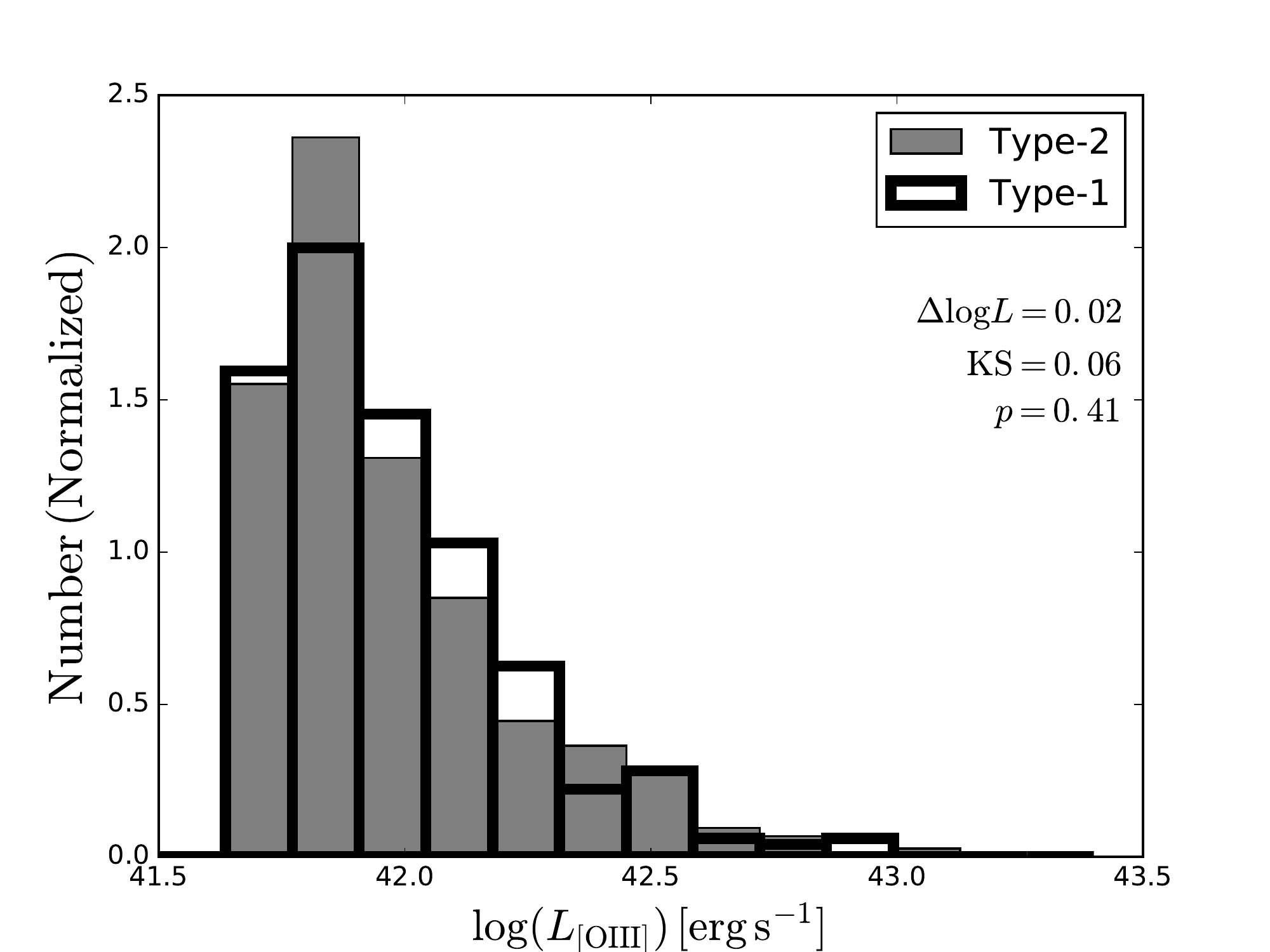}
	}	
    \caption{
    Comparison of the \emph{WISE} mid-IR luminosities between type 1 (outlined histogram) and type 2 (shaded gray histogram) AGN at three rest-frame wavelengths -- 8 \micron{} ({\it upper-left}), 15 \micron{} ({\it upper-right}), and 22 \micron{} ({\it lower-left}). 
    The sample comprises spectroscopically selected luminous AGN from \citet{Mullaney2013} that have high \oiii{} luminosity of \loiii{}$>5\times10^{41}$ \ergs{}. The type 1 and type 2 AGN have similar \loiii{} distributions with a KS-test p-value of 0.41 ({\it lower-right}). 
    The type 1 AGN have higher 8 \micron{} luminosities than type 2 AGN by 0.21 dex on average, which is statistically significant with a KS-test $p$-value of $4\times10^{-10}$. Such a discrepancy becomes less significant towards longer wavelengths as shown in the 15 and 22 \micron{} panels, see Appendix \ref{sec:append:WISE}. 
 	}
    \label{fig:KStest}
\end{figure*}

%
%
%
\begin{figure*}[h]
	\hbox{
	\includegraphics[scale=0.45]{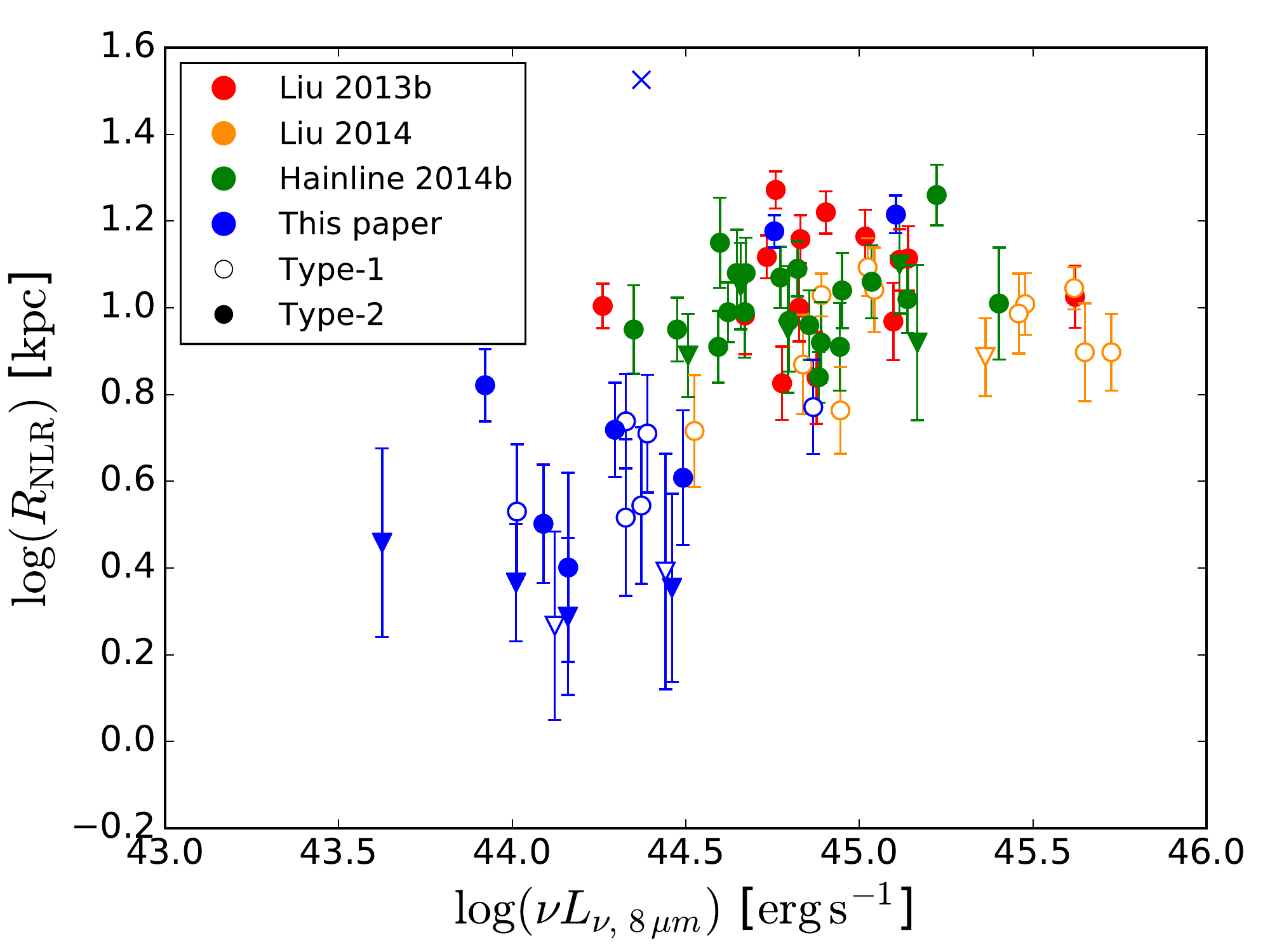}
	\includegraphics[scale=0.45]{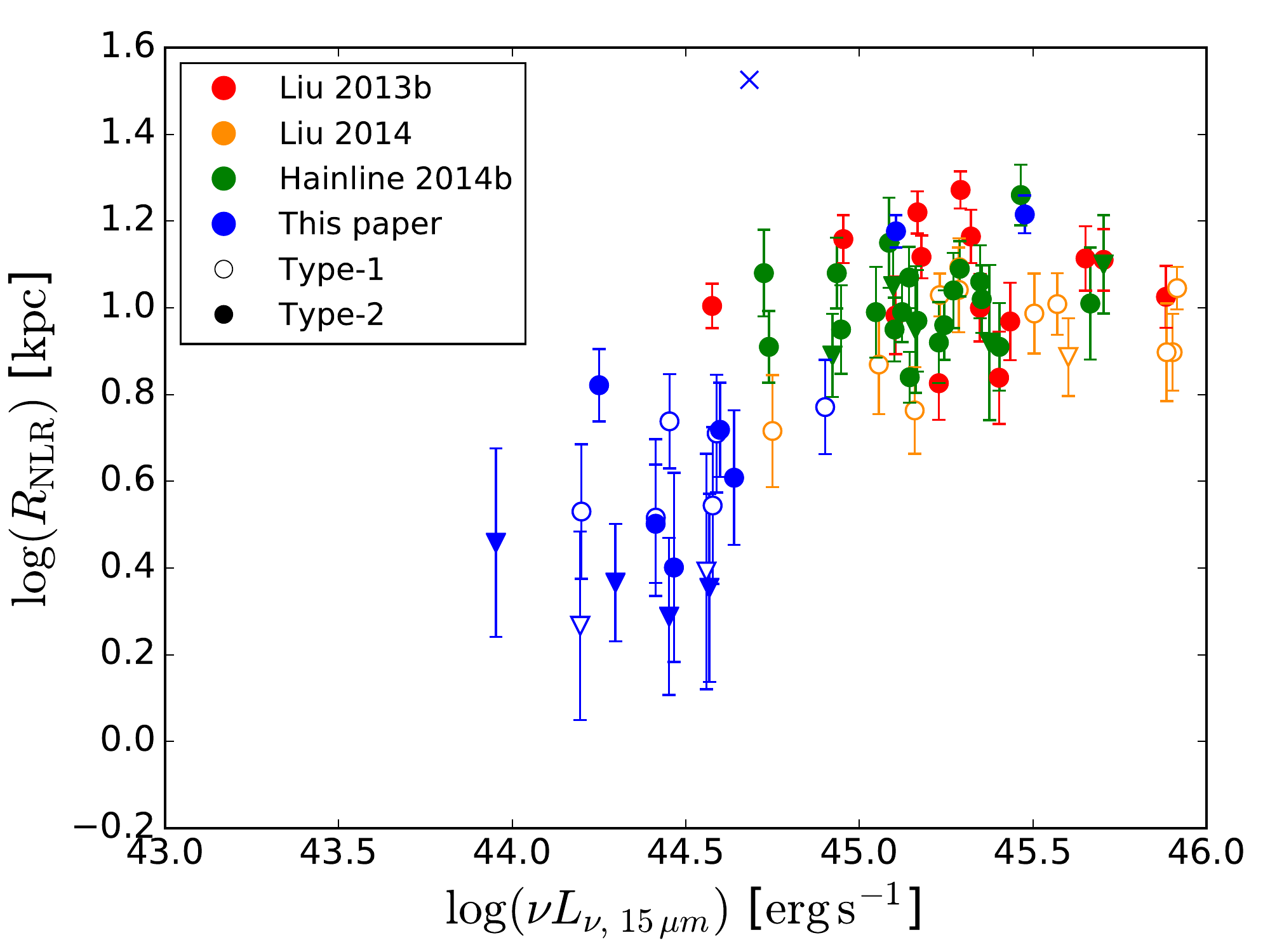}
	}
    \caption{
    The narrow line region size \riso{} and mid-IR luminosity relations based on  8 \micron{} ({\it left}) and 15 \micron{} ({\it right}). 
    The type 1 (empty symbols) and type 2 (solid symbols) AGN follow different size luminosity relations, as type 1 AGN tend to have higher mid-IR luminosities, especially at 8 \micron{}, resulting in an apparent flattening of the relation at the high luminosity end. 
    Four samples are used: this paper (blue), \citet{Liu2013} (red), \citet{Liu2014a} (yellow), and \citet{Hainline2014d} (green). Circles and triangles are size measurements and upper-limits. The blue cross is the object with ionized tidal tails SDSS J1255$-$0339. 
 	}
    \label{fig:sizelum_whichWISE}
\end{figure*}

\clearpage

\setcounter{figure}{0}
\renewcommand{\thefigure}{D\arabic{figure}}
%
%
\begin{figure*}[h]
	\hbox{
	\includegraphics[scale=0.45]{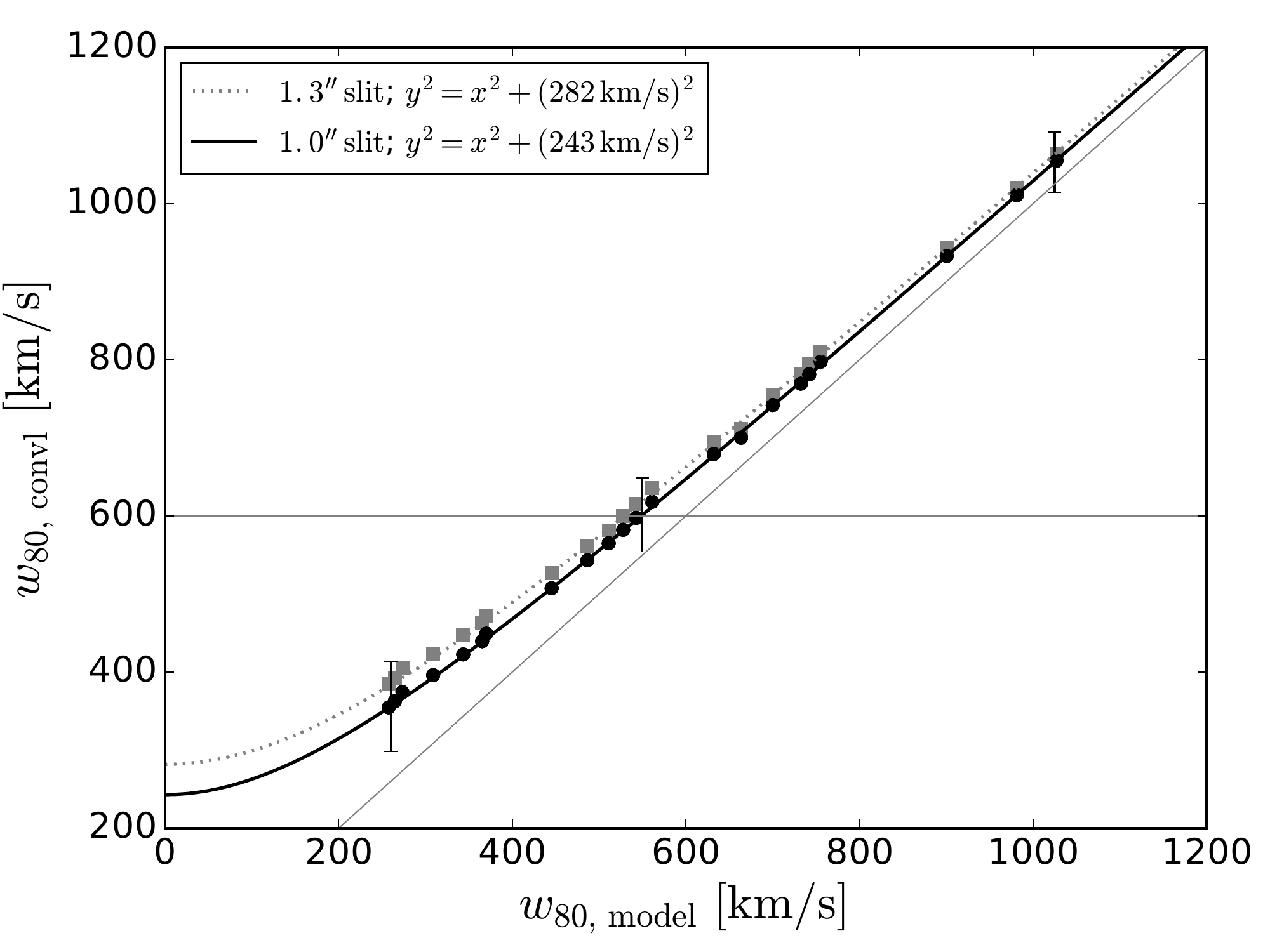}
	\includegraphics[scale=0.45]{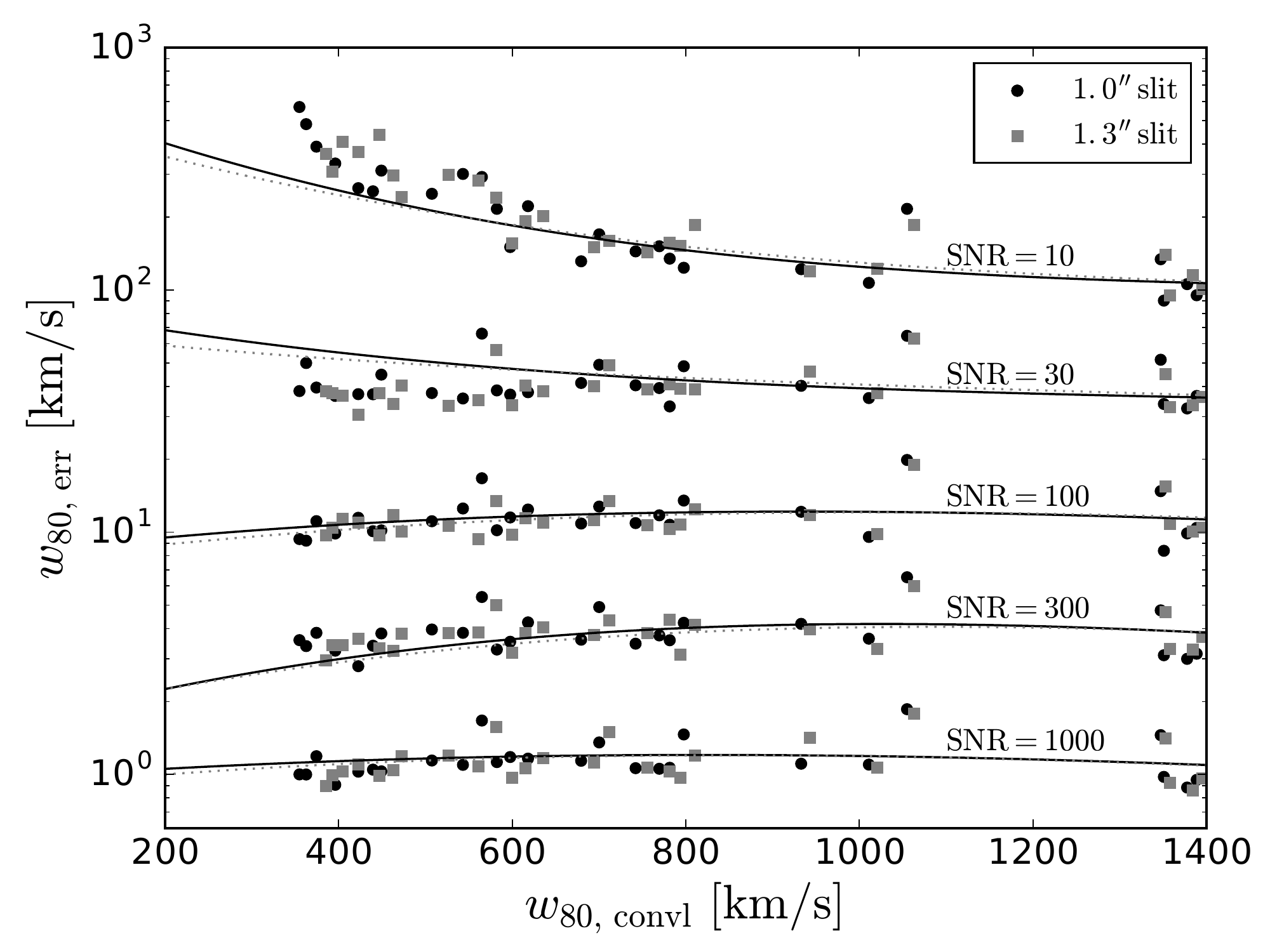}
	}
    \caption{
    Simulated uncertainties in \weighty{} due to the PSF ({\it left}) and the noise ({\it right}) as described in Appendix \ref{sec:append:sim_w80}. 
    Both effects are about 10\% for a typical line of \weighty{} $=$ 600 \kms{} and a signal-to-noise ratio of 30. {\it Left}: the relation between the \weighty{} of the PSF convolved line ($w_{80, \mathrm{convl}}$) and that of the original model ($w_{80, \mathrm{model}}$). Each point corresponds to one simulated line profile and the relation is well fitted by a quadratic mean function. Three representative error bars are shown for a signal-to-noise ratio of 30. {\it Right}: The simulated errors (biases and random uncertainties) on \weighty{} due to the noise as a function of $w_{80, \mathrm{convl}}$ and the signal-to-noise ratio. The lines show the best fit polynomial function. 
    The 1\farcs{}0 slit (black circles and black solid lines) and the 1\farcs{}3 slit (gray squares and gray dotted lines) give similar results. 
    }
    \label{fig:sim_w80}
\end{figure*}

\clearpage

\setcounter{figure}{0}
\renewcommand{\thefigure}{E\arabic{figure}}
%
%
\begin{figure*}[h]
	\vbox{
    \centering
	\includegraphics[scale=0.40]{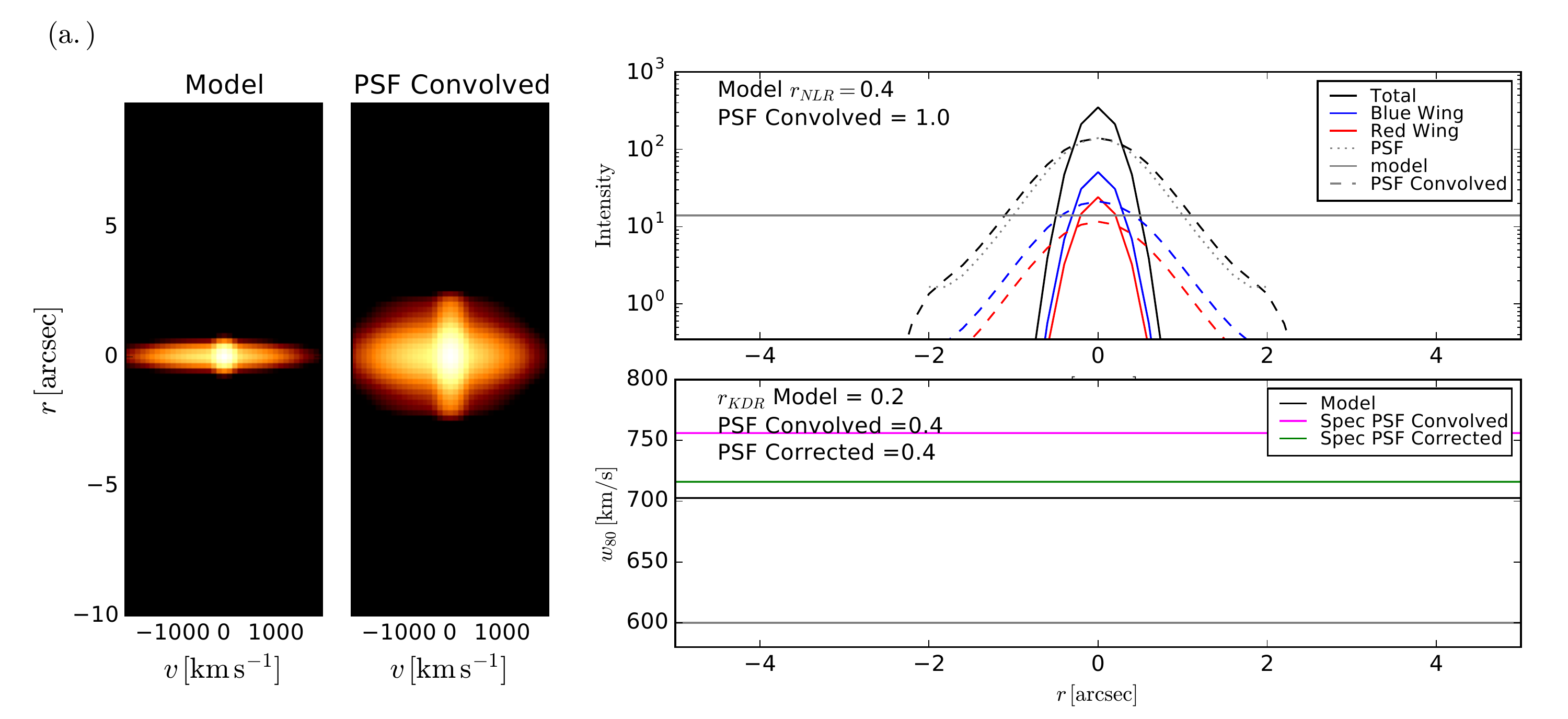}
    \includegraphics[scale=0.40]{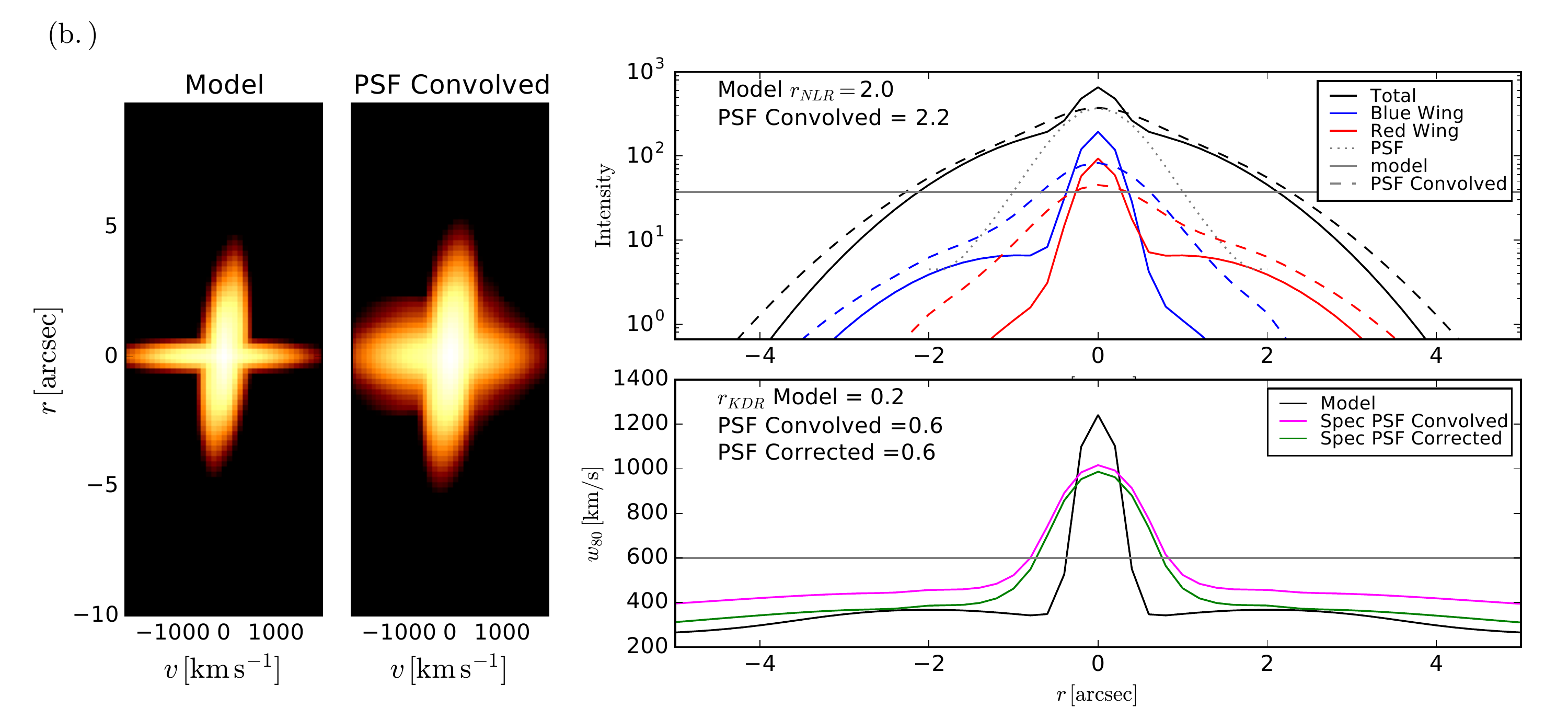}
    \includegraphics[scale=0.40]{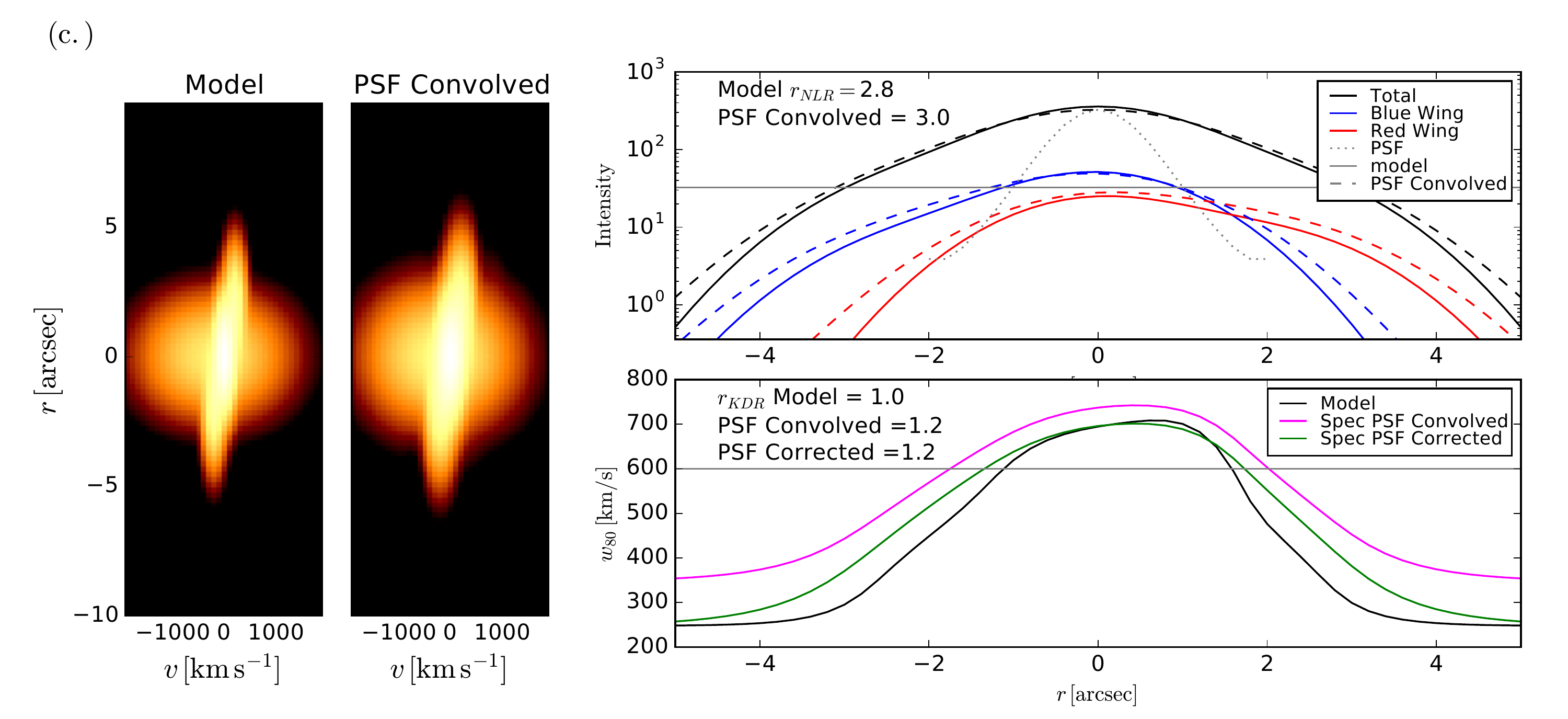}
	}
    \caption{
    Four examples of the 2-D spectrum ($pv$-diagram) simulations to estimate the size biases due to the PSF, described in Appendix \ref{sec:append:sim_pv}. 
    In each of these examples, the left panels show the original model and PSF convolved (spatial and spectral) 2-D spectrum. The upper right panel shows the surface brightness profiles of the total \oiiil{} line, the blue, and the red wings ($|v|>$ 300 \kms{}) in arbitrary units. The lower right panel shows the \weighty{} profiles before (black) and after (magenta) the convolution, as well as after the spectral PSF correction (green, see Appendix \ref{sec:append:sim_w80}). 
    (a.) The case where both the broad and the narrow components are compact ($\sigma = $ 0\farcs2). The surface brightness profiles are consistent with the PSF and the \weighty{} profiles are flat. Both \riso{} and \rv{} are overestimated and should be treated as upper-limits. 
    (b.) Compact broad component plus extended and rotating narrow components. The core of the surface brightness profiles of the high velocity gas (blue and red) are consistent with the PSF but the rotating narrow components contribute to the extended tails of the surface brightness profiles. 
    (c.) Both the broad and the narrow (rotating) components are spatially resolved and the narrow components are more extended than the broad one. The sizes are overestimated by a small amount. 
    (d.) The broad and the narrow components are extended and of the same size. The high \weighty{} is propagated to large radii by the PSF and \rv{} would be arbitrarily overestimated if there were no surface brightness requirements. 
    The spatial PSF is taken from a star observed with a seeing of 1\arcsec. The spectral PSF is measured from the 1\farcs0 slit. 
    }
    \label{fig:sim_pv}
\end{figure*}

\setcounter{figure}{0}
\renewcommand{\thefigure}{E\arabic{figure}  (Cont.)}

\begin{figure*}[h]
    \vbox{
    \centering
    \includegraphics[scale=0.40]{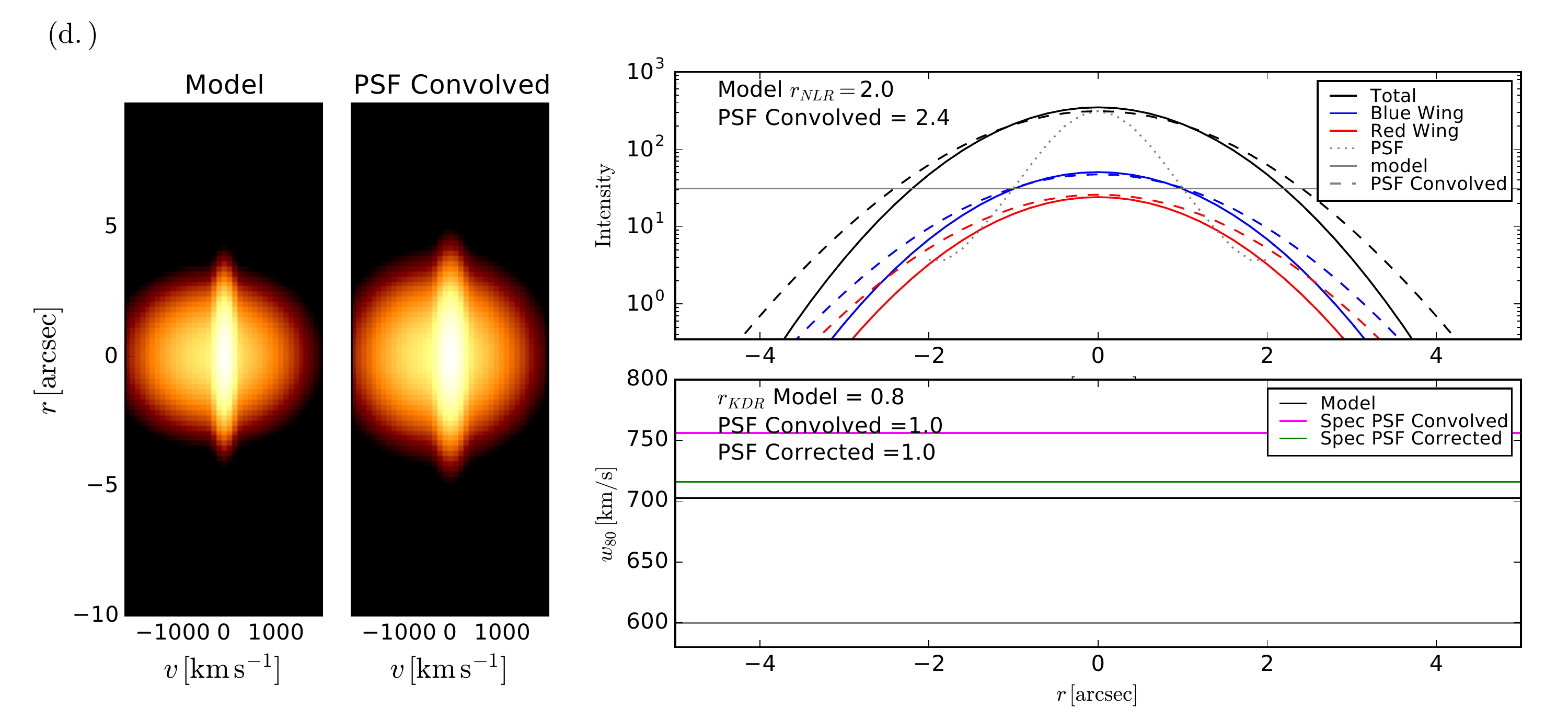}
    }
    \caption{Figure continued. 
    }
\end{figure*}

\clearpage

\end{document}